\documentclass[preprints,article,accept,moreauthors,pdftex]{Definitions/mdpi}
\firstpage{1} 
\pubvolume{1}
\issuenum{1}
\articlenumber{0}
\pubyear{2021}
\copyrightyear{2020}
\datereceived{} 
\dateaccepted{} 
\datepublished{} 
\hreflink{https://doi.org/} % If needed use \linebreak
\pdfoutput=1

\newcommand{\Del}{\Delta}

\newcommand{\al}{\alpha}
\newcommand{\ba}{\beta}
\newcommand{\Lam}{\Lambda}
\newcommand{\lam}{\lambda}

\newcommand{\ga}{\gamma}
\newcommand{\vth}{\vartheta}
\newcommand{\vsa}{\varsigma}
\newcommand{\vphi}{\varphi}

\newcommand{\sech}{\mbox{ sech}}
\newcommand{\bea}{\begin{eqnarray}}
\newcommand{\eea}{\end{eqnarray}}
\newcommand{\bes}{\begin{subequations}}
\newcommand{\ees}{\end{subequations}}
\newcommand{\ds}{\displaystyle}
\Title{Nondegenerate bright solitons in coupled nonlinear Schr\"{o}dinger systems: Recent developments on optical vector solitons}

\TitleCitation{Nondegenerate optical vector solitons in certain coupled nonlinear Schr\"{o}dinger systems}

\Author{S Stalin $^{1,\dagger}$\orcidA{},  R Ramakrishnan $^{1,\dagger}$ and M Lakshmanan $^{1,\dagger}$*\orcidB{}}

\AuthorNames{Stalin, S.; Ramakrishnan, R.; Lakshmanan, M.}

\AuthorCitation{Stalin, S.; Ramakrishnan, R.; Lakshmanan, M.}
\address{%
	$^{1}$ \quad Department of Nonlinear Dynamics, School of Physics, Bharathidasan University, Tiruchirappalli - 620024, Tamil Nadu, India}

\corres{Correspondence: lakshman.cnld@gmail.com; lakshman@cnld.bdu.ac.in}

\firstnote{These authors contributed equally to this work.}
\abstract{Nonlinear dynamics of an optical pulse or a beam continue to be one of the active areas of research in the field of optical solitons. Especially, in multi-mode fibers or fiber arrays and photorefractive materials,  the vector solitons display rich nonlinear phenomena. Due to their fascinating and intriguing  novel properties, the theory of optical vector solitons has been developed considerably both from theoretical and experimental points of view leading to soliton based promising  potential applications. Mathematically, the dynamics of vector solitons can be understood from the framework of coupled nonlinear Schr\"{o}dinger (CNLS) family of equations. In the recent past, many types of vector solitons have been identified  both in the integrable and non-integrable CNLS framework. 	In this article, we review some of the recent progress in understanding the dynamics of the so called nondegenerate vector bright solitons in nonlinear optics, where the fundamental soliton can have more than one propagation constant. We address this theme by considering the integrable two coupled nonlinear Schr\"{o}dinger family of equations, namely Manakov system, mixed 2-CNLS system (or focusing-defocusing CNLS system), coherently coupled nonlinear Schr\"{o}dinger (CCNLS) system, generalized coupled nonlinear Schr\"{o}dinger (GCNLS) system and two-component long-wave short-wave resonance interaction (LSRI) system. In these models, we discuss the existence of  
	nondegenerate vector solitons and their associated novel multi-hump geometrical profile nature by deriving their analytical forms through the Hirota bilinear method.  Then we reveal  the novel collision properties of the nondegenerate solitons in the  Manakov system as an example. The asymptotic  analysis shows that the nondegenerate solitons, in general, undergo three types of elastic collisions without any energy redistribution among the modes. Further, we show that the energy sharing collision exhibiting vector solitons arises as a special case of the newly reported nondegenerate vector solitons.  Finally, we point out the possible further developments in this subject and potential applications.  }


\begin{document}
	%%%%%%%%%%%%%%%%%%%%%%%%%%%%%%%%%%%%%%%%%%
	\setcounter{section}{0} %% Remove this when starting to work on the template.
	%\section{How to Use this Template}
	%%%%%%%%%%%%%%%%%%%%%%%%%%%%%%%%%%%%%%%
	\section{Introduction}
	Solitons are stable localized nonlinear wave packets which can propagate without distortion over long distances. After the discovery of solitons in the numerical experiments on Fermi-Pasta-Ulam-Tsingou anharmonic lattice problem \cite{zabusky-prl,dauxois-physicstoday}, the field of solitons and related nonlinear phenomena flourished and advanced by the invaluable discoveries in nonlinear optics. The concept of solitons is not only reserved for nonlinear optics, but it ubiquitously appears in many branches of physics, including hydrodynamics, Bose-Einstein condensates, plasma physics, particle physics, and even astrophysics apart from the mathematical interest in the theory of integrable nonlinear partial differential equations. In optics, in general, an optical pulse or a beam has a natural property to spread while it propagates in a linear medium because the Fourier components of the pulse or the beam start to travel with distinct velocities. The spreading occurs in the temporal domain because of the material dispersion while in the spatial domain it is due to diffraction. In some cases, the spreading takes place due to the combined effects of dispersion and diffraction. However, a stable localized wave packet forms when this linear effect is balanced by the nonlinear response of the medium. Such a stable light wave envelope is known as the optical soliton. Optical soliton can be further classified as (i) spatial soliton, (ii) temporal soliton and (iii) spatio-temporal soliton depending on the nature of formation mechanism \cite{kivshar-book1}. The evolution of optical soliton, whether it is a spatial or temporal one, in $(1+1)$-dimensional setting is described by the ubiquitous nonlinear Schr\"{o}dinger (NLS) equation. For instance, the dimensionless NLS equation, derived from the Maxwell's equations under slowly varying envelope approximation, for the optical field propagation in a single mode optical fiber turns out to be \cite{agrawal-book2} 
	
	\begin{equation} 
	iq_z-\text{sgn(K'')}q_{tt}+2|q|^2q=0,~ K''=\big(\frac{\partial ^2K}{\partial \omega^2}\big)_{\omega=\omega_0}=\frac{1}{v_g^2}.\label{nls}
	\end{equation}
	In the temporal soliton case, where the soliton evolution is confined along the optical fiber, $q(z,t)$ is the complex wave amplitude and the independent variables $z$ and $t$ denote normalized distance along the fiber and retarded time, respectively. Also $q_z=\frac{\partial q}{\partial z}$ and $q_{tt}=\frac{\partial^2 q}{\partial t^2}$. Here, the sign of the group velocity dispersion (GVD) or simply the coefficient of the second derivative in time, in Eq. (\ref{nls}), characterizes the nature of the fiber dispersion. If $K''<0$, then the dispersion is anomalous whereas the dispersion is normal for $K''>0$. The nonlinearity in Eq. (\ref{nls}) arises due to the self phase modulation (SPM), where the intensity of light induces a change in the refractive index of the medium $\Delta n(I)=n_0(\omega)+n_2|E|^2=n_0+n_2I$, where $n_0$ refers to the linear  refractive index and $n_2$ is the nonlinear refractive index of the medium due to Kerr effect, which gives rise to an intensity-dependent phase modulation. On the other hand, the spatial soliton is a self-trapped optical beam that guides itself by inducing a waveguide during the stable propagation in a photorefractive medium without diffraction. Here, the diffraction is exactly balanced by the nonlinearly induced self-focusing effect.  In this context, the independent variables, $z$ and $t$ in Eq. (\ref{nls}), correspond to transverse spatial coordinates. Since this review will focus on the theoretical aspects of vector bright solitons of certain coupled integrable field models that emerge in optical fiber systems, the readers can find a detailed discussion on the development and advancement of both spatial and spatio-temporal solitons in the interesting review articles by Chen et al \cite{chen-repprogphys} and by Malomed et al \cite{malomed-joptb}, respectively.
	
	In 1973, Hasegawa and Tappert theoretically demonstrated  that the lossless fibers can admit bright soliton structure, which exhibits an intensity maximum in the time domain when the GVD regime is anomalous \cite{hasegawa-apl1}. They have also  shown that the dark soliton, with the intensity minimum or dip on a constant wave background field, arises in the normal GVD regime \cite{hasegawa-apl2}. After this theoretical work, in 1980, Mollenauer and his coworkers succeeded experimentally in observing the optical soliton in a fiber \cite{mollenauer-prl}. These discoveries clearly demonstrated how an abstract mathematical concept can turn into a practical use. Both these theoretical and experimental works have opened up a new possibility of using the ultra-short optical pulses in long distance communication applications \cite{mollenauer-book3}. On the other hand, the mathematical interest in understanding the analytical structure of the underlying integrable models intensified after the NLS equation was solved by Zakharov and Shabat through a more sophisticated inverse scattering transform (IST) method \cite{zakharov-jetp}, developed earlier by Gardner et al for the celebrated Korteweg-deVries equation \cite{gardner-prl}. Now, it is well known that the NLS equation (\ref{nls}) is a completely integrable infinite dimensional Hamiltonian system having special mathematical properties like an infinite number of conserved quantities and Lax pair \cite{ablowitz-book4}. We note that in \cite{zakharov-jetp} the authors had derived a double-pole solution, which has recently received attention in the theory of rogue-waves for describing the Peregrine breather on the zero background field of the NLS equation \cite{chabchoub-arxiv}, by considering the merging of two simple poles in the complex plane. The interesting fact of the temporal bright solitons of the scalar NLS equation is that they exhibit particle-like elastic collision.
	
	Apart from the above fundamental aspects, in 1983, Gordon had predicted that when two or more light pulses propagate in a nonlinear optical fiber, they exert forces, either attractive or repulsive, on their neighbors \cite{gordon-ol}.  This has been experimentally verified by Mitschke and Mollenauer in \cite{mitschke-ol}. Such a study brought out a special kind of soliton state, namely bound soliton state or soliton molecule \cite{stratmann-prl}. A soliton molecule is a bound soliton state that can be formed when two solitons persist at a stable equilibrium separation distance, where the interaction force is zero among the individuals. Such a stable equilibrium  manifests as this bound state structure,  reminiscent of a diatomic molecule in chemical physics.   The binding force arises between the constituents of the soliton composite due to the Kerr nonlinearity \cite{gordon-ol,mitschke-ol} and the detailed mechanism can be found in Ref. \cite{hause-pra}. This special kind of soliton state has been extensively studied in non-dispersion managed fibers \cite{akhmediev-optcomm,malomed-springer,malomed-pre,khawaja-pre,grelu-joptb,tang-pra,akhmediev-joptsocb}. Recently,  the existence of soliton molecules in dispersion-managed fiber \cite{stratmann-prl} and their usefulness in optical telecommunications with enhanced data carrying capacity have been pointed out \cite{melchert-prl,Rohrmann-scirep}. However, in order to elevate the transmission capacity of the optical telecommunication systems, it is necessary to consider multichannel bit-parallel wavelength fiber networks and wavelength division multiplexing schemes, where the light pulses propagate in multi channels simultaneously. In fact, practically even in a single mode fiber the bending and strains or birefringence induce two orthogonal polarization modes.  To pursue this kind of practical applications, one has to essentially understand the problem of the intermodal interaction of solitons. Therefore the contribution of the interaction of copropagating modes must be taken into account.  In fact, there is no surprise other than the standard elastic collision of the bright solitons in single mode optical fibers. In contrast to this, the bright soliton structure  in two mode fibers or in a single mode fiber with birefringence property or even in multimode fibers display rich propagation and collisional properties. Due to these fascinating features and intriguing collision dynamics, vector solitons are receiving intense attention among  researchers. Apart from the several interesting properties, vector solitons have also been found in a variety of applications, including soliton based optical computing \cite{jakubowski-pre,steiglitz-pre}, multi-level optical communication with enhanced bit-rate transmission \cite{akhmediev-chaos}, soliton based signal processing systems \cite{mitschke-springer} and so on.
	
	Vector solitons are fascinating nonlinear objects in which a given soliton is split among two or more components. In other words, a vector soliton with two or more polarization components coupled together maintains its shape during propagation. Such vector solitons are also named as multicolour solitons. The dynamics of vector solitons is usually understandable within the framework of coupled nonlinear Schr\"{o}dinger (CNLS) equations. In general, the CNLS equations are non-integrable and they become integrable for specific choices of parameters \cite{Radhakrishnan-csf}. Therefore, mathematically vector solitons arise as solutions of the CNLS equations. Like in the scalar NLS equation, the optical vector solitons are formed due to an exact balance between the dispersion/diffraction and the self-phase modulation and cross-phase modulation. This interesting class of optical solitons has been first predicted by Manakov in 1974, where he has derived the one-soliton solution and made an asymptotic analysis for the two-soliton solution through the IST method, by introducing a set of two CNLS equations for the nonlinear interaction of the two orthogonally polarized optical waves in birefringent fibers \cite{manakov-jetp}. The Manakov system is essentially an integrable system, where the strength of the nonlinear interactions within and between the components are equal. Vector optical solitary wave propagation in birefringent fiber has been first theoretically studied by Menyuk by considering a pair of non-integrable CNLS equations \cite{menyuk-ieee}. Very interestingly one of the present authors (ML) along with Radhakrishnan and Hietarinta theoretically predicted that the bright solitons of the  Manakov model exhibit novel energy sharing collision through intensity redistribution \cite{Radhakrishnan-pre}. They have explicitly demonstrated this fascinating collision scenario by analysing the two bright soliton solution derived through the Hirota bilinear method. Then this study has been extended to $N$-CNLS equations by Kanna and Lakshmanan in \cite{kanna-prl}, where there is a lot of exciting possibilities for the occurrence of energy redistribution among the $N$-modes that have been reported. This theoretical development was experimentally verified in \cite{anastassiou-prl,kang-prl,rand-prl} and subsequently, it gave rise to the possibility of constructing all optical logic gates \cite{jakubowski-pre,steiglitz-pre,soljacic-prl,kanna-pre,vijayajayanthi-preR}. The  discovery of photorefractive solitons \cite{segev-prl1,duree-prl,segev-prl2,christodoulides-josocamb} and the subsequent experimental developments  \cite{christodoulides-apl,chen-ol1,chen-ol2,mitchell-prl} have substantially enriched our knowledge on vector solitons. It is known that a set of $N$-CNLS equations describes the beam propagation in a Kerr-like photorefractive medium \cite{akhmediev-prl,sukhorukov-prl,ankiewicz-pre,krolikowski-pre}. Further, the experimental studies on vector solitons in photorefractive media as well as in dispersive media during the past three decades demand investigation of physical and mathematical aspects of CNLS equations even more rigorously. 
	
	It is very important to point out there exist many types of vector solitons that have been reported so far for both integrable and non-integrable CNLS type equations. For instance, in the non-integrable cases, a temporal light pulse composed of orthogonally polarized components propagate with common group velocity and it is called group velocity-locked soliton \cite{christodoulides-ol-gvls}. On the other hand, if the two polarization components of the soliton are locked in phase, then such vector soliton has been called a phase-locked soliton \cite{akhmediev-jopsocab} whereas for the polarization-locked vector soliton \cite{collings-jopsocab}, the relative phase between the components is locked at $\pm\frac{\pi}{2}$ but across the pulse, the polarization state profile is not uniform. However, that profile is invariant with propagation. Apart from the above, other types of vector solitary waves have been reported in birefringent fibers \cite{tratnik-pra,haelterman-ol,ostrovskaya-ol,yang-physicad} and in saturable nonlinear medium \cite{ostrovskaya-prl,pelinovsky-studaplmath}, where the stability of multi-hump solitons has been reported. In the integrable cases, bright-bright solitons \cite{manakov-jetp,Radhakrishnan-pre,kanna-prl,kanna-pre2006}, bright-dark or dark-bright solitons \cite{vijayajayanthi-pra,sheppard-pre,Radhakrishnan-pre2007,Radhakrishnan-pre2015,feng-jphysa} and dark-dark solitons  \cite{Radhakrishnan-jphysa-1995,ohta-studapplmath} were documented in the context of nonlinear optics and their novel properties in multicomponent BECs have also been investigated considerably \cite{kevrekidis-revphys}. In a photorefractive medium, partially coherent solitons or soliton complexes were identified in the $N$-CNLS system, and their special properties were revealed by Akhmediev and his collaborators in \cite{akhmediev-prl,ankiewicz-pre,sukhorukov-prl,akhmediev-chaos,krolikowski-pre}. Apart from the above, during the last decade, a large volume of work has been dedicated to the temporal optical solitons (both theoretically and experimentally) by considering the fiber lasers, which has been reported as a very useful nonlinear system to study the dynamics and formation of temporal optical solitons \cite{song-applphysrev}. There exist different types of  optical solitons in dissipative systems too and their various properties have been explored in \cite{akhmediev-book5}.  
	
	From the above studies on vector solitons, especially in integrable coupled nonlinear Schr\"{o}dinger models, we have identified that there exists a degeneracy in the structure of the bright solitons as we have explained below in Section 3. That is, the solitons in two-mode fibers or in multi-mode fibers propagate  with identical wave numbers. In order to avoid this degeneracy, we introduce two non-identical propagation constants appropriately in the structure of the fundamental bright solitons of the 2-CNLS equation to start with. Consequently, the degeneracy is removed and it leads to a new class of fundamental bright solitons, namely nondegenerate fundamental vector bright solitons \cite{stalin-prl}. For the first time, we have shown that
	such an inclusion of additional distinct propagation constants brings out a general form of vector bright soliton solution to the several integrable CNLS systems \cite{ramakrishnan-pre,stalin-pla}, namely Manakov system or 2-CNLS system, mixed 2-CNLS system  (with one mode in the anomalous dispersion regime and the other mode in the normal dispersion regime), 2-component coherently coupled NLS system, generalized CNLS system, and 2-component long-wave short-wave resonance interaction system \cite{stalin-pla}. We note that very recently the nondegenerate solitons have also been studied in other contexts as well. For instance, in multi-component BECs \cite{qin-pre} using the Darboux transformation method, in coupled Fokas-Lenells system \cite{zhang-mathphys} and in AB-system \cite{ding-csf} such nondegenerate solitons have been identified. We also note that multi-valley dark nondegenerate soliton has been studied in the context of multicomponent repulsive BECs \cite{qin-arxiv}.	In this paper, we critically review, the existence and their salient novel features of the general form of nondegenerate vector bright solitons in the above class of 2-component nonlinear Schr\"{o}dinger systems. Then we also critically analyse their novel collision properties with the Manakov system as an example. Further, we also discuss in detail the corresponding already known degenerate vector bright solitons and their intriguing collisional properties. Additionally, we also illustrate the multi-hump nature of the nondegenerate fundamental bright solitons in $N$-CNLS system \cite{ramakrishnan-jpa}. 
	
	The outline of this review paper is as follows. In Section 2, we quickly point out the derivation of 2-CNLS equations in the context of multi-mode fibers and introduce the various coupled integrable models and their physical importance. In Section 3, we clearly distinguish how the vector bright soliton reported so far in the literature for the integrable coupled NLS family type equations may be considered as a special case of the fundamental nondegenerate bright soliton solution derived recently by us. In Section 4, we discuss the nondegenerate soliton solutions of the Manakov system and analyse their underlying novel collision dynamics. In this section, we also describe the degenerate soliton solutions and their interesting energy sharing collision apart from mentioning the possible experimental realization and the multi-hump nature of the nondegenerate fundamental bright solitons in the $N$-CNLS system. Then in Section 5, we describe the properties and the existence of nondegenerate fundamental bright soliton of the mixed CNLS system. We also discuss the collision dynamics of the degenerate solitons by pointing out their explicit analytical forms. In Section 6, we discuss the existence of both nondegenerate and degenerate fundamental bright solitons in the coherently coupled NLS system and point out the energy switching collision scenario of degenerate bright solitons.  Further, we illustrate the existence of nondegenerate bright soliton in the generalized coupled nonlinear Schr\"{o}dinger system and point out its degenerate limit in Section 7. Then, in Section 8, we also elucidate the existence of nondegenerate soliton in the two-component (1+1)-dimensional LSRI system. Finally, in Section 9, we summarize the results and provide a possible future outlook.
	%%%%%%%%%%%%%%%%%%%%%%%%%%%%%%%%%%%%%%%%%%%%%%%%%%%
	\section{Derivation of CNLS equations and other integrable CNLS type models }
	In general, the interaction between two or more co-propagating optical modes is governed by the coupled nonlinear Schr\"{o}dinger family of equations. The derivation of one such CNLS equations starts from the Maxwell's equations for electromagnetic wave propagation in a dielectric medium, 
	\begin{equation}
	\nabla^2 \vec{E}-\frac{1}{c^2}\frac{\partial^2 \vec{E}}{\partial t^2}=-\mu_0\frac{\partial^2 \vec{P}}{\partial t^2},
	\end{equation} 
	where $\vec{E}(\vec{r},t)$ is the electric field, $\vec{P}(\vec{r},t)$ is the induced polarization,  $\mu_0$ is the permeability of free space and $c$ is the velocity of light. The induced polarization $\vec{P}(\vec{r},t)$ contains both a linear part and a nonlinear part. That is $\vec{P}(\vec{r},t)=\vec{P}_L(\vec{r},t)+\vec{P}_{NL}(\vec{r},t)$. The linear and nonlinear induced polarizations  can be further written as
	\bes\begin{eqnarray}
	\hspace{-1.5cm}&&\vec{P}_L(\vec{r},t)=\epsilon_0\int_{-\infty}^{+\infty}\chi^{(1)}(t-t')\vec{E}(\vec{r},t')dt',\label{3a}\\
	\hspace{-1.5cm}&&\vec{P}_{NL}(\vec{r},t)=\epsilon_0\int_{-\infty}^{+\infty}\int_{-\infty}^{+\infty}\int_{-\infty}^{+\infty}\chi^{(3)}(t-t_1,t-t_2,t-t_3)\vec{E}(\vec{r},t_1)\vec{E}(\vec{r},t_2)\vec{E}(\vec{r},t_3)dt_1dt_2dt_3.~\label{3b}
	\end{eqnarray}   \ees
	Here, $\epsilon_0$ is the permitivity of the free space and $\chi^{(j)}$ is the $j$th order suceptibility tensor of rank $(j+1)$ \cite{agrawal-book2,lakshmanan-pramana}. For elliptically birefringent fibers, the electric field $\vec{E}(\vec{r},t)$ can be written as
	\begin{eqnarray}
	\vec{E}(\vec{r},t)=\frac{1}{2}\bigg(\hat{e}_1E_1(z,t)+\hat{e}_2E_2(z,t)\bigg)e^{-i\omega_0 t}+c.c.\label{4}	
	\end{eqnarray}
	In the above, the variables $z$ and $t$ denote the direction of propagation and retarded time, respectively and $c.c$ stands for complex conjugation. The orthonormal vectors $\hat{e}_1$ and $\hat{e}_2$  are expressed as,  $\hat{e}_1=\frac{\hat{x}+ir\hat{y}}{\sqrt{1+r^2}}$ and $\hat{e}_2=\frac{r\hat{x}-i\hat{y}}{\sqrt{1+r^2}}$, where $r$ is a measure of the extent of ellipticity and $\hat{x}$ and $\hat{y}$ are unit polarization vectors along $x$ and $y$ directions, respectively. In Eq. (\ref{4}), $E_1$ and $E_2$ are complex amplitudes of the polarization components at frequency $\omega_0$. The nonlinear polarization can be obtained by substituting the expression of the electric field $\vec{E}(\vec{r},t)$ from Eq. (\ref{4}) in Eqs. (\ref{3a})  and (\ref{3b}). The electric-field components are written under slowly varying approximation as 
	\begin{equation}
	E_j(z,t)=F_j(x,y)Q_j(z,t)e^{iK_{0j}z}, ~j=1,2,
	\end{equation}       
	where $F_j(x,y)$ are the fiber distribution function in the transverse directions $x$ and $y$ and $K_{0j}$, $j=1,2$ are the propagation constants for the two modes. By doing so, the following coupled equations are obtained for $Q_j(z,t)$:
	\bes\begin{eqnarray}
	iQ_{1,z}+\frac{i}{v_{g1}}Q_{1,t}-\frac{k''}{2}Q_{1,tt}+\mu(|Q_1|^2+B|Q_2|^2)Q_1=0,\\
	iQ_{2,z}+\frac{i}{v_{g2}}Q_{2,t}-\frac{k''}{2}Q_{2,tt}+\mu(|Q_1|^2+B|Q_2|^2)Q_2=0.
	\end{eqnarray} \ees
	Here, $k''=\big(\frac{\partial ^2k}{\partial \omega^2}\big)_{\omega=\omega_0}$ accounts for the group velocity dispersion, $\mu$ is the nonlinearity coefficient and $v_{g1}$ and $v_{g2}$ are the group velocities of the two co-propagating modes, respectively. The constant $B=\frac{2+2\sin^2\theta}{2+\cos^2\theta}$ is the cross-phase modulation coupling parameter, where $\theta$ is the angle of ellipticity which varies between $0$ and $\frac{\pi}{2}$. Here, we have assumed that the fiber is having a strong birefringent nature. Under three sets of consecutive transformations (detailed derivation  can be found in \cite{lakshmanan-pramana}), we obtain the following dimensionless 2-CNLS equation with the integrability restriction $B=1$ \cite{Radhakrishnan-csf}, which is obtained from the Painlev\'{e} analysis,
	\bes\begin{eqnarray}
	iq_{1,z}+q_{1,tt}+2\mu(|q_1|^2+|q_2|^2)q_1=0,\label{7a}\\
	iq_{2,z}+q_{2,tt}+2\mu(|q_1|^2+|q_2|^2)q_2=0.\label{7b} 
	\end{eqnarray}    \ees
	The above set of CNLS equations constitute the completely integrable system introduced by Manakov to describe the propagation of an intense electromagnetic pulse  in a birefringent fiber \cite{manakov-jetp}. The system (\ref{7a})-(\ref{7b}) is well discussed in nonlinear optics and in other areas of physics. In this review, we also wish to consider another 2-CNLS equation which is a variant of the Manakov system, namely the mixed coupled nonlinear Schr\"{o}dinger system or  Zakharov and Schulman system \cite{zakharov-physicad,kanna-pre2006}. One can write both the mixed CNLS equation and Manakov equation in a unified form as given below:        
	\bea
	&&i q_{j,z}+ q_{j,tt}+2 \left(\sigma_1|q_1|^2+\sigma_2 |q_2|^2\right) q_j =0, \quad ~~j=1,2.
	\label{cnls}
	\eea
	In Eq. (\ref{cnls}), $\sigma_1$ and $\sigma_2$ are the strength of the SPM and cross-phase modulation (XPM) nonlinearities. If $\sigma_1=\sigma_2=+1$, the above equation becomes the Manakov equation (focusing type 2-CNLS equations), where the two optical fields $q_1$ and $q_2$ propagate in the anomalous dispersion regimes \cite{manakov-jetp}, whereas for $\sigma_1=\sigma_2=-1$, they propagate in the normal dispersion regimes or in other words, the resultant model (\ref{cnls}) turns out to be the defocusing Manakov system \cite{Radhakrishnan-jphysa-1995}. For the other choice, $\sigma_1=+1$ and $\sigma_2=-1$, the system (\ref{cnls}) becomes the mixed-CNLS system \cite{kanna-pre2006}, in which the SPM is positive and the XPM is negative in both the modes, where the first mode $q_1$ is propagating in the anomalous dispersion regime while the second mode $q_2$ is propagating in the normal dispersion regime. Both the focusing and defocusing Manakov models also find applications in attractive and repulsive multicomponent BECs \cite{kevrekidis-revphys}. We note that the soliton trapping and daughter wave (shadow) formation have been reported \cite{kaup-pre} using the bright soliton solutions of the Manakov system. Radhakrishnan and Lakshmanan have derived the dark-dark soliton solution \cite{Radhakrishnan-jphysa-1995} and Sheppard and Kivshar have obtained bright-dark soliton solution  \cite{sheppard-pre} to the above system. In the latter case, the authors have pointed out the existence of breathing bound states. Further, it has been shown that the mixed CNLS system models the electromagnetic pulse propagation in isotropic and homogeneous nonlinear left handed materials \cite{lazarides-pre}.  By taking into account the electron-phonon interaction and in the long-wavelength approximation, the mixed-CNLS system can also be obtained as the modified Hubbard model (Lindner-Fedyanin system) \cite{makhankov-pla1,makhankov-pla2,lindner-phys}. The mixed CNLS system is also realized in two species BECs for a suitable choice of interspecies and intraspecies interactions \cite{garcia-pra}. We point out that the IST method and Darboux transformation method have been rigorously developed to obtain the bright-bright, dark-dark and bright-dark soliton solutions of the multicomponent focusing, defocusing and mixed CNLS systems \cite{ablowitz-invprob,prinari-jmp1,biondini-jmp,prinari-studapplmath,biondini-siam,prinari-jmp,biondini-communmathphys,park-pre,degasperis-jpa1,degasperis-jpa2,ling-nonlinearity,ling-cnsns,tsuchida-arxiv}. 
	
	Next, we consider the two-component coherently coupled nonlinear Schr\"{o}dinger equation, which arises due to the coherent effects of the coupling among the copropagating optical fields. In general, an ultrashort pulse propagation in non-ideal weakly birefringent multimode fibers and optical beam propagation in low anisotropic Kerr type nonlinear media are described by the following two-component non-integrable CCNLS system, 
	\cite{kivshar-book1,crosignani-joptsocam,park-pre2};
	\bes\bea
	i q_{1,z}+ \delta q_{1,tt}-\mu q_{1}+ (|q_{1}|^2+\sigma |q_{2}|^2)q_{1}+\lambda q_2^2 q_{1}^*=0,\\
	i q_{2,z}+\delta q_{2,tt}+\mu q_{2}+ (\sigma |q_{1}|^2+|q_{2}|^2)q_2+\lambda q_{1}^2 q_{2}^*=0.
	\eea \label{e1} \ees
	The above equation also appears in isotropic Kerr-type nonlinear gyrotropic medium \cite{akhmediev-optcommun}. In the above $q_1$ and $q_2$ are two coherently coupled orthogonally polarized  modes, $z$ and $t$ are the propagation direction and transverse direction, respectively, $\mu$ is the degree of birefringence, $\sigma$ and  $\lambda$ are the incoherent and coherent coupling parameters, respectively, and $\delta$ is the group velocity dispersion. The nonlinearities arise in Eq. (\ref{e1}) due to SPM ($|q_j|^2q_j$, $j=1,2$), XPM ($\sigma |q_k|^2q_j$, $j,k=1,2$, $j\neq k$) and four-wave mixing effect ( $\lambda q_k^2 q_j^*$, $j,k=1,2$, $j\neq k$). Equation (\ref{e1}) is shown to be integrable for a specific choice of system parameters ($\delta,~\mu,~\sigma$ and $\lambda$) \cite{park-pre2} and soliton solutions were derived by linearly superposing the soliton solutions of the two nonlinear Schr\"{o}dinger equations through a transformation. The corresponding integrable two-component CCNLS system (2-CCNLS system) is 
	\bes \bea
	i q_{1,z}+ q_{1,tt}+ \gamma (|q_{1}|^2+2 |q_{2}|^2)q_{1}-\gamma q_2^2 q_{1}^*=0,\\
	i q_{2,z}+ q_{2,tt}+ \gamma (2 |q_{1}|^2+|q_{2}|^2)q_2-\gamma q_{1}^2 q_{2}^*=0.
	\eea \label{2ccnlseqn} \ees
	Interestingly, Kanna et al \cite{kanna-jpa-2010} have derived the fundamental and two bright soliton solutions of (\ref{2ccnlseqn}) and its multicomponent version \cite{kanna-jpa-2011} by developing a non-standard Hirota bilinearization procedure. They have classified the fundamental bright soliton as incoherently coupled soliton (ICS) and coherently coupled soliton (CCS)  based on a condition on the parameters in the auxiliary function.  A novel double-hump soliton profile arises in these CCNLS systems due to the coherent coupling among the two copropagating optical fields. Further, they have also demonstrated a fascinating energy switching collision during the interaction of ICS and CCS \cite{kanna-jpa-2010,kanna-jpa-2011}. We remark that the  CCNLS type equations are useful in studying the dynamics of solitons in spinor BECs and coherently coupled BECs \cite{kasamatsu-prl,congy-pra,babu-pla}also. A similar type of CCNLS equation has been identified in the context of spinor BEC and is shown to be integrable \cite{ieda-prl,prinari-studinapplmath,li-pra}.
	
	Next, we wish to examine the bright soliton solutions of the general coupled nonlinear Schr\"{o}dinger (GCNLS) system \cite{wang-stud}, namely 
	\bes\begin{eqnarray}
	iq_{1,z}+q_{1,tt}+2(a|q_1|^2+c|q_2|^2+bq_1q_2^*+b^*q_1^*q_2)q_1=0,\label{11a}\\\
	iq_{2,z}+q_{2,tt}+2(a|q_1|^2+c|q_2|^2+bq_1q_2^*+b^*q_1^*q_2)q_2=0\label{11b}. 
	\end{eqnarray}    \ees
	In the above GCNLS equations, $a$ and $c$ account for the strength of the SPM and XPM nonlinearities whereas the complex parameter $b$ in the phase dependent terms, $bq_1q_2^*+b^*q_1^*q_2$, describes the four-wave mixing effect that arises in multichannel communication systems \cite{agrawal-book2}. When $a=c$ and $b=0$ the system (\ref{11a})-(\ref{11b}) reduces to the Manakov system (or Eq. (\ref{cnls}) with $\sigma_1=\sigma_2=+1$). Then, if $a=-c$ and $b=0$ the GCNLS system becomes the mixed-CNLS model.
	This GCNLS system has received considerable attention recently in both mathematical and physical aspects \cite{wang-stud,lu-nonlindyn,vpriya-cnsns,agalarov}. The integrability properties of the system (\ref{11a}) and (\ref{11b}) have been studied in \cite{wang-stud} in which the $N$-soliton solution was obtained through the Riemann-Hilbert method. The GCNLS system is shown to be integrable through Weiss-Tabor-Carnevale (WTC) test \cite{lu-nonlindyn}. In \cite{vpriya-cnsns}, bright and dark-soliton solutions were obtained through the Hirota bilinear method. By relating the GCNLS system with the Manakov and Makhankov vector models using a transformation ($q_1=\psi_1-b^*\psi_2$ and $q_2=a\psi_2$), the authors in \cite{agalarov} have constructed bright-bright, dark-dark and a quasibreather-dark soliton solutions.
	
	Finally, for our investigation, we also wish to take into account the following coupled nonlinear Schr\"{o}dinger type equations, namely the two-component long-wave short-wave resonance interaction system,  
	\begin{eqnarray}
	\hspace{-0.5cm} iS_t^{(1)}+S_{xx}^{(1)}+LS^{(1)}=0,~iS_t^{(2)}+S_{xx}^{(2)}+LS^{(2)}=0,~L_t=\sum_{l=1}^{2} (|S^{(l)}|^2)_x.\label{lsri}
	\end{eqnarray} 
	In the above, $S^{(l)}$'s, $l=1,2$, are short-wave (SW) components, $L$ is the long-wave (LW) component and suffixes $x$ and $t$ denote partial derivatives with respect to spatial and temporal coordinates, respectively. The above LSRI system arises whenever the phase velocity of the low-frequency long-wave  matches with the group velocity of the high-frequency short-waves \cite{zakharov-jetp1,benny-studapplmath}. In Eq. (\ref{lsri}), the formation of soliton in the SW components is due to the exact balance between its dispersion by the nonlinear interaction of the LW with the SW. At the same time, the formation and evolution of the soliton in the LW components is determined by the self-interaction of the SWs. The above LSRI system (\ref{lsri}) has considerable physical relevance in nonlinear optics \cite{kivshar-ol-1992,chowdhury-prl,ablowitz-pre-2001,sazonov-jetp}, plasma physics \cite{nishikawa-prl,oikawa-ptp}, hydrodynamics \cite{benny-studapplmath,kawahara-jphysocjap1,kawahara-jphysocjap2,grimshaw-studapplmath,kopp-jfluidmech,boyd-jphysocean} and BECs \cite{zabolotskii-jetp2009,aguero-jpa,niztazakis-pra}. The LSRI system originally arose from the pioneering study of nonlinear resonant interaction of  the plasma waves by Zakharov \cite{zakharov-jetp1}, where generalized Zakharov equations were deduced to describe Langmuir waves. Such generalized Zakharov equations were reduced to ($1+1$)-dimensional Yajima-Oikawa equation for describing the one-dimensional two-layer fluid flow \cite{oikawa-ptp} for which soliton solutions were obtained through the IST method. Benney has also derived a single-component LSRI system for modelling the dynamics of short capillary gravity waves and gravity waves in deep water \cite{benny-studapplmath}. After these works, there have been a large amount of work in the direction of LSRI involving ($1 + 1$) and ($2 + 1$)-dimensional single component and multi-component cases \cite{ma-studapplmath,kanna-pre2013,chen-jphyssocjpn1,chen-jphyssocjpn2,funakoshi-jphyssocjpn,ohta-jpa-2dlsri,radha-jpa,kanna-jpa-lsri,sakkaravarthi-pre-lsri,kanna-pre-lsri,chen-NonlinearDyn,chow-jphyssocjpn,chen-pre-lsri,chan-nonlineardyn,chen-pre-2dlsri,chen-pla,rao-procrsoca,yang-epjplus}. In nonlinear optics, the single component LSRI system was deduced from the coupled nonlinear Schr\"{o}dinger equations describing the interaction of two optical modes under small amplitude asymptotic expansion \cite{kivshar-ol-1992}. In the negative refractive index media, the LSRI process has been investigated \cite{chowdhury-prl}. We wish to point out that the bright soliton solutions for the general multi-component LSRI system have been derived through the Hirota bilinear method \cite{kanna-pre2013}. In this paper, the authors have demonstrated two types of energy sharing collisions for two different choices of nonlinearity coefficients. Considering the collisions of solitons in these cases one finds that the solitons appearing in the LW component always exhibit elastic collision whereas the solitons in the SW components always undergo energy sharing collisions.
	
	In this review, we investigate the existence of nondegenerate vector bright solitons and their novel properties in the above described five interesting integrable coupled field models. 
	%%%%%%%%%%%%%%%%%%%%%%%%%%%%%%%%%%%%%%%%%%
	\section{Statement of the problem}
	As we have pointed out in Section 1, the  fundamental (and even higher order ) bright soliton solutions which have been already reported for the integrable coupled nonlinear Schr\"{o}dinger family of equations are degenerate. Here, by degenerate, we mean that the fundamental bright soliton nature is characterized by a  single wave number in all the modes or components. The presence of identical wave number in all the modes restricts the motion as well as the structure of the fundamental bright soliton in most of the CNLS type equations. Thus, the bright solitons propagate in all the modes with identical velocity apart from the distinct polarization vector constants. Such a constrained motion always persists in most of the fundamental bright soliton solutions of various CNLS systems. As a consequence of this degeneracy, a single-hump structure only emerges in the fundamental bright soliton profile.  In order to demonstrate this clearly, in the following, we consider the fundamental bright soliton solution of the Manakov system:
	\begin{eqnarray}
	q_{j}=\frac{\alpha_{1}^{(j)}e^{\eta_1}}{1+e^{\eta_1+\eta_1^*+R}}\equiv A_jk_{1R}e^{i\eta_{1I}}\sech(\eta_{1R}+\frac{R}{2}),~j=1,2.
	\end{eqnarray}
	Here $A_j$'s are the unit polarization vectors,  $A_j=\frac{\alpha_1^{(j)}}{(|\alpha_1|^2+|\beta_1|^2)^{1/2}}$, $j=1,2$, the wave variable $\eta_1$  ($=\eta_{1R}+i\eta_{1I}$),   $\eta_{1R}=k_{1R}(t-2k_{1I}z)$, $\eta_{1I}=k_{1I}t+(k_{1R}^2-k_{1I}^2)z$ and $e^R=\frac{(|\alpha_1|^2+|\beta_1|^2)}{(k_1+k_1^*)^2}$. From the above expression for the one-soliton solution, it is evident that the fundamental soliton is described by only one complex wave number $k_1$. Consequently, the single-hump soliton propagates in the two modes, $q_1$ and $q_2$, with identical velocity $v=2k_{1I}$. A similar situation always persists in the other coupled field models mentioned above and their generalizations. For instance, the $N$-component  Manakov type system \cite{kanna-prl}, the mixed $N$-CNLS system \cite{kanna-pre2006}, the GCNLS system \cite{wang-stud,vpriya-cnsns}, and the multi-component LSRI system \cite{oikawa-ptp,kanna-pre2013} are such cases. However, in  contrast to such cases, the coherent coupling among the copropagating optical fields induces a special  type of double-hump vector bright soliton in the CCNLS system \cite{kanna-jpa-2010,kanna-jpa-2011}. In this four wave mixing physical situation also the  coherently coupled soliton is governed by an identical propagation constant in all the modes. Therefore it is clear that the above mentioned degeneracy in propagation constants always persist in all the previously reported vector bright solitons.
	
	In order to differentiate the above class of vector bright solitons from more general fundamental solitons, we classify them as degenerate and nondegenerate solitons based on the absence or presence of more than one wave numbers in the multi-component soliton solution. We call the solitons which propagate in all the modes with identical wave number as degenerate vector solitons whereas the solitons with nonidentical wave numbers as nondegenerate vector solitons. From the above literature, it is clear that the vector bright solitons with identical wave numbers have been well understood. However, the studies on solitons with non-identical propagation constants in all the modes have not been considered until recently. Therefore one would like to investigate the role of additional wave number(s) on the vector bright soliton structures and collision scenario as well. With this motivation, we plan to look for a class of fundamental soliton solutions, in a more general form, which possesses more than one distinct propagation constants. Recently, we have successfully identified such a general class of fundamental vector bright soliton solutions for a wide class of physically important CNLS type equations using the Hirota bilinear method. In this review, we briefly describe the novel properties, including the various collision properties, associated with the nondegenerate vector bright solitons of the Manakov system by deriving their analytical forms through the bilinearization method. Then we point out the existence of such nondegenerate solitons in other coupled systems, namely $N$-CNLS system, mixed 2-CNLS system, 2-CCNLS system, GCNLS system and two-component LSRI system. In these systems, we also specify how the degenerate bright soliton solution arises as a special case of the nondegenerate soliton solution and point out their fascinating energy sharing collisions.
	%%%%%%%%%%%%%%%%%%%%%%%%%%%%%%%%%%%%%%%%%%
	\section{Nondegenerate solitons and their collisions in Manakov system}
	To begin, we derive the nondegenerate bright soliton solutions of the Manakov system (Eq. (\ref{cnls}) with $\sigma_1=\sigma_2=1$) using the Hirota bilinear method. In order to obtain such new class of soliton solutions, we first bilinearize the Manakov system with the bilinearizing transformation, $q_j=\frac{g^{(j)}(z,t)}{f(z,t)}$, $j=1,2$, where $g^{(j)}$'s are complex functions and $f$ is a real function. It leads to the following bilinear forms of Eq. (\ref{cnls}), namely $(iD_z+D^2_t)g^{(j)} \cdot f=0$, $j=1,2$, $D^2_t f \cdot f=2 \sum_{n=1}^{2}g^{(n)}g^{(n)*}$, where $*$ denotes complex conjugation. Here, the Hirota's bilinear operators $D_z$ and $D_t$ are defined \cite{hirota-book} as
	$D_z^mD_t^n (a\cdot b)=\bigg(\frac{\partial}{\partial z}-\frac{\partial}{\partial z'}\bigg)^m\bigg(\frac{\partial}{\partial t}-\frac{\partial}{\partial t'}\bigg)^n a(z,t)b(z',t')_{|z=z',~t=t'}$. Substituting the standard Hirota series expansions for the unknown functions $g^{(j)}=\epsilon g_1^{(j)}+\epsilon^3 g_3^{(j)}+...$, $j=1,2,$ and $f=1+\epsilon^2 f_2+\epsilon^4 f_4+...$
	in the above bilinear equations, one can get a system of linear partial differential equations (PDEs).  Here $\epsilon$ is the series expansion parameter. These linear PDEs arise after collecting the coefficients of the same powers of  $\epsilon$,  and they can be solved recursively for every order of $\epsilon$ with the general forms of seed solutions. The resultant associated explicit expressions for $g^{(j)}$'s and $f$ constitute the soliton solutions to the underlying Manakov system (\ref{cnls}).
	%%%%%%%%%%%%%%%%%%%%%%%%%%%%%%%%%%%%%%%%%%
	\subsection{Nondegenerate fundamental soliton solution of the Manakov system}
	The exact form of the nondegenerate fundamental soliton solution can be obtained by considering the two different seed solutions for the two modes as 
	\begin{eqnarray}
	g_1^{(1)}=\alpha_{1}^{(1)}e^{\eta_1},~ g_1^{(2)}=\alpha_{1}^{(2)}e^{\xi_1},~ \eta_{1}=k_{1}t+ik_{1}^{2}z, ~\xi_{1}=l_{1}t+il_{1}^{2}z,\label{14}
	\end{eqnarray}
	to the following lowest order linear PDEs, $ ig_{1z}^{(j)}+g_{1tt}^{(j)}=0$, $j=1,2$. In the above $k_1$, $l_1$, $\alpha_1^{(j)}$, $j=1,2$, are distinct complex parameters. The presence of two distinct complex wave numbers $k_1$ and $l_1$ ($k_1\neq l_1$, in general) in Eq. (\ref{14}) makes the final solution as nondegenerate one. However, the identical seed solutions, that is the solutions (\ref{14}) with $k_1=l_1$ but different $\alpha_1^{(j)}$'s $j=1,2$, have been used so far to derive the vector bright soliton solutions \cite{Radhakrishnan-pre}.  With the general forms of starting solutions (\ref{14}), we allow the series expansions of the unknown functions $g^{(j)}$ and $f$ to terminate themselves while solving the system of linear PDEs. We find that the series expansions get truncated as $g^{(j)}=\epsilon g_1^{(j)}+\epsilon^3 g_3^{(j)}$ and  $f=1+\epsilon^2 f_2+\epsilon^4 f_4$. With  the explicit forms of unknown functions $g_3^{(j)}$, $f_2$ and $f_4$, finally we obtain the following  a new fundamental one-soliton solution for the Manakov system,  
	\bes
	\begin{eqnarray}
	q_{1}=\frac{g_1^{(1)}+g_3^{(1)}}{1+f_2+f_4}=\frac{1}{D}(\alpha_{1}^{(1)} e^{\eta_{1}}+e^{\eta_{1}+\xi_{1}+\xi_{1}^*+\Delta_{1}^{(1)}}), \label{15a}\\ 
	q_{2}=\frac{g_1^{(2)}+g_3^{(2)}}{1+f_2+f_4}=\frac{1}{D}(\alpha_{1}^{(2)} e^{\xi_{1}}+e^{\eta_{1}+\eta_{1}^*+\xi_{1}+\Delta_{1}^{(2)}}).
	\label{15b}
	\end{eqnarray}\ees
	Here $D=1+e^{\eta_{1}+\eta_{1}^{*}+\delta_{1}}+e^{\xi_{1}+\xi_{1}^{*}+\delta_{2}}+e^{\eta_{1}+\eta_{1}^{*}+\xi_{1}+\xi_{1}^{*}+\delta_{11}}$, $e^{\Delta_{1}^{(1)}}=\frac{(k_{1}-l_{1})\alpha_{1}^{(1)}|\alpha_{1}^{(2)}|^2}{(k_{1}+l_{1}^*)(l_{1}+l_{1}^*)^{2}}$,\\ $e^{\Delta_{1}^{(2)}}=-\frac{(k_{1}-l_{1})|\alpha_{1}^{(1)}|^2\alpha_{1}^{(2)}}{(k_{1}+k_{1}^*)^{2}(k_{1}^*+l_{1})}$,  $e^{\delta_{1}}=\frac{|\alpha_{1}^{(1)}|^2}{(k_{1}+k_{1}^{*})^{2}}$, $e^{\delta_{2}}=\frac{ |\alpha_{1}^{(2)}|^2}{(l_{1}+l_{1}^*)^{2}}$ and 
	$e^{\delta_{11}}=\frac{|k_{1}-l_{1}|^{2} |\alpha_{1}^{(1)}|^{2}|\alpha_{1}^{(2)}|^{2}}{(k_{1}+k_{1}^*)^{2}(k_{1}^*+l_{1})(k_{1}+l_{1}^*)(l_{1}+l_{1}^*)^{2}}$.  The above one-soliton solution possesses two distinct complex wave numbers, $k_1$ and $l_1$, which appear in both the expressions of $q_1$ and $q_2$ simultaneously. This confirms that the obtained soliton solution is nondegenerate. The fundamental soliton solution (\ref{15a}) and (\ref{15b}) can also be rewritten using Gram determinant forms as well \cite{ablowitz-pla,vijayajayanthi-epjst},
	\begin{subequations}
		\begin{eqnarray}
		g^{(1)} =
		\begin{vmatrix}
		\frac{e^{\eta_1+\eta_1^*}}{(k_1+k_1^*)} & \frac{e^{\eta_1+\xi_1^*}}{(k_1+l_1^*)} & 1 & 0 & e^{\eta_1} \\ 
		\frac{e^{\xi_1+\eta_1^*}}{(l_1+k_1^*)} & \frac{e^{\xi_1+\xi_1^*}}{(l_1+l_1^*)}  & 0 & 1 &   e^{\xi_1}\\
		-1 & 0 & \frac{|\al_1^{(1)}|^2}{(k_1+k_1^*)} & 0 & 0\\
		0 & -1 & 0 &  \frac{|\al_1^{(2)}|^2}{(l_1+l_1^*)} & 0\\
		0 & 0& -\al_1^{(1)} & 0 &0
		\end{vmatrix},
		\end{eqnarray}
		\begin{eqnarray}
		\hspace{-0.3cm}g^{(2)} =
		\begin{vmatrix}
		\frac{e^{\eta_1+\eta_1^*}}{(k_1+k_1^*)} & \frac{e^{\eta_1+\xi_1^*}}{(k_1+l_1^*)} & 1 & 0 & e^{\eta_1} \\ 
		\frac{e^{\xi_1+\eta_1^*}}{(l_1+k_1^*)} & \frac{e^{\xi_1+\xi_1^*}}{(l_1+l_1^*)}  & 0 & 1 &   e^{\xi_1}\\
		-1 & 0 & \frac{|\al_1^{(1)}|^2}{(k_1+k_1^*)} & 0 & 0\\
		0 & -1 & 0 &  \frac{|\al_1^{(2)}|^2}{(l_1+l_1^*)} & 0\\
		0 & 0& 0 & -\al_1^{(2)} &0
		\end{vmatrix},
		\end{eqnarray}	
		\begin{eqnarray}
		f=
		\begin{vmatrix}
		\frac{e^{\eta_1+\eta_1^*}}{(k_1+k_1^*)} & \frac{e^{\eta_1+\xi_1^*}}{(k_1+l_1^*)} & 1 & 0  \\ 
		\frac{e^{\xi_1+\eta_1^*}}{(l_1+k_1^*)} & \frac{e^{\xi_1+\xi_1^*}}{(l_1+l_1^*)}  & 0 & 1 \\
		-1 & 0 & \frac{|\al_1^{(1)}|^2}{(k_1+k_1^*)} & 0 \\
		0 & -1 & 0 &  \frac{|\al_1^{(2)}|^2}{(l_1+l_1^*)}
		\end{vmatrix}.
		\end{eqnarray}
	\end{subequations}
	The above Gram determinant forms indeed satisfy the bilinear equations as well as the Manakov Eq. (\ref{cnls}).
	
\begin{figure}[H]
		\begin{center}
			\includegraphics[width=13.0cm]{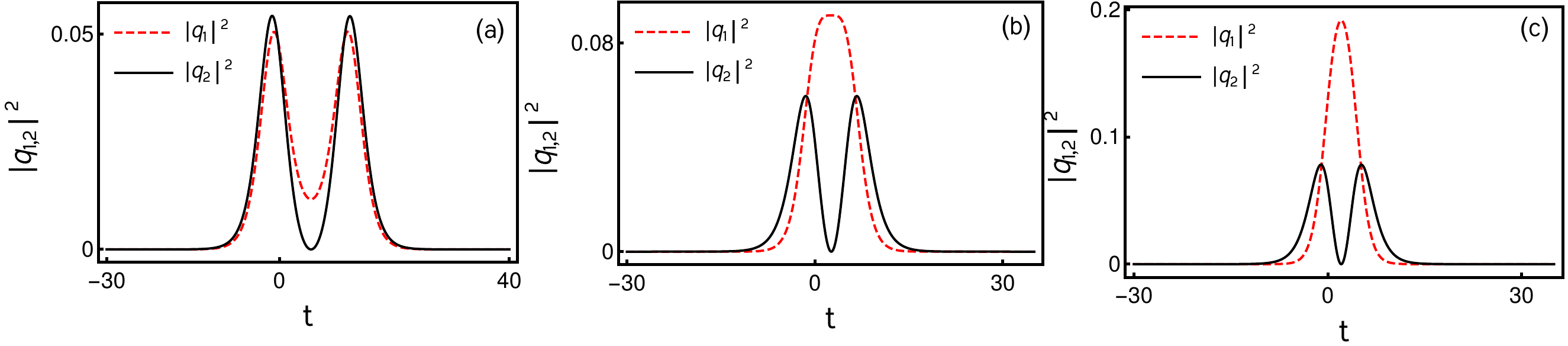}
			\caption{Symmetric intensity profiles of nondegenerate fundamental bright soliton solution (\ref{18a})-(\ref{18b}): While (a) denotes double-hump soliton in both the modes (b) represents flat-top in $q_1$ mode and double-hump in $q_2$ mode and (c) denotes 
				single-hump $q_1$ mode and double-hump in $q_2$ mode. The parameter values of each figures are: (a): $k_1=0.333+0.5i$, $l_1=0.315+0.5i$,  $\alpha_{1}^{(1)}=0.45+0.45i$, $\alpha_{1}^{(2)}=0.49+0.45i$. (b): $k_1=0.425+0.5i$, $l_1=0.3+0.5i$,  $\alpha_{1}^{(1)}=0.44+0.51i$, $\alpha_{1}^{(2)}=0.43+0.5i$. (c): $k_1=0.55+0.5i$, $l_1=0.333+0.5i$,  $\alpha_{1}^{(1)}=0.5+0.5i$, $\alpha_{1}^{(2)}=0.5+0.45i$.\label{fig1}}
		\end{center}	
	\end{figure}

	To explain  the properties associated with the solution (\ref{15a}) and (\ref{15b}), we rewrite it in hyperbolic form as
	\bes\begin{eqnarray}
	q_1=\frac{2k_{1R}A_1e^{i\eta_{1I}}[\cosh(\xi_{1R}+\phi_{1R})\cos\phi_{1I}+i\sinh(\xi_{1R}+\phi_{1R})\sin\phi_{1I}]}{\big[{a_{11}}\cosh(\eta_{1R}+\xi_{1R}+\phi_1+\phi_2+c_1)+\frac{1}{a_{11}^*}\cosh(\eta_{1R}-\xi_{1R}+\phi_2-\phi_1+c_2)\big]},\label{17a}\\
	q_2=\frac{2l_{1R}A_2e^{i\xi_{1I}}[\cosh(\eta_{1R}+\phi_{2R})\cos\phi_{2I}+i\sinh(\eta_{1R}+\phi_{2R})\sin\phi_{2I}]}{\big[{a_{12}}\cosh(\eta_{1R}+\xi_{1R}+\phi_1+\phi_2+c_1)+\frac{1}{a_{12}^*}\cosh(\eta_{1R}-\xi_{1R}+\phi_2-\phi_1+c_2)\big]},\label{17b}
	\end{eqnarray}\ees 
	where $a_{11}=\frac{(k_{1}^{*}-l_{1}^{*})^{\frac{1}{2}}}{(k_{1}^{*}+l_{1})^{\frac{1}{2}}}$,  $a_{12}=\frac{(k_{1}^{*}-l_{1}^{*})^{\frac{1}{2}}}{(k_{1}+l_{1}^{*})^{\frac{1}{2}}}$, $c_1=\frac{1}{2}\log\frac{(k_1^*-l_1^*)}{(l_1-k_1)}$, $c_2=\frac{1}{2}\log\frac{(k_1-l_1)(k_1^*+l_1)}{(l_1-k_1)(k_1+l_1^*)}$, $\phi_1=\frac{1}{2}\log\frac{(k_1-l_1)|\alpha_{1}^{(2)}|^2}{(k_1+l_1^*)(l_1+l_1)^2}$, $\phi_2=\frac{1}{2}\log\frac{(l_1-k_1)|\alpha_{1}^{(1)}|^2}{(k_1^*+l_1)(k_1+k_1)^2}$, $\eta_{1R}=k_{1R}(t-2k_{1I}z)$, $\eta_{1I}=k_{1I}t+(k_{1R}^2-k_{1I}^2)z$, $\xi_{1R}=l_{1R}(t-2l_{1I}z)$, $\xi_{1I}=l_{1I}t+(l_{1R}^2-l_{1I}^2)z$,
	$A_1=[\alpha_{1}^{(1)}/\alpha_{1}^{(1)*}]^{1/2}$, $A_2=i[\alpha_{1}^{(2)}/\alpha_{1}^{(2)*}]^{1/2}$. Here, 
	$\phi_{1R}$, $\phi_{2R}$, $\phi_{1I}$ and $\phi_{2I}$ are real and imaginary parts of $\phi_1$ and  $\phi_2$,  respectively and $k_{1R}$, $l_{1R}$, $k_{1I}$ and $l_{1I}$ denote the real and imaginary parts of $k_1$ and $l_1$, respectively. The geometrical structure of the solution (\ref{17a}) and (\ref{17b}) is described by the four complex parameters $k_1$, $l_1$, $\alpha_1^{(j)}$, $j=1,2$.  The nondegenerate fundamental bright soliton solution (\ref{17a}) and (\ref{17b}) either propagates with identical velocity $k_{1I}=l_{1I}$ or with non-identical velocities $k_{1I}\neq l_{1I}$ in the two modes $q_1$ and $q_2$.
	In the identical velocity case, the quantity $\phi_{jI}=0$, $j=1,2$ in  (\ref{17a}) and(\ref{17b}) when $k_{1I}=l_{1I}$.
	This results in the forms 
	\bes\begin{eqnarray}
	q_1=\frac{2k_{1R}A_1e^{i\eta_{1I}}\cosh(\xi_{1R}+\phi_{1R})}{\big[{a_{11}}\cosh(\eta_{1R}+\xi_{1R}+\phi_1+\phi_2+c_1)+\frac{1}{a_{11}^*}\cosh(\eta_{1R}-\xi_{1R}+\phi_2-\phi_1+c_2)\big]},\label{18a}\\
	q_2=\frac{2l_{1R}A_2e^{i\xi_{1I}}\cosh(\eta_{1R}+\phi_{2R})}{\big[{a_{12}}\cosh(\eta_{1R}+\xi_{1R}+\phi_1+\phi_2+c_1)+\frac{1}{a_{12}^*}\cosh(\eta_{1R}-\xi_{1R}+\phi_2-\phi_1+c_2)\big]},\label{18b}
	\end{eqnarray}\ees
	where  $\eta_{1R}=k_{1R}(t-2k_{1I}z)$, $\eta_{1I}=k_{1I}t+(k_{1R}^2-k_{1I}^2)z$, $\xi_{1R}=l_{1R}(t-2k_{1I}z)$, $\xi_{1I}=k_{1I}t+(l_{1R}^2-k_{1I}^2)z$. The amplitude, velocity and central position of the nondegenerate fundamental soliton in the first mode are found from Eq. (\ref{18a}) as $2k_{1R}$, $2k_{1I}$ and $\frac{\phi_{1R}}{l_{1R}}$, respectively. Similarly they are found for the soliton in the second mode from Eq. (\ref{18b}) as $2l_{1R}$, $2k_{1I}$ and $\frac{\phi_{2R}}{k_{1R}}$, respectively.
	The solution (\ref{18a}) and (\ref{18b}) admits  both the symmetric and asymmetric profiles, including a double-hump, a flat top and a single-hump profiles. We have displayed a combination of these three types of symmetric profiles (and their corresponding asymmetric profiles also) in our recent paper \cite{ramakrishnan-pre}. However, here, we display a typical novel double-hump, a flat top and a single-hump profile in Figure \ref{fig1}.    

	The symmetric and asymmetric nature of the solution (\ref{18a}) and (\ref{18b}) can be confirmed by calculating either the relative separation distance between the minima of the two modes or by finding the corresponding extremum points from it.  We remark that the double-hump formation occurs in the structure of nondegenerate one-bright soliton solution (\ref{17a}) and (\ref{17b}) when the relative velocity of the solitons in the two modes tends to zero. That is $\Delta v=v_1-v_2=2(l_{1I}-k_{1I})\rightarrow 0$. One can find the various special features associated with the obtained nondegenerate fundamental soliton solution (\ref{17a}) and (\ref{17b}) further in Ref. \cite{ramakrishnan-pre}.
	%%%%%%%%%%%%%%%%%%%%%%%%%%%%%%%%%%%%%%%%%%%
	\subsection{Nondegenerate two-soliton solution}
	To get the nondegenerate two-soliton solution of Manakov Eq. (\ref{cnls}) we proceed with the procedure given in the previous subsection along with the following seed solutions, $g_1^{(1)}=\al_1^{(1)}e^{\eta_1}+\al_2^{(1)}e^{\eta_2}$ and $g_1^{(2)}=\al_1^{(2)}e^{\xi_1}+\al_2^{(2)}e^{\xi_2}$, $\eta_{j}=k_{j}t+ik_{j}^{2}z$ and $\xi_{j}=l_{j}t+il_{j}^{2}z$, $j=1,2$. We find that the series expansions for $g^{(j)}$, $j=1,2,$ and $f$ get terminated as $g^{(j)}=\epsilon g_1^{(j)}+\epsilon^3 g_3^{(j)}+\epsilon^5 g_5^{(j)}+\epsilon^7 g_7^{(j)}$ and $f=1+\epsilon^2 f_2+\epsilon^4 f_4+\epsilon^6 f_6+\epsilon^8 f_8$.  Here we assume that all the $k_j$'s and $l_j$'s, $j=1,2$, are distinct. The explicit forms of the obtained unknown functions in the truncated series expansions constitute the following nondegenerate two-soliton solution and it  can be expressed using Gram determinants in the following way:
	\begin{eqnarray}
	g^{(N)}=\begin{vmatrix}
	A & I & \phi \\ 
	-I &B & {\bf 0}^T \\ 
	{\bf 0}  & C_N & 0
	\end{vmatrix},~~f=\begin{vmatrix}
	A & I \\ 
	-I &B \\ 
	\end{vmatrix},~~N=1,2.\label{3}
	\end{eqnarray}
	Here the matrices $A $ and $B$ are of the order $(4\times 4)$ defined as $
	A=\begin{pmatrix}
	A_{mm'} & A_{mn} \\ 
	A_{nm}& A_{nn'}\\ 
	\end{pmatrix}$, $B=\begin{pmatrix}
	\kappa_{mm'} & \kappa_{mn} \\ 
	\kappa_{nm} &\kappa_{nn'} \\ 
	\end{pmatrix}$, $m, m' ,n ,n'=1,2$.
	The various elements of the matrix $A$ can be obtained from the following, 
	$A_{mm'}=\frac{e^{\eta_m+\eta_{m'}^*}}{(k_m+k_{m'}^*)},~ A_{mn}=\frac{e^{\eta_m+\xi_{n}^*}}{(k_m+l_{n}^*)}$,
	$	A_{nn'}=\frac{e^{\xi_n+\xi_{n'}^*}}{(l_n+l_{n'}^*)}, ~A_{nm}=\frac{e^{\eta_n^*+\xi_{m}}}{(k_n^*+l_{m})},~m,m',n,n'=1,2$. 	The elements of the matrix $B$ are
	$\kappa_{mm'}=\frac{\psi_m^{\dagger}\sigma\psi_{m'}}{(k_m^*+k_{m'})},~\kappa_{mn}=\frac{\psi_m^{\dagger}\sigma\psi'_{n}}{(k_m^*+k_{n})},~\kappa_{nm}=\frac{\psi_n^{'\dagger}\sigma\psi_{m}}{(l_n^*+k_{m})},~\kappa_{nn'}=\frac{\psi_n^{'\dagger}\sigma\psi'_{n'}}{(l_n^*+l_{n'})}$.
	In the latter, the column matrices are defined as
	$\psi_j=\begin{pmatrix}
	\alpha_j^{(1)}\\
	0
	\end{pmatrix}$, ~$\psi'_j=\begin{pmatrix}
	0\\
	\alpha_j^{(2)}
	\end{pmatrix}$, $j=m,m',n,n'=1,2$, $\eta_j=k_jt+ik_j^2z$ and $\xi_j=l_jt+il_j^2z$, $j=1,2$.
	The other matrices in Eq. (3) are defined as
	$\phi=\begin{pmatrix}
	e^{\eta_1} & e^{\eta_2}  & e^{\xi_1}  & e^{\xi_2}
	\end{pmatrix}^T$, $C_1=-\begin{pmatrix}
	\alpha_1^{(1)} & \alpha_2^{(1)} & 0 &0
	\end{pmatrix}$, $C_2=-\begin{pmatrix}
	0 & 0 &\alpha_1^{(2)} & \alpha_2^{(2)}  \end{pmatrix}$, ${\bf 0} =\begin{pmatrix}
	0 &0 & 0& 0 \end{pmatrix}$ and $\sigma=I$ is a $(4\times 4)$ identity matrix. The presence of eight arbitrary complex parameters $k_{j}$, $l_{j}$, $\alpha_{1}^{(j)}$ and $\alpha_{2}^{(j)}$, $j=1,2$, define the profile shapes of the nondegenerate two solitons and their interesting collision scenarios. In addition to the above, we also find that the Manakov system also admits degenerate and nondegenerate solitons simultaneously under  the wave numbers restriction $k_1=l_1$ (or $k_2=l_2$) but $k_2\neq l_2$ (or $k_1\neq l_1$). Such a special kind of partially nondegenerate two-soliton solution can be deduced by fixing the latter wave number restriction in the completely nondegenerate two-soliton solution (\ref{3}). This partially nondegenerate soliton solution can also be derived through the Hirota bilinear method. To derive this solution one has to assume the following seed solutions, $g_1^{(1)}=\al_1^{(1)}e^{\eta_1}+\al_2^{(1)}e^{\eta_2}$ and $g_1^{(2)}=\al_1^{(2)}e^{\eta_1}+\al_2^{(2)}e^{\xi_2}$, $\eta_{j}=k_{j}t+ik_{j}^{2}z$ and $\xi_{2}=l_{2}t+il_{2}^{2}z$, $j=1,2$, in the solution construction process. The resultant coexistence soliton solution and its dynamics are characterized by only seven complex parameters $k_{j}$, $l_{2}$, $\alpha_{1}^{(j)}$ and $\alpha_{2}^{(j)}$, $j=1,2$.
	%%%%%%%%%%%%%%%%%%%%%%%%%%%%%%%%%%%%%%%%
	\subsection{Various types of collision dynamics of nondegenerate solitons}
	In order to understand the interesting collision properties associated with the nondegenerate solitons, one has to  analyze the asymptotic forms of the complete nondegenerate two-soliton solution (\ref{3}) of the Manakov equation.  By doing so, we observe that the nondegenerate solitons in general exhibit three types of collision scenarios, namely shape preserving, shape altering and shape changing collision behaviours, for either of the two cases (i) Equal velocities: $k_{1I}=l_{1I}$, $k_{2I}=l_{2I}$ and  (ii) Unequal velocities: $k_{1I}\neq l_{1I}$, $k_{2I}\neq l_{2I}$. To facilitate the understanding of these collision properties, here we present the asymptotic analysis for the case of equal velocities only and it can be performed for unequal velocities case also in a similar manner.
	%%%%%%%%%%%%%
	\subsubsection{Asymptotic analysis}
	We perform a careful asymptotic analysis for the nondegenerate two soliton solution (\ref{3}) in order to understand the interaction dynamics of  the nondegenerate solitons completely. We deduce the explicit expressions for the individual solitons at the aymptotic limits $z\rightarrow \pm\infty$. To explore this, we consider as a typical example
	$k_{jR},l_{jR}>0$, $j=1,2$, $k_{1I}>k_{2I}$, $l_{1I}>l_{2I}$,  $k_{1I}=l_{1I}$ and $k_{2I}=l_{2I}$, that corresponds to head-on collision between the two nondegenerate solitons. In this situation the two fundamental solitons $S_1$ and $S_2$ are well separated and subsequently the asymptotic forms of the individual nondegenerate solitons can be deduced from the solution (\ref{3}) by incorporating the following asymptotic nature of the wave variables $\xi_{jR}=l_{jR}(t-2l_{jI}z)$ and $\eta_{jR}=k_{jR}(t-2k_{jI}z)$,  $j=1,2$, in it. The wave variables $\eta_{jR}$ and $\xi_{jR}$ behave asymptotically as (i) Soliton 1 ($S_1$): $\eta_{1R}$, $\xi_{1R}\simeq 0$, $\eta_{2R}$, $\xi_{2R}\rightarrow\mp \infty$ as $z\mp\infty$ and (ii) Soliton 2 ($S_2$): $\eta_{2R}$, $\xi_{2R}\simeq 0$, $\eta_{1R}$, $\xi_{1R}\rightarrow\mp \infty$ as $z\pm\infty$. Correspondingly these results lead to the following asymptotic expressions of nondegenerate individual solitons.  \\
	\underline{(a) Before collision}: $z\rightarrow -\infty$\\
	\underline{Soliton 1}: In this limit, the asymptotic forms of $q_1$ and $q_2$ are deduced from the two soliton solution (\ref{3}) for soliton 1 as below:
	\begin{subequations}
		\begin{eqnarray}
		&&\hspace{-1.1cm}q_{1}\simeq \frac{2A_1^{1-}k_{1R}e^{i\eta_{1I}}\cosh(\xi_{1R}+\phi_1^-)}{\big[a_{11}\cosh(\eta_{1R}+\xi_{1R}+\phi_1^-+\phi_2^-+c_1)+\frac{1}{a_{11}^*}\cosh(\eta_{1R}-\xi_{1R}+\phi_2^--\phi_1^-+c_2)\big]},\label{20a}~~~\end{eqnarray}	\begin{eqnarray}
		&&\hspace{-1.1cm}q_{2}\simeq\frac{2A_2^{1-}l_{1R}e^{i\xi_{1I}}\cosh(\eta_{1R}+\phi_2^-)}{\big[a_{12}\cosh(\eta_{1R}+\xi_{1R}+\phi_1^-++\phi_2^-+c_1)+\frac{1}{a_{12}^*}\cosh(\eta_{1R}-\xi_{1R}+\phi_2^--\phi_1^-+c_2)\big]}.~~~\label{20b}
		\end{eqnarray}\end{subequations}\\
	Here, $a_{11}={\frac{(k_{1}^{*}-l_{1}^{*})^{\frac{1}{2}}}{(k_{1}^{*}+l_{1})^{\frac{1}{2}}}}$, $a_{12}=\frac{(k_{1}^{*}-l_{1}^{*})^{\frac{1}{2}}}{(k_{1}+l_{1}^{*})^{\frac{1}{2}}}$,  $\phi_1^-=\frac{1}{2}\log\frac{(k_1-l_1)|\alpha_1^{(2)}|^2}{(k_1+l_1^*)(l_1+l_1^*)^2}$,  $\phi_2^-=\frac{1}{2}\log\frac{(l_1-k_1)|\alpha_1^{(1)}|^2}{(k_1^*+l_1)(k_1+k_1^*)^2}$,  $A_{1}^{1-}=[\alpha_{1}^{(1)}/\alpha_{1}^{(1)^*}]^{1/2}$ and $A_{2}^{1-}=i[\alpha_{1}^{(2)}/\alpha_{1}^{(2)^*}]^{1/2}$. In the latter, superscript ($1-$) represents soliton  $S_1$ before collision and subscript $(1,2)$ denotes the two modes $q_1$ and $q_2$ respectively.     \\
	\underline{Soliton 2}: The asymptotic expressions for  soliton 2 in the two modes before collision turn out to be
	\begin{subequations}
		\begin{eqnarray}
		&&\hspace{-1.1cm}q_{1}\simeq \frac{2k_{2R}A_1^{2-}e^{i(\eta_{2I}+\theta_1^-)}\cosh(\xi_{2R}+\varphi_1^-)}{\big[a_{21}\cosh(\eta_{2R}+\xi_{2R}+\varphi_1^-+\varphi_2^-+c_3)+\frac{1}{a_{21}^*}\cosh(\eta_{2R}-\xi_{2R}+\varphi_2^--\varphi_1^-+c_4)\big]},\label{21a}~~~\\
		&&\hspace{-1.1cm}q_2\simeq \frac{2l_{2R}A_2^{2-}e^{i(\xi_{2I}+\theta_2^-)}\cosh(\eta_{2R}+\varphi_2^-)}{\big[a_{22}\cosh(\eta_{2R}+\xi_{2R}+\varphi_1^-+\varphi_2^-+c_3)+\frac{1}{a_{22}^*}\cosh(\eta_{2R}-\xi_{2R}+\varphi_2^--\varphi_1^-+c_4)\big]}.\label{21b}~~~
		\end{eqnarray} \end{subequations}
	In the above, $a_{21}=\frac{(k_{2}^{*}-l_{2}^{*})^{\frac{1}{2}}}{(k_{2}^{*}+l_{2})^{\frac{1}{2}}}$, $a_{22}=\frac{(k_{2}^{*}-l_{2}^{*})^{\frac{1}{2}}}{(k_{2}+l_{2}^*)^{\frac{1}{2}}}$, $c_3=\frac{1}{2}\log\frac{(k_2^*-l_2^*)}{(l_2-k_2)}$, $c_4=\frac{1}{2}\log\frac{(k_2-l_2)(k_2^*+l_2)}{(l_2-k_2)(k_2+l_2^*)}$,
	$\vphi_1^-=\frac{1}{2}\log\frac{(k_2-l_2)|\alpha_{2}^{(2)}|^2}{(k_2+l_2^*)(l_2+l_2^*)^2}+\Psi_1$,  $\Psi_1=\frac{1}{2}\log\frac{|k_1-l_2|^2|l_1-l_2|^4}{|k_1+l_2^*|^2|l_1+l_2^*|^4}$,
	$\vphi_2^-=\frac{1}{2}\log\frac{(l_2-k_2)|\alpha_{2}^{(1)}|^2}{(k_2^*+l_2)(k_2+k_2^*)^2}+\Psi_2$, ~ $\Psi_2=\frac{1}{2}\log\frac{|k_2-l_1|^2|k_1-k_2|^4}{|k_2+l_1^*|^2|k_1+k_2^*|^4}$,
	$e^{i\theta_1^-}=\frac{(k_{1}-k_{2})(l_{1}-l_{2})(l_{1}^*+l_{2})(k_{2}-l_{1})^{\frac{1}{2}}(k_{1}+k_{2}^{*})(k_{2}^{*}+l_{1})^{\frac{1}{2}}}{(k_{1}^{*}-k_{2}^{*})(l_{1}+l_{2}^*)(l_{1}^*-l_{2}^*)(k_{2}^{*}-l_{1}^{*})^{\frac{1}{2}}(k_{1}^{*}+k_{2})(k_{2}+l_{1}^{*})^{\frac{1}{2}}}$, $A_{1}^{2-}=[\alpha_{2}^{(1)}/\alpha_{2}^{(1)^*}]^{1/2}$, $A_{2}^{2-}=[\alpha_{2}^{(2)}/\alpha_{2}^{(2)^*}]^{1/2}$, 
	$e^{i\theta_2^-}=\frac{(l_{1}-l_{2})(k_{1}-l_{2})^{\frac{1}{2}}(k_{1}+l_{2}^{*})^{\frac{1}{2}}(l_{1}+l_{2}^{*})}{(k_{1}^{*}-l_{2}^{*})^{\frac{1}{2}}(l_{1}^{*}-l_{2}^{*})(k_{1}^{*}+l_{2})^{\frac{1}{2}}(l_{1}^{*}+l_{2})}$. Here, superscript ($2-$) refers to soliton $S_2$ before collision. 
	\\
	\underline{(b) After collision}: $z\rightarrow +\infty$\\
	\underline{Soliton 1}: The asymptotic form for soliton 1 after collision is deduced as,
	\begin{subequations}
		\begin{eqnarray}
		&&\hspace{-1.1cm}q_{1}\simeq \frac{2k_{1R}A_1^{1+}e^{i(\eta_{1I}+\theta_1^+)}\cosh(\xi_{1R}+\phi_1^+)}{\big[a_{11}\cosh(\eta_{1R}+\xi_{1R}+\phi_1^++\phi_2^++c_1)+\frac{1}{a_{11}^*}\cosh(\eta_{1R}-\xi_{1R}+\phi_2^+-\phi_1^++c_2)\big]},\label{22a}~~~\\
		&&\hspace{-1.1cm}q_2\simeq \frac{2l_{1R}A_1^{2+}e^{i(\xi_{1I}+\theta_2^+)}\cosh(\eta_{1R}+\phi_{2}^+)}{\big[a_{12}\cosh(\eta_{1R}+\xi_{1R}+\phi_1^++\phi_2^++c_1)+\frac{1}{a_{12}^*}\cosh(\eta_{1R}-\xi_{1R}+\phi_2^+-\phi_1^++c_2)\big]}.~~~\label{22b}
		\end{eqnarray} \end{subequations}
	Here,~ 
	$\phi_1^+=\phi_1^-+\psi_1$, ~$\psi_1=\frac{1}{2} \log\frac{|k_2-l_1|^2|l_1-l_2|^4}{|k_2+l_1^*|^2|l_1+l_2^*|^4}$,~ $\phi_2^+=\phi_2^-+\psi_2$,~$\psi_2=\frac{1}{2}\log\frac{|k_1-l_2|^2|k_1-k_2|^4}{|k_1+l_2^*|^2|k_1+k_2^*|^4}$,\\ $e^{i\theta_1^+}=\frac{(k_{1}-k_{2})(k_{1}-l_{2})^{\frac{1}{2}}(k_{1}^*+k_{2})(k_{1}^{*}+l_{2})^{\frac{1}{2}}}{(k_{1}^{*}-k_{2}^{*})(k_{1}^{*}-l_{2}^{*})^{\frac{1}{2}}(k_{1}+k_{2}^*)(k_{1}+l_{2}^{*})^{\frac{1}{2}}}$, $e^{i\theta_2^+}=\frac{(l_{1}-l_{2})(k_{2}-l_{1})^{\frac{1}{2}}(k_{2}+l_{1}^{*})^{\frac{1}{2}}(l_{1}^*+l_{2})}{(k_{2}^{*}-l_{1}^{*})^{\frac{1}{2}}(l_{1}^{*}-l_{2}^{*})(k_{2}^{*}+l_{1})^{\frac{1}{2}}(l_{1}+l_{2}^*)}$, $A_{1}^{1+}=[\alpha_{1}^{(1)}/\alpha_{1}^{(1)^*}]^{1/2}$ and $A_{2}^{1+}=[\alpha_{1}^{(2)}/\alpha_{1}^{(2)^*}]^{1/2}$,  in which superscript ($1+$) denotes soliton $S_1$ after collision.
	\underline{Soliton 2}: The expression for soliton 2 after collision deduced from the two soliton solution is
	\begin{subequations}
		\begin{eqnarray}
		&&\hspace{-1.1cm}q_{1}\simeq\frac{2A_2^{1+}k_{2R}e^{i\eta_{2I}}\cosh(\xi_{2R}+\varphi_1^+)}{\big[a_{21}\cosh(\eta_{2R}+\xi_{2R}+\varphi_1^++\varphi_2^++c_3)+\frac{1}{a_{21}^*}\cosh(\eta_{2R}-\xi_{2R}+\varphi_2^+-\varphi_1^++c_4)\big]},\label{23a}~~~\\
		&&\hspace{-1.1cm}q_{2}\simeq\frac{2A_2^{2+}l_{2R}e^{i\xi_{2I}}\cosh(\eta_{2R}+\varphi_2^+)}{[a_{22}\cosh(\eta_{2R}+\xi_{2R}+\varphi_1^++\varphi_2^++c_3)+\frac{1}{a_{22}^*}\cosh(\eta_{2R}-\xi_{2R}+\varphi_2^+-\varphi_1^++c_4)\big]},~~~\label{23b}
		\end{eqnarray} \end{subequations}
	where $\varphi_1^+=\frac{1}{2}\log\frac{(k_2-l_2)|\alpha_{2}^{(2)}|^2}{(k_2+l_2^*)(l_2+l_2^*)^2}$, $\varphi_2^+=\frac{1}{2}\log\frac{(l_2-k_2)|\alpha_{2}^{(1)}|^2}{(k_2^*+l_2)(k_2+k_2^*)^2}$, $\varphi_3^+=\frac{1}{2}\log\frac{|k_2-l_2|^2|\alpha_{2}^{(1)}|^2|\alpha_{2}^{(2)}|^2}{|k_2+l_2^*|^2(k_2+k_2^*)^2(l_2+l_2^*)^2}$, $\varphi_4^+=\frac{1}{2}\log\frac{|\alpha_{2}^{(1)}|^2(l_2+l_2^*)^2}{|\alpha_{2}^{(2)}|^2(k_2+k_2^*)^2}$,   $A_{1}^{2+}=[\alpha_{2}^{(1)}/\alpha_{2}^{(1)^*}]^{1/2}$ and $A_{2}^{2+}=i[\alpha_{2}^{(2)}/\alpha_{2}^{(2)^*}]^{1/2}$. In the latter, superscript ($2+$) represents soliton $S_2$ after collision. 
	
	In the above, $\eta_{jI}=k_{jI}t+(k_{jR}^{2}-k_{jI}^{2})z$,  $\xi_{jI}=l_{jI}t+(l_{jR}^{2}-l_{jI}^{2})z$, $j=1,2,$ and that the phase terms $\varphi_j^-$, $j=1,2,$ can also be rewritten as
	$
	\vphi_1^-=\vphi_1^++\Psi_1$,  $\vphi_2^-=\vphi_2^++\Psi_2$. The above asymptotic analysis clearly shows that there is a definite drastic alteration in the phase terms only. It can be identified from the following relations among the phase terms before and after collisions. That is,  \begin{equation} 
	\phi_1^+=\phi_1^-+\psi_1, ~\phi_2^+=\phi_2^-+\psi_2,~ \varphi_1^+=\varphi_1^--\Psi_1,~ \varphi_2^+=\varphi_2^--\Psi_2. \label{8}\end{equation}
	The above relations imply that the initial structures of the nondegenerate two solitons are preserved except for the phase terms. From this, we infer that they undergo either shape preserving collision with zero phase shift or shape changing collision with a finite phase shift. In addition to this, a special shape altering collision can also occur with a small phase shift. 
	The zero phase shift condition, deduced from Eq. (\ref{8}), turns out to be
	\begin{equation}
	\phi_j^+=\phi_j^-,~ \varphi_j^+=\varphi_j^-,~j=1,2.\label{25}
	\end{equation}
	In order to follow the above condition, the additional phase constants $\psi_j's$ and $~\Psi_j$'s should be maintained as zero. That is, 
	\bes\begin{eqnarray}
	\psi_1=\frac{1}{2} \log\frac{|k_2-l_1|^2|l_1-l_2|^4}{|k_2+l_1^*|^2|l_1+l_2^*|^4}=0, ~\psi_2=\frac{1}{2}\log\frac{|k_1-l_2|^2|k_1-k_2|^4}{|k_1+l_2^*|^2|k_1+k_2^*|^4}=0.\\
	\Psi_1=\frac{1}{2}\log\frac{|k_1-l_2|^2|l_1-l_2|^4}{|k_1+l_2^*|^2|l_1+l_2^*|^4}=0,~\Psi_2=\frac{1}{2}\log\frac{|k_2-l_1|^2|k_1-k_2|^4}{|k_2+l_1^*|^2|k_1+k_2^*|^4}=0.
	\end{eqnarray}\ees
	From the above,  we deduce the following criterion, corresponding to the conditions (\ref{25}), for the occurrence of shape preserving collision with zero phase shift,
	\begin{equation}
	\frac{|k_2+l_1^*|^2}{|k_2-l_1|^2}|-\frac{|k_1+l_2^*|^2}{|k_1-l_2|^2}=0. \label{27}
	\end{equation}
	As a result, whenever the conditions (\ref{25}) or equivalently the criterion (\ref{27}), are satisfied the nondegenerate bright  solitons exhibit shape preserving collision with a zero phase shift. Otherwise, they undergo shape altering and shape changing collisions, as discussed in the following. Further, the shape changing (and altering) collision scenario also belongs to the elastic collision as we describe below.
	
	The above analysis clearly demonstrates that during the collision process the initial phase of each of the soliton gets changed. The total phase shift of soliton $S_1$ in the two modes after collision becomes
	\bes\begin{eqnarray}
	\Delta \Phi_1&=&(\phi_1^++\phi_2^+)-(\phi_1^-+\phi_2^-)=\psi_1+\psi_2\nonumber\\
	&&\hspace{-0.6cm}=\frac{1}{2}\log\frac{|k_2-l_1|^2|l_1-l_2|^4|k_1-l_2|^2|k_1-k_2|^4}{|k_2+l_1^*|^2|l_1+l_2^*|^4|k_1+l_2^*|^2|k_1+k_2^*|^4}.\label{25a}
	\end{eqnarray} 
	Similarly the total phase shift suffered by soliton $S_2$ in the two modes is
	\begin{eqnarray}
	\hspace{-0.5cm}\Delta \Phi_2&=&(\varphi_1^++\varphi_2^+)-(\varphi_1^-+\varphi_2^-)=-(\Psi_1+\Psi_2)\nonumber\\
	&&\hspace{-0.6cm}=-\frac{1}{2}\log\frac{|k_1-l_2|^2|l_1-l_2|^4|k_2-l_1|^2|k_1-k_2|^4}{|k_1+l_2^*|^2|l_1+l_2^*|^4|k_2+l_1^*|^2|k_1+k_2^*|^4}=-(\psi_1+\psi_2)=-\Delta \Phi_1.\label{25b}
	\end{eqnarray} \ees
	From the above expressions, we conclude that the phases of all the solitons are mainly influenced by the wave numbers $k_j$ and $l_j$, $j=1,2$, and not by the complex parameters $\al_1^{(j)}$'s and $\al_2^{(j)}$'s, $j=1,2$. This peculiar property of nondegenerate solitons is different in the case of degenerate vector bright solitons \cite{Radhakrishnan-pre,kanna-prl}, see also Section. 4.2.1 below, where the complex parameters $\al_1^{(j)}$'s and $\al_2^{(j)}$'s, associated with polarization constants, play a crucial role in shifting the position of solitons after the collision.
		\begin{figure}[H]
		\begin{center}
			\includegraphics[width=8.5 cm]{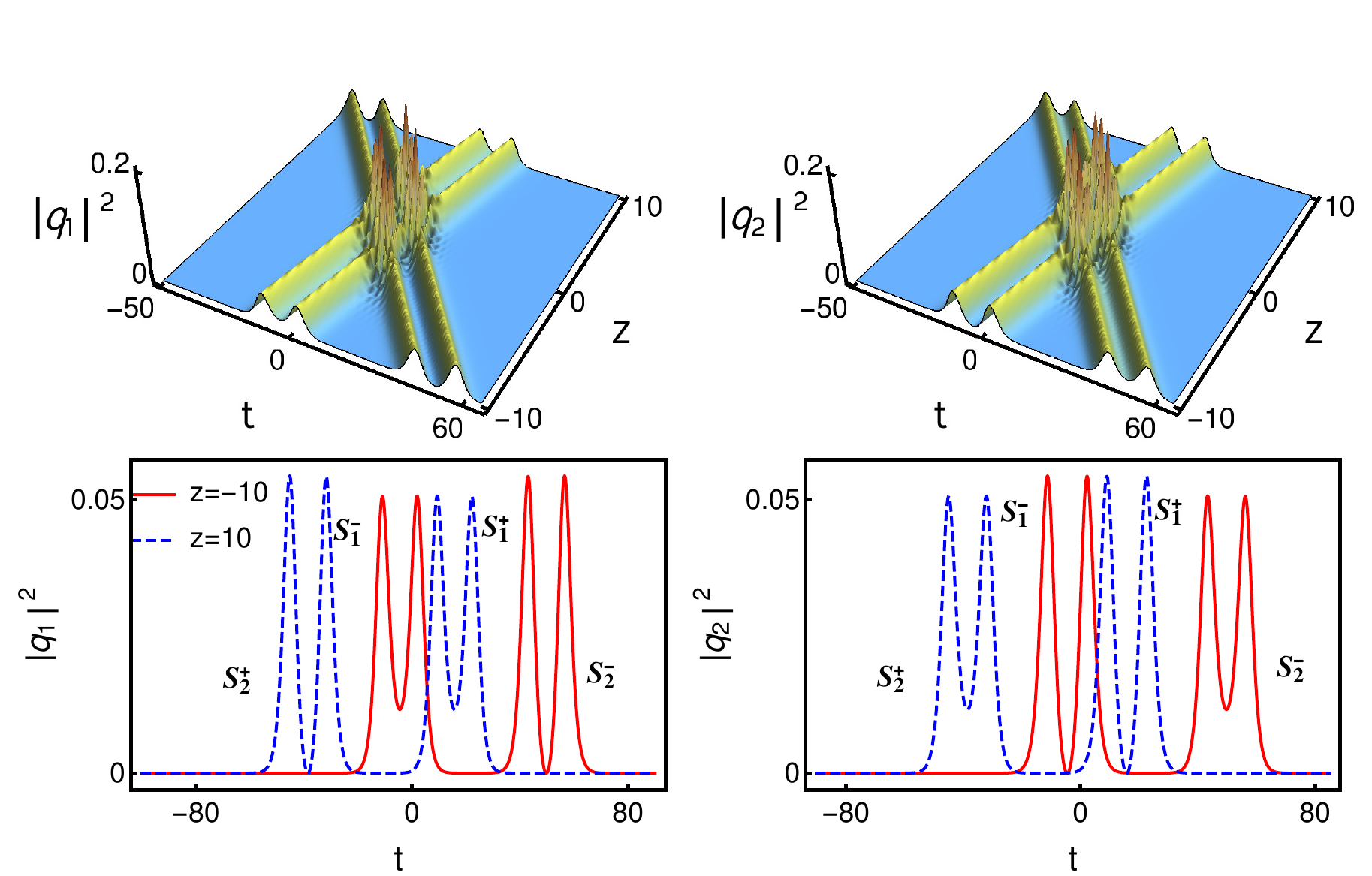}		
			\caption{Shape preserving collision of two symmetric double-hump solitons - The energy does not get exchanged among the nondegenerate solitons during the collision process. The parameter values are $k_{1}=0.333+0.5i$, $l_{1}=0.315+0.5i$, $k_{2}=0.315-2.2i$, $l_{2}=0.333-2.2i$, $\alpha_{1}^{(1)}=0.45+0.45i$, $\alpha_{2}^{(1)}=0.49+0.45i$, $\alpha_{1}^{(2)}=0.49+0.45i$ and $\alpha_{2}^{(2)}=0.45+0.45i$.\label{fig2}}
		\end{center}
	\end{figure}
	%%%%%%%%%%%%%%%%%%%%%%%%%%%%%%%%%%%%%
	\subsubsection{Elastic collision: shape preserving, shape altering and shape changing collisions}
\begin{figure}[H]
		\begin{center}
			\includegraphics[width=8.5cm]{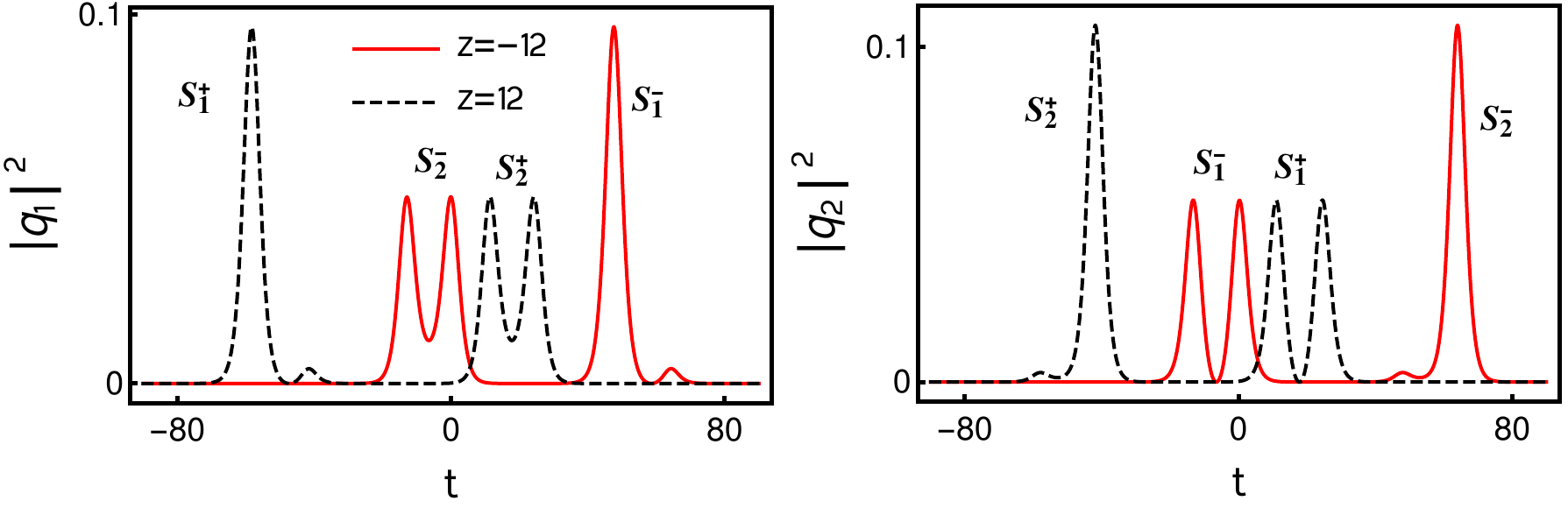}	
			\caption{Shape preserving collision between a symmetric double-hump soliton and a asymmetric double-hump soliton: The parameter values are $k_{1}=0.333+0.5i$, $l_{1}=0.315+0.5i$, $k_{2}=0.315-2.2i$, $l_{2}=0.333-2.2i$, $\alpha_{1}^{(1)}=0.45+0.45i$, $\alpha_{2}^{(1)}=2.49+2.45i$, $\alpha_{1}^{(2)}=0.49+0.45i$ and $\alpha_{2}^{(2)}=0.45+0.45i$. \label{fig3}}
		\end{center}
	\end{figure}

	From the above asymptotic analysis, we observe that the intensities of nondegenerate solitons $S_{1}$ and $S_{2}$ in the two modes are the same before and after collision in the equal velocities case, $k_{1I}=l_{1I}$ and $k_{2I}=l_{2I}$. To confirm this, we calculate the transition intensities (using the expressions for the transition amplitudes $T_j^i=\frac{A_j^{i+}}{A_j^{i-}}$, $i,j=1,2$),  $|T_{1}^{1}|^2=\frac{|A_{1}^{1+}|^2}{|A_{1}^{1-}|^2}$, $|T_{2}^{1}|^2=\frac{|A_{2}^{1+}|^2}{|A_{2}^{1-}|^2}$, $|T_{1}^{2}|^2=\frac{|A_{1}^{2+}|^2}{|A_{1}^{2-}|^2}$ and $|T_{2}^{2}|^2=\frac{2|A_{2}^{2+}|^2}{2|A_{2}^{2-}|^2}$. The various expressions deduced for the different $A_j^i$'s previously confirm that the transition intensities are unimodular. That is,  $|T_{j}^{l}|^2=1$, $j,l=1,2$.  Thus, the collision scenario that occurs among the nondegenerate solitons, in general, is always elastic. So, the nondegenerate solitons, for $k_{1I}=l_{1I}$, $k_{2I}=l_{2I}$, (but $k_1 \neq l_1$, $k_2 \neq l_2$) corresponding to two distinct wave numbers in general undergo elastic collision without any intensity redistribution between the modes $q_1$ and $q_2$. However, it is clear from Eq. (\ref{8}), the changes that occur in the phase terms do alter the structure of the nondegenerate solitons during the collision scenario. Consequently, there is a possibility of  shape altering and shape changing collisions occurring, without violating the unimodular conditions of transition intensities, in the equal velocities case, apart from the earlier mentioned shape preserving collision. A typical shape-preserving collision is displayed in Figure 2, in which we set two well separated symmetric double-hump soliton profiles as initial profiles in both the modes at $z=-10$.  The initial structures of the two double-hump solitons are preserved after the collision. It is evident from the dashed red curves drawn at $z=+10$ in Figure \ref{fig2}. In addition to this, we have also verified that the wave parameters $k_j$ and $l_j$, $j=1,2$, that are given in the caption of Figure \ref{fig2}, satisfy the zero phase shift criterion (\ref{27}). The obtained numerical value from Eq. (\ref{27}) is equal to $-0.0064$ (nearly equal to) $0$. This value physically implies that during the collision the two double-humped nondegenerate bright solitons pass through one another without a phase shift and emerge from the collision unaltered in shape, amplitude and velocity. This remarkable property has not been observed earlier in the cases of scalar NLS bright solitons as well as in the degenerate vector bright solitons \cite{Radhakrishnan-pre,kanna-prl}. Very interestingly, a similar zero phase shift shape preserving collision also occurs even when the symmetric double-hump soliton interacts with an asymmetric double-hump soliton. Such collision is illustrated in Figure \ref{fig3}.   
 
	In this case, the total intensity of each soliton is conserved which can be verfied from the relations $|A_j^{l-}|^2=|A_j^{l+}|^2$, $j,l=1,2$. In addition to this, the total intensity in each of the modes is also conserved, that is $|A_j^{1-}|^2+|A_j^{2-}|^2=|A_j^{1+}|^2+|A_j^{2+}|^2=\text{constant}$.
	
	\begin{figure}[H]
		\begin{center}
			\includegraphics[width=9.0 cm]{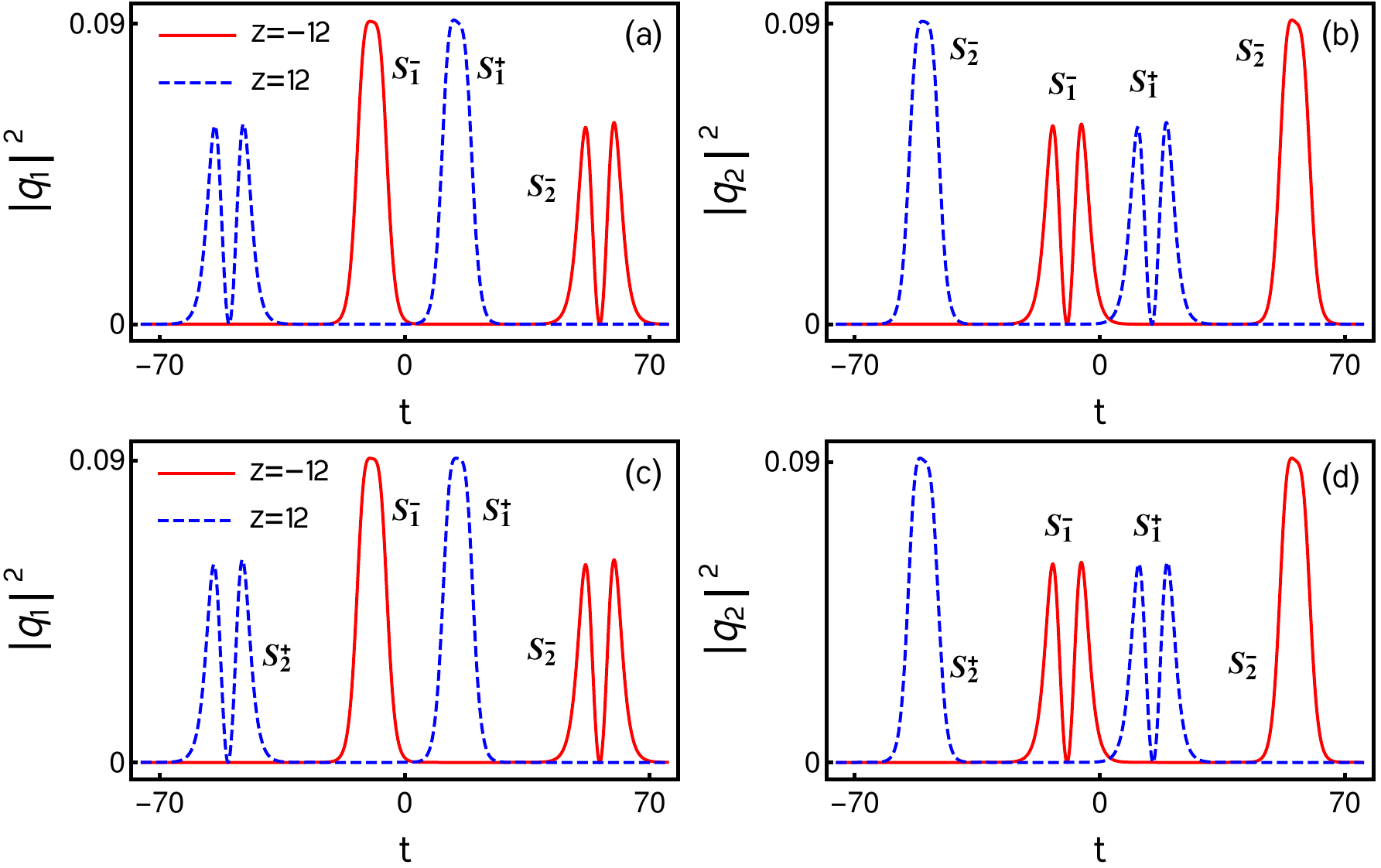}	
			\caption{A typical shape altering collision is displayed in the top panels. Their corresponding shape preserving nature is brought out in the bottom panels after taking a pair of postion shifts, ($z'=z-\frac{\psi_1}{2l_{1R}k_{1I}}=12.3053$,  $z'=z-\frac{\psi_2}{2k_{1R}k_{1I}}=12.27$) and ($z'=z+\frac{\Psi_1}{2l_{2R}k_{2I}}=12.0614$,  $z'=z+\frac{\Psi_2}{2k_{2R}k_{2I}}=12.0694$) in the expressions (\ref{22a})-(\ref{22b}) of soliton 1  and the expressions (\ref{23a})-(\ref{23b}) of soliton 2, respectively.  \label{fig4}}
		\end{center}
	\end{figure}
	Then, we also come across another type of elastic collision, namely shape altering collision for certain sets of parametric choices again with $k_{1I}=l_{1I}$ and $k_{2I}=l_{2I}$. To demonstrate such collision scenario in Figure 4, we fix the parameter values as $k_{1}=0.425+0.5i$, $l_{1}=0.3+0.5i$, $k_{2}=0.3-2.2i$, $l_{2}=0.425-2.2i$, $\alpha_{1}^{(1)}=\alpha_{2}^{(2)}=0.5+0.5i$ and $\alpha_{2}^{(1)}=\alpha_{1}^{(2)}=0.45+0.5i$. From this figure, one can observe that a symmetric (or asymmetric) flattop soliton collides with an asymmetric (or symmetric) double-hump soliton in the $q_1$ (or $q_2$) component. As a result, the symmetric flattop profile in the $q_1$ mode  gets  modified slightly as the asymmetric flattop profile and slightly asymmetric double-hump soliton $S_2^-$ becomes symmetric double-hump soliton. Similarly, while the symmetric double-hump soliton $S_1^-$ in the $q_2$ mode changes slightly into an asymmetric structure, the asymmetric flattop soliton    
	$S_2^-$ becomes symmetric. As we have pointed out earlier,  this kind of shape alteration essentially arises in the structures of nondegenerate bright solitons is due to the phase conditions (\ref{8}). However, the shape preserving nature of the nondegenerate solitons can be brought out by taking appropriate position shifts based on the expressions (\ref{22a})-(\ref{22b}) and (\ref{23a})-(\ref{23b}). For example, the expressions (\ref{22a}) and (\ref{22b}) of soliton 1 after collision exactly coincides with the expressions (\ref{20a}) and (\ref{20b})  after substituting $z'=z-\frac{\psi_1}{2l_{1R}k_{1I}}$ and   $z'=z-\frac{\psi_2}{2k_{1R}k_{1I}}$, respectively, in it. Similarly, for the soliton 2, the expressions (\ref{23a})-(\ref{23b}) exactly  matches with the expressions (\ref{21a}) and (\ref{21b}) after taking the position shifts $z'=z+\frac{\Psi_1}{2l_{2R}k_{2I}}$ and  $z'=z+\frac{\Psi_2}{2k_{2R}k_{2I}}$, respectively. Correspondingly the shapes of the nondegenerate solitons are preserved. A typical example of this transition is illustrated in Figures 4(c) and 4(d), where the initial profiles are retained after taking the shifts in the positions of solitons. This is also true in the case of shape changing collision. Here, we have not displayed the shape changing collision and their corresponding position shift plots for brevity.
	%%%%%%%%%%%%%%%%%%%%%%%%%%%%%%%%%%%%%%%%%%%%%%
	\subsection{Collision between nondegenerate and degenerate solitons}
	In this sub-section, we discuss the collision among the degenerate and nondegenerate solitons admitted by the two-soliton solution (\ref{3}) of the Manakov system (\ref{cnls}) in the partial nondegenerate limit $k_1=l_1$ and $k_2\neq l_2$. The following asymptotic analysis assures that there is a definite energy redistribution occurs among the modes $q_1$ and $q_2$.
	\subsubsection{Asymptotic analysis}
	To elucidate this new kind of collision behaviour, we analyze the partially nondegenerate two-soliton solution (\ref{3}) in the asymptotic limits $z\rightarrow \pm \infty$. The resultant action yields the asymptotic forms corresponding to degenerate and nondegenerate solitons. To obtain the asymptotic forms for the present case we incorporate the asymptotic nature of the wave variables $\eta_{jR}=k_{jR}(t-2k_{Ij}z)$ and $\xi_{2R}=l_{2R}(t-2l_{2I}z)$, $j=1,2$, in the solution (\ref{3}). Here the wave variable $\eta_{1R}$ corresponds to the degenerate soliton and  $\eta_{2R}$, $\xi_{2R}$ correspond to the nondegenerate soliton.  In order to find the asymptotic behaviour of these wave variables we consider the parametric choice as $k_{1R},k_{2R},l_{2R}>0$, ~$k_{1I}>0$,~ $k_{2I},l_{2I}<0$, ~$k_{1I}>k_{2I}$,~$k_{1I}>l_{2I}$. For this choice, the wave variables behave asymptotically as follws: (i) degenerate soliton $S_1$: $\eta_{1R}\simeq0$, $\eta_{2R}$,$\xi_{2R}\rightarrow \mp\infty$ as $z\rightarrow \mp\infty$ (ii) nondegenerate soliton $S_2$: $\eta_{2R},\xi_{2R}\simeq 0$, $\eta_{1R}\rightarrow\pm \infty$ as $z\rightarrow\pm \infty$. By incorporating these asymptotic behaviours of wave variables in the solution (\ref{3}), we deduce the following asymptotic expressions for degenerate and nondegenerate solitons. \\
	\underline{(a) Before collision}: $z\rightarrow -\infty$\\
	\underline{Soliton 1}: In this limit, the asymptotic form for the degenerate soliton deduced from the partially nondegenerate two soliton solution (\ref{3}) is
	\begin{eqnarray}
	q_j\simeq\begin{pmatrix}
	A_1^{1-}\\ \\
	A_2^{1-}
	\end{pmatrix} k_{1R}e^{i\eta_{1I}}\sech(\eta_{1R}+\frac{R}{2}), ~j=1,2,
	\label{5.1}
	\end{eqnarray}
	where $A_j^{1-}=\al_1^{(j)}/(|\al_1^{(1)}|^2+|\al_1^{(2)}|^2)^{1/2}$, $j=1,2$, $R=\ln\frac{(|\al_1^{(1)}|^2+|\al_1^{(2)}|^2)}{(k_1+k_1^*)^2}$. Here, in $A_j^{1-}$ the superscript $1-$ denotes soliton $S_1$ before collision and subscript $j$ refers to the mode number.  \\
	\underline{Soliton 2}: The asymptotic expressions for the nondegenerate soliton $S_2$ which is present in the two modes before collision are obtained as
	\begin{subequations}
		\begin{eqnarray}
		&&\hspace{-1.2cm}q_1\simeq\frac{2k_{2R}A_1^{2-}}{D}\bigg(	e^{i\xi_{2I}+\Lam_1}\cosh(\eta_{2R}+\frac{\Phi_{21}-\Del_{21}}{2})+e^{i\eta_{2I}+\Lam_{2}}\cosh(\xi_{2R}+\frac{\lam_{2}-\lam_{1}}{2})\bigg),\label{5.2}	\\
		&&\hspace{-1.2cm}q_2\simeq\frac{2l_{2R}A_2^{2-}}{D}\bigg(e^{i\eta_{2I}+\Lam_7}\cosh(\xi_{2R}+\frac{\Gamma_{21}-\ga_{21}}{2})+e^{i\xi_{2I}+\Lam_{6}}\cosh(\eta_{2R}+\frac{\lam_{7}-\lam_{6}}{2})\bigg),\label{5.3}
		\end{eqnarray}\begin{eqnarray}
		&&\hspace{-1.2cm}D=e^{\Lam_5}\cosh(\eta_{2R}-\xi_{2R}+\frac{\lam_3-\lam_4}{2})+e^{\Lam_3}\cosh(i(\eta_{2I}-\xi_{2I})+\frac{\vth_{12}-\varphi_{21}}{2})\nonumber\\
		&&\hspace{-0.6cm}	+e^{\Lam_4}\cosh(\eta_{2R}+\eta_{3R}+\frac{\lam_5-R}{2}).\nonumber
		\end{eqnarray}
	\end{subequations}
	Here,  $A_{1}^{2-}=[\alpha_{2}^{(1)}/\alpha_{2}^{(1)^*}]^{1/2}$, $A_{2}^{2-}=[\alpha_{2}^{(2)}/\alpha_{2}^{(2)^*}]^{1/2}$. In the latter the superscript $2-$ denote nondegenerate soliton $S_2$ before collision. The various other constants appearing in Eq. (30) are defined in the Appendix A.
	
	\underline{(b) After collision}: $z\rightarrow +\infty$\\
	\underline{Soliton 1}: The asymptotic forms for degenerate soliton $S_1$ after collision deduced from the solution (\ref{3}) (with $k_1=l_1$ and $k_2\neq l_2$) as,
	\begin{eqnarray}
	q_j\simeq\begin{pmatrix}
	A_1^{1+}\\ \\
	A_2^{1+}
	\end{pmatrix}e^{i(\eta_{1I}+\theta_j^+)}k_{1R}\sech(\eta_{1R}+\frac{R'-\vsa_{22}}{2}),~j=1,2,\label{5.4}
	\end{eqnarray}
	where  $A_1^{1+}=\al_1^{(1)}/(|\al_1^{(1)}|^2+\chi|\al_1^{(2)}|^2)^{1/2}$, $A_2^{1+}=\al_1^{(2)}/(|\al_1^{(1)}|^2\chi^{-1}+|\al_1^{(2)}|^2)^{1/2}$, $\chi=(|k_1-l_2|^2|k_1+k_2^*|^2)/(|k_1-k_2|^2|k_1+l_2^*|^2)$, $e^{i\theta_1^+}=\frac{(k_1-k_2)(k_1^*+k_2)(k_1-l_2)^{\frac{1}{2}}(k_1^*+l_2)^{\frac{1}{2}}}{(k_1^*-k_2^*)(k_1+k_2^*)(k_1^*-l_2^*)^{\frac{1}{2}}(k_1+l_2^*)^{\frac{1}{2}}}$,\\ $e^{i\theta_2^+}=\frac{(k_1-k_2)^{\frac{1}{2}}(k_1^*+k_2)^{\frac{1}{2}}(k_1-l_2)(k_1^*+l_2)}{(k_1^*-k_2^*)^{\frac{1}{2}}(k_1+k_2^*)^{\frac{1}{2}}(k_1^*-l_2^*)(k_1+l_2^*)}$. Here $1+$ in $A_1^{1+}$ refers to degenerate soliton $S_1$ after collision. \\
	\underline{Soliton 2}: Similarly the expression for the nondegenerate soliton, $S_2$, after collision deduced from the two soliton solution (\ref{3}) (with $k_1=l_1$ and $k_2\neq l_2$) is
	\bes\begin{eqnarray}
	&&\hspace{-1cm}q_{1}\simeq \frac{2k_{2R}A_1^{2+}e^{i\eta_{2I}}\cosh(\xi_{2R}+\frac{\Lam_{22}-\rho_1}{2})}{\big[\frac{(k_{2}^{*}-l_{2}^{*})^{\frac{1}{2}}}{(k_{2}^{*}+l_{2})^{\frac{1}{2}}}\cosh(\eta_{2R}+\xi_{2R}+\frac{\vsa_{22}}{2})+\frac{(k_{2}+l_{2}^{*})^{\frac{1}{2}}}{(k_{2}-l_{2})^{\frac{1}{2}}}\cosh(\eta_{2R}-\xi_{2R}+\frac{R_3-R_6}{2})\big]},\label{32a}\\
	&&\hspace{-1cm}q_2\simeq \frac{2l_{2R}A_2^{2+}e^{i\xi_{2I}}\cosh(\eta_{2R}+\frac{\mu_{22-\rho_2}}{2})}{\big[\frac{(k_{2}^{*}-l_{2}^{*})^{\frac{1}{2}}}{(k_{2}+l_{2}^*)^{\frac{1}{2}}}\cosh(\eta_{2R}+\xi_{2R}+\frac{\vsa_{22}}{2})+\frac{(k_{2}^*+l_{2})^{\frac{1}{2}}}{(k_{2}-l_{2})^{\frac{1}{2}}}\cosh(\eta_{2R}-\xi_{2R}+\frac{R_3-R_6}{2})\big]}. \label{32b}
	\end{eqnarray}\ees
	where $\rho_j=\log\al_2^{(j)}$, $j=1,2$, $A_{1}^{2+}=[\alpha_{2}^{(1)}/\alpha_{2}^{(1)^*}]^{1/2}$, $A_{2}^{2+}=i[\alpha_{2}^{(2)}/\alpha_{2}^{(2)^*}]^{1/2}$. The explicit expressions of all the undefined constants are given in Appendix A.
	%%%%%%%%%%%%%%%%%%%%%%%%%%%%%%%%%%%%%%%%%%%%%%%
	\subsection{Degenerate soliton collision induced shape changing scenario of nondegenerate soliton}
\begin{figure}[H]
		\begin{center}
			\includegraphics[width=10.0 cm]{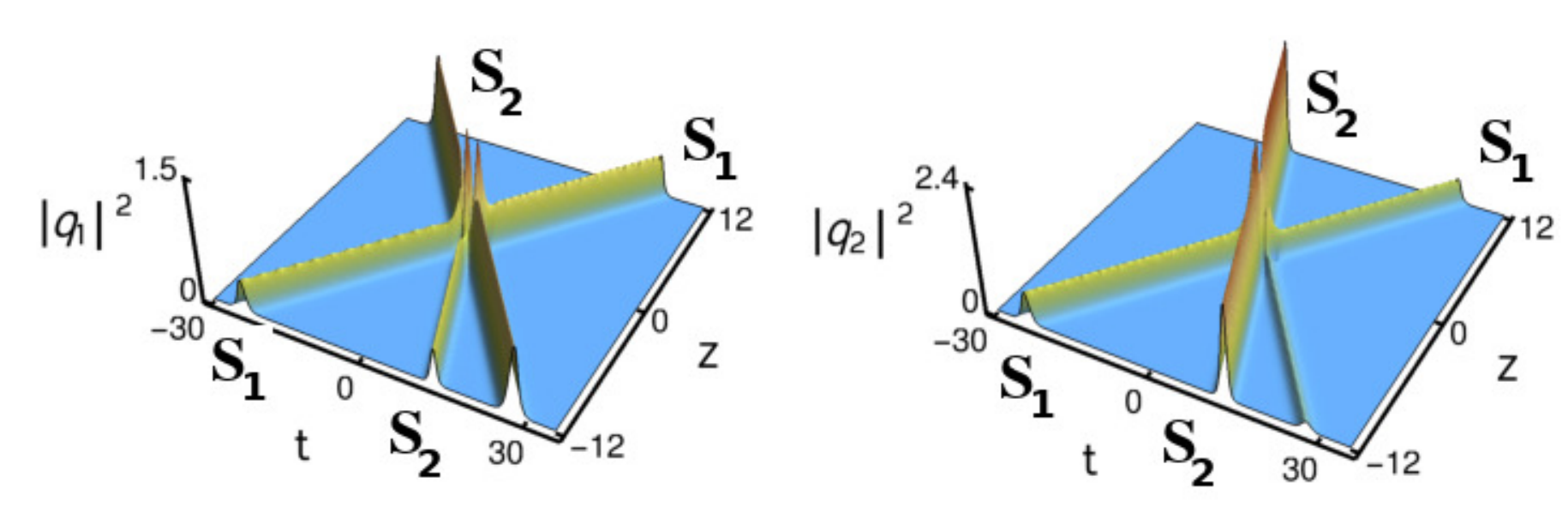}
			\caption{Shape changing collision between degenerate and nondegenerate soliton: $k_{1}=l_1=1+i$,  $k_{2}=1-i$, $l_{2}=1.5-0.5i$, $\alpha_{1}^{(1)}=0.8+0.8i$, $\alpha_{2}^{(2)}=0.6+0.6i$, $\alpha_{2}^{(1)}=0.25+0.25i$, $\alpha_{1}^{(2)}=1+i$.\label{fig6}}
		\end{center}
	\end{figure}

	The coexistence of nondegenerate and degenerate solitons can be realized from the partially nondegenerate limit of the soliton solution (\ref{3}) (with $k_1=l_1$ and $k_2\neq l_2$). Such coexisting solitons undergo a novel collision property, that has been illustrated in Figure. \ref{fig6}. From this figure, one can observe that the intensity of the degenerate soliton $S_1$ is enhanced after collision in the $q_1$ mode and it gets suppressed in the $q_2$ mode. As we expected, like in the complete degenerate case \cite{Radhakrishnan-pre,kanna-pre}, the degenerate soliton undergoes energy redistribution among both the modes.  In this case, the polarization vectors, $A_j^{l}=\alpha_l^{(j)}/(|\alpha_1^{(1)}|^2+|\alpha_1^{(2)}|^2)^{1/2}$, $l,j=1,2$, play a crucial role in changing the shape of the degenerate solitons under collision,  where the intensity redistribution occurs between the modes $q_1$ and $q_2$. As we have pointed out below in the next subsection, the shape preserving collision arises in the pure degenerate case when the polarization parameters obey the condition, $\frac{\alpha_1^{(1)}}{\alpha_2^{(1)}}=\frac{\alpha_1^{(2)}}{\alpha_2^{(2)}}$, where $\alpha _i^{(j)}$'s, $i, j = 1,2$, are complex parameters related to the polarization vectors as given above. However, this collision property is not true in the case of nondegenerate solitons as we have depicted in Figure. \ref{fig6}. As a result, the nondegenerate soliton $S_2$ switches its asymmetric double-hump profile into a single-hump profile along with a phase shift. In addition, we also noticed from the asymptotic expressions (\ref{5.2})-(\ref{5.3}) and (\ref{32a})-(\ref{32b}) the asymmetric double-hump profile of nondegenerate soliton gets  transformed into another form of an asymmetric double-hump profile when it interacts with a degenerate soliton for a specific choice of parameter values.  In the nondegenerate case, the relative separation distances (or phases) are in general not preserved during the collision. Therefore the mechanism behind the occurrence of shape preserving and shape changing collisions in the nondegenerate solitons is quite new. These novel collision properties can be understood from the corresponding asymptotic analysis given in the previous subsection. The analysis reveals that energy redistribution occurs between the modes $q_1$ and $q_2$. In order to confirm the shape changing nature of this interesting collision scenario we obtain the following expression for the transition amplitudes, 
	\begin{eqnarray}
	T_1^1=\frac{(|\al_1^{(1)}|^2+|\al_1^{(2)}|^2)^{1/2}}{(|\al_1^{(1)}|^2+\chi|\al_1^{(2)}|^2)^{1/2}},~
	T_2^1=\frac{(|\al_1^{(1)}|^2+|\al_1^{(2)}|^2)^{1/2}}{(|\al_1^{(1)}|^2\chi^{-1}+|\al_1^{(2)}|^2)^{1/2}}.
	\end{eqnarray}
	In general, the transition amplitudes are not equal to unity. If the quantity $T_j^l$ is not unimodular (for this case the constant $\chi\neq 1$)  then the degenerate and nondegenerate solitons always exhibit shape changing collision.
	The standard elastic collision can be recovered when $\chi=1$. One can calculate the shift in the positions of both degenerate and nondegenerate solitons after collision from the asymptotic analysis.  This new kind of collision property has not been observed in the degenerate vector bright solitons of the Manakov system \cite{Radhakrishnan-pre,kanna-pre}.
	%%%%%%%%%%%%%%%%%%%%%%%%%%%%%%%%%%%%%%%%%%%%%%%%%%%%%%
	\subsection{Degenerate bright solitons and their shape changing/energy redistribution collision in Manakov system}
\begin{figure}[H]
		\begin{center}
			\includegraphics[width=5.0 cm]{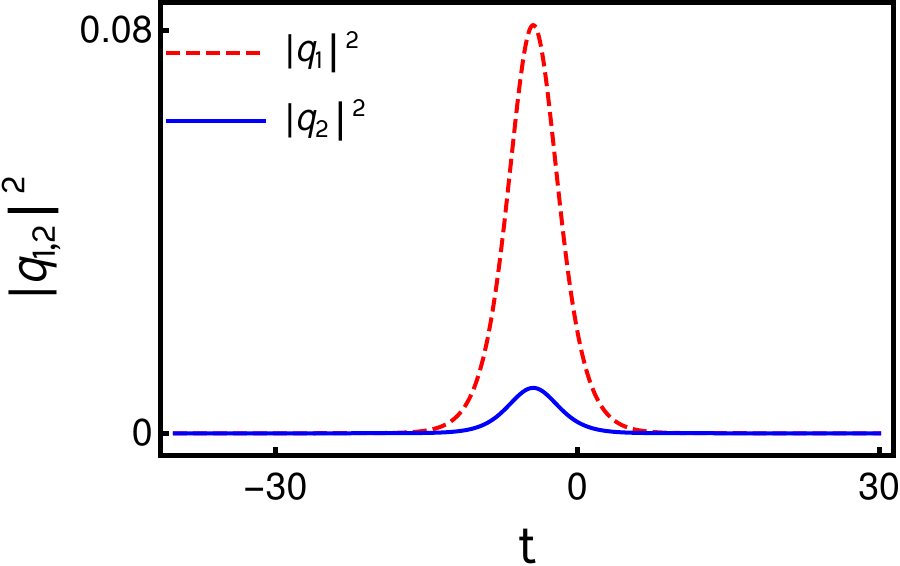}
			\caption{Degenerate one-soliton of the Manakov equation: The values of the parameters are $k_1=0.3+0.5i$, $\alpha_{1}^{(1)}=1.5+1.5i$, $\alpha_{1}^{(2)}=0.5+0.5i$. \label{fig7}}
		\end{center}
	\end{figure}

	The already reported degenerate vector one-bright soliton solution of the Manakov system (\ref{e1}) can be deduced from the one-soliton solution (\ref{15a})-(\ref{15b}) by imposing the condition $k_{1}=l_{1}$ in it. The forms of $q_j$ given in Eq. (\ref{15a})-(\ref{15b}) degenerate into the standard bright soliton form \cite{Radhakrishnan-pre,kanna-pre}   
	\begin{equation}
	q_{j}=\frac{\alpha_{1}^{(j)}e^{\eta_1}}{1+e^{\eta_1+\eta_1^*+R}}, ~j=1,2,\label{6.1}\\
	\end{equation}
	which can be rewritten as
	\begin{equation}
	q_{j}=k_{1R} \hat{A_{j}}e^{i\eta_{1I}}\sech(\eta_{1R}+\frac{R}{2}),\label{6.2}
	\end{equation}
	where $\eta_{1R}=k_{1R}(t-2k_{1I}z)$, $\eta_{1I}=k_{1I}t+(k_{1R}^{2}-k_{1I}^{2})z$, $\hat{A_{j}}=\frac{\alpha_{1}^{(j)}}{\sqrt{(|\alpha_{1}^{(1)}|^2+|\alpha_{1}^{(2)}|^2)}}$, $e^{R}=\frac{(|\alpha_{1}^{(1)}|^2+|\alpha_{1}^{(2)}|^2)}{(k_{1}+k_{1}^*)^2}$, $j=1,2$. 
	Note that the above fundamental bright soliton always propagates in both the modes $q_{1}$ and $q_{2}$ with the same velocity $2k_{1I}$. The polarization vectors $(\hat{A}_1,\hat{A}_2)^{\dagger}$ have different amplitudes and phases, unlike the case of nondegenerate solitons where they have only different unit phases. The presence of a single wave number $k_1$ in the solution (\ref{6.2}) restricts the degenerate soliton to have a single-hump form only.  A typical profile of the degenerate soliton is shown in Figure \ref{fig7}. As already pointed out in \cite{Radhakrishnan-pre,kanna-pre} the amplitude and central position of the degenerate vector bright soliton are obtained as $2k_{1R}\hat{A}_j$, $j=1,2$ and $\frac{R}{2k_{1R}}$, respectively.
	\begin{figure}[H]
	\begin{center}
		\includegraphics[width=9.5 cm]{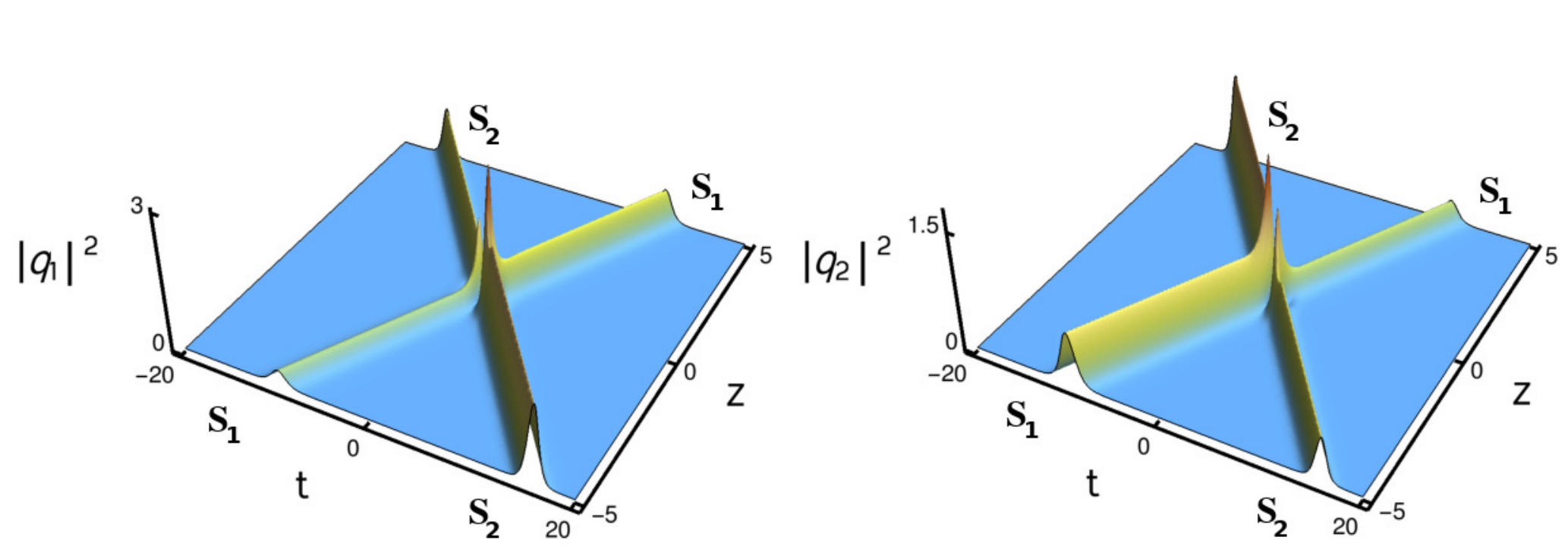}
		\caption{Shape changing collision of degenerate two-solitons: $k_{1}=l_1=1+i$, $k_{2}=l_2=1.51-1.51i$,  $\alpha_{1}^{(1)}=0.5+0.5i$, $\alpha_{2}^{(1)}=\alpha_{1}^{(2)}=\alpha_{2}^{(2)}=1$.\label{fig8}}
	\end{center}	
\end{figure}
	%%%%%%%%%%%%%%%%%%%%%%%%%%%%%%%%%%%%%%%%%%%%%%%%%%%
	\subsubsection{Degenerate two soliton solution and its energy sharing collision}
	Further, the degenerate two-soliton solution can be deduced from the nondegenerate two-soliton solution (\ref{3}) by applying the degenerate limits $k_{1}=l_{1}$ and $k_{2}=l_{2}$. Such degenerate two-soliton solution of the Manakov system is obtained in \cite{Radhakrishnan-pre}. The two-soliton solution can be compactly written in terms of Gram determinants as    
	\bes \bea
	q_j&&=\frac{g^{(j)}}{f}, \quad j=1, 2,
	\label{trans}
	\eea
	where
	\bea
	\hspace{-0.5cm}g^{(j)}=
	\left|
	\begin{array}{ccccc}
		A_{11} & A_{12}&1&0& e^{\eta_1}\\
		A_{21} & A_{22}&0&1& e^{\eta_2}\\
		-1&0 & B_{11} &B_{12} & 0\\
		0&-1 & B_{21} &B_{22} & 0\\
		0 &0& -\alpha_1^{(j)}&-\alpha_2^{(j)} & 0
	\end{array}
	\right|, \quad \quad f= \left|
	\begin{array}{cccc}
		A_{11} &A_{12}& 1&0\\
		A_{21} &A_{22}& 0&1\\
		-1&0 & B_{11}&B_{12} \\
		0&-1 & B_{21}&B_{22} \\
	\end{array}
	\right|,
	\label{2sol-cnls}
	\eea \ees
	in which $A_{ij}=\ds{\frac{e^{\eta_i+\eta_j^*}}{k_i+k_j^*}}$, and $B_{ij}=\kappa_{ji}=\ds{\frac{\left(\alpha_j^{(1)}\alpha_i^{(1)*}+\alpha_j^{(2)}\alpha_i^{(2)*}\right)}{(k_j+k_i^*)}}$, \;\;\;$i,j=1,2$. The above degenerate bright two-soliton solution is characterized by six arbitrary complex parameters $k_1$, $k_2$, $\alpha_1^{(j)}$ and
	$\alpha_2^{(j)}$, $j=1,2$.

	By fixing the wave numbers as $k_{i}=l_{i}, i=1,2,...,N$, the $N$ degenerate vector bright soliton solution can be recovered from the nondegenerate $N$-soliton solutions. In passing we  also note that the nondegenerate one-soliton solution (\ref{15a})-(\ref{15b}) can arise when we fix the parameters $\alpha_{2}^{(1)}=\alpha_{1}^{(2)}=0$ in Eqs. (\ref{trans}) and (\ref{2sol-cnls})  and rename the constants $k_2$ as $l_1$ and $\alpha_{2}^{(2)}$ as $\alpha_{1}^{(2)}$ in the resultant solution. We also note that the above degenerate two-soliton solution (\ref{trans})-(\ref{2sol-cnls}) can also be rewritten from the Gram determinant forms of nondegenerate two-soliton solution (\ref{3}). 
	
	As reported in \cite{Radhakrishnan-pre,kanna-prl,kanna-pre}, the degenerate fundamental solitons ($k_i=l_i$, $i=1,2$) in the Manakov system undergo  shape changing collision due to the intensity redistribution among the modes. The energy redistribution occurs in the degenerate case because of the polarization vectors of the two modes combine with each other in a specific way. This shape changing collision illustrated in Figure \ref{fig8} where the intensity redistribution occurs because of the enhancement of soliton $S_1$ in the first mode and the corresponding suppression of the intensity of the same soliton in the second mode.   
	To hold the conservation of energy between the modes, the intensity of the soliton $S_{2}$ gets suppressed in the first mode and it is enhanced in the second mode. The standard elastic collision occurs (as already noted) for the very special choice of parameters, namely  $\frac{\alpha_1^{(1)}}{\alpha_2^{(1)}}=\frac{\alpha_1^{(2)}}{\alpha_2^{(2)}}$ \cite{Radhakrishnan-pre,kanna-prl}.
	%%%%%%%%%%%%%%%%%%%%%%%%%%%%%%%%%%%%%%%%%%%%%%%%%%%% 
	\subsection{Possible experimental realization of nondegenerate solitons}
	To experimentally observe the nondegenerate vector solitons (single hump/double hump solitons) in the Manakov system one may adopt the mutual-incoherence method that has been used to observe the multi-hump multi-mode solitons experimentally (Ref. \cite{mitchell-prl}). The Manakov solitons (degenerate solitons) can also  be  observed by the same experimental procedure with appropriate modifications (Ref. \cite{anastassiou-prl}). In the following, we briefly envisage how the procedure is given in Ref. \cite{mitchell-prl} can be redesigned to generate the double-humped  nondegenerate soliton as it has been discussed in our work \cite{ramakrishnan-pre}.  
	
	To observe the nondegenerate vector solitons experimentally it is essential to consider two laser sources with different properties so that the wavelength of the second laser beam is different from the first one. Using polarizing beam splitters, each one of the laser beams can be split into ordinary and extraordinary beams. The extraordinary beam coming out from the first source can be further split into two individual fields $F_{11}$ and $F_{12}$ by allowing it to fall on a beam splitter. These two fields are nothing but the reflected and transmitted extraordinary beams coming out from the beam splitter. The intensities of these two fields are different. Similarly, the second beam which is coming out from the second source  can also be split into two fields $F_{21}$ and $F_{22}$ by passing through another beam splitter. The intensities of these two fields are also different. As a result, one can generate four fields that are incoherent to each other. To set the incoherence in phase among these four fields one should allow them to travel a sufficient distance before the coupling is performed. The fields  $F_{11}$ and $F_{12}$ now become nondegenerate two individual solitons in the first mode whereas $F_{21}$ and $F_{22}$ form another set of two nondegenerate solitons in the second mode. The coupling between the fields $F_{11}$ and $F_{21}$ can be performed by combining them using another beam splitter. Similarly, by suitably locating another beam splitter, one can combine the fields $F_{12}$ and $F_{22}$, respectively.  After appropriate coupling is performed the resultant optical field beams can now be focused through two individual cylindrical lenses and the output may be recorded in an imaging system, which consists of a crystal and CCD camera. The collision between the nondegenerate two-solitons in both the modes can now be seen from the recorded images.
	
	To observe the elastic collision between double-humped  nondegenerate solitons, one must make arrangements to vanish the mutual coherence property between the solitons $F_{11}$ and $F_{12}$ in the first mode $q_1$ and $F_{21}$ and $F_{22}$ in the second mode $q_2$ (Ref. \cite{anastassiou-prl}). The four optical beams are now completely independent and incoherent with one another. The collision angle at which the nondegenerate solitons interact should be sufficiently large enough. Under this situation, no energy exchange is expected to occur between the nondegenerate solitons of the two modes. This experimental procedure can also be used to realize multi-humped nondegenerate vector solitons in $N$-CNLS system but with appropriate modification in the initial conditions.
	%%%%%%%%%%%%%%%%%%%%%%%%%%%%%%%%%%%%%
	\subsection{Multi-humped nondegenerate fundamental bright soliton solution in N-CNLS system}
	In this sub-section, we explore the existence of nondegenerate fundamental bright soliton solution for coupled multi-component nonlinear Schr\"{o}dinger equations of Manakov type \cite{kanna-prl,ramakrishnan-jpa}. Here,  we intend to point out the multi-hump nature of the nondegenerate fundamental solitons in the following system of multi-component nonlinear Schr\"{o}dinger equations,  
	\begin{equation}
	\label{eq1}
	iq_{j,z}+q_{j,tt}+2 \sum_{p=1}^{N}|q_{p}|^{2}q_{j}=0,~~~j=1,2,...,N. 
	\end{equation}
	Here, straightaway we provide the nondegenerate fundamental soliton solution of the above $N$-CNLS system, which is derived through the Hirota bilinear method. We note that for detailed derivation one can refer to our recent paper \cite{ramakrishnan-jpa}. The nondegenerate fundamental bright soliton solution $q_j=\frac{g^{(j)}}{f}$, $j=1,2,...,N$, of the $N$-CNLS system written in a more compact form using the following Gram determinants as 
	\begin{eqnarray}
	\label{eq2}
	g^{(N)} =
	\left|
	\begin{array}{ccc}
	A & I & \phi \\
	-I & B & \bf{0}^{T} \\
	\bf{0} & C_{N} & 0
	\end{array}
	\right|,~~~
	%%%%%%%%%%%%%%%%%%%%%%%%%%%
	f =
	\left|
	\begin{array}{cc}
	A & I  \\
	-I & B   
	\end{array}
	\right|,
	\end{eqnarray}
	where the elements of the matrices $A$ and $B$ are
	\begin{eqnarray}
	&&\hspace{-1.7cm} A_{ij}=\frac{e^{\eta_{i}+\eta_{j}^{*}}}{(k_{i}+k_{j}^{*})}, ~ B_{ij}=\kappa_{ji}=\frac{\psi_{i}^{\dagger} \sigma \psi_{j}}{(k_{i}^{*}+k_{j})},~
	%%%%%%%%%%%%%%%%%%%%%%%%%%%
	C_{N} = - 
	\left(
	\begin{array}{ccccc}
	\alpha_{1}^{(1)}, & \alpha_{1}^{(2)}, & .~.~.~,& \alpha_{1}^{(N)} 
	\end{array}
	\right),\nonumber \\
	&&\hspace{-1.7cm}\psi_{j}=
	\left(
	\begin{array}{ccccc}
	\alpha_{1}^{(1)}, &\alpha_{1}^{(2)},& .~ .~ . ~,&\alpha_{1}^{(j)} 
	\end{array}
	\right)^T,
	\phi=
	\left(
	\begin{array}{ccccc}
	e^{\eta_{1}},&	e^{\eta_{2}},&. ~.~.~, &e^{\eta_{n}}  
	\end{array}
	\right)^T, j,n=1,2,..,N.\nonumber
	\end{eqnarray}
	In the above, $g^{(N)}$ and $f$ are $((2^2N)+1)$ and  $(2^2N)$th order determinants, respectively. When $j \neq i$ the elements $\kappa_{ji}$'s in the square matrix $B$  do not exist ($\kappa_{ji}=0$). Then, in the above fundamental soliton solution $T$  denotes the transpose of the matrices $\psi_{j}$ and $\phi$, $\dagger$ represents transpose complex conjugate,  $\sigma=I$ is an (n $\times$ n) identity matrix, $\phi$ is a (n $\times$ 1) column matrix, $\bf{0}$ is a (1 $\times$ n) null matrix, $C_{N}$ is a (1 $\times$ n) row matrix  and  $\psi$ represents a (n $\times$ 1) column matrix. Further, for a given set of $N$ and $j$ values the corresponding elements only exist and all the other elements are equal to zero in  $\psi_{j}$  and  $C_{N}$ matrices. We have verified the reliability  of the nondegenerate fundamental soliton solution (\ref{eq2}) by substituting it in the bilinear equations of $N$-CNLS system along with the following derivative formula of the determinants, $\frac{\partial M}{\partial x}	= \sum_{1\le i,j\le n} \frac{\partial a_{i,j}}{\partial x} \frac{\partial M }{\partial a_{i,j}} = \sum_{1\le i,j\le n} \frac{\partial a_{i,j}}{\partial x} \Delta_{i,j}$, where $\Delta_{i,j}$'s are the cofactors of the matrix $M$, the elementary properties of the determinants  and the bordered determinant properties \cite{hirota-book,vijayajayanthi-epjst}. This action results a pair of Jacobi identities and thus their occurrence confirms the validity of the obtained soliton solution. Multi-hump profile nature is a special feature of the obtained nondegenerate fundamental soliton solution (\ref{eq2}). Such multi-hump structures and their propagation are characterized by $2N$ arbitrary complex wave parameters. The funamental nondegenerate soliton admits a very interesting $N$-hump profile in the present $N$-CNLS system. The number of peaks or humps in the intensity profile of the nondegenerate fundamental soliton solution of the $N$-CNLS system is essentially equal to the number of wave numbers or equivalently the number of components involved.	
	In this system, in general, the nondegenerate solitons propagate with different velocities in different modes but one can make them to propagate with identical velocity by restricting the imaginary parts of all the wave numbers $k_{j}$, $j=1,2,...,N$, to be equal. We wish to note that the degenerate fundamental bright soliton solution of the $N$-CNLS system can be obtained by setting all the wavenumbers $k_j$ , $j = 1, 2, ..., N$ ,
	as identical, $k_j=k_1 $, $j = 1, 2, ..., N$. It leads to single-hump intensity profiles only in all the modes \cite{kanna-prl}. Very interestingly, the $N$-CNLS system (\ref{e1}) also admits a special kind of multi-humped partially nondegenerate fundamental soliton solution for a lesser number of restrictions on the wave numbers, as we have explained in \cite{ramakrishnan-jpa}. Consequently, in this partially nondegenerate case, the number of humps is not equal to the number of components.
	
	In order to indicate the multi-hump nature of the nondegenerate soliton, here we demonstrate such special feature in the case of $3$-CNLS and $4$-CNLS systems.  As a specific example, we can easily check that such multi-parameter solution admits a novel asymmetric triple-hump profile in the case of $3$-CNLS system when we fix the velocity as $k_{1I}=k_{2I}=k_{3I}=0.5$. The other parameter values are chosen as $k_{1R} = 0.53$, $k_{2R} = 0.5$, $k_{3R} = 0.45$, $\alpha_{1}^{(1)} = 0.65 + 0.65i$, $\alpha_{1}^{(2)} = 0.45 - 0.45i$ and $\alpha_{1}^{(3)} = 0.35 + 0.35i$. In Figure 9(a), we display the  asymmetric triple-hump profiles in all the components for the above choice of parameter values. Then, the nondegenerate one-soliton  solution in the $4$-CNLS system exhibits asymmetric quadruple-hump profile in all the modes. Such novel quadruple-hump profile is displayed in Figure 9(b) for the parameter values $k_1 = 0.48 + 0.5i$, $k_2 = 0.5 + 0.5i$, $k_3 = 0.53 + 0.5i$, $k_4 = 0.55 + 0.5i$, $\alpha_{1}^{(1)} = 0.65 + 0.65i$, $\alpha_{1}^{(2)} = 0.55 - 0.55i$, $\alpha_{1}^{(3)} = 0.45 + 0.45i$ and $\alpha_{1}^{(4)} = 0.35 - 0.35i$. We remark that the nondegenerate fundamental soliton solution reduces to a double-humped partially nondegenerate soliton by considering a restriction $k_1=k_2$ (or $k_2=k_3$) \cite{ramakrishnan-jpa}.
	\begin{figure}[H]
		\begin{center}
			\includegraphics[width=5.0 cm]{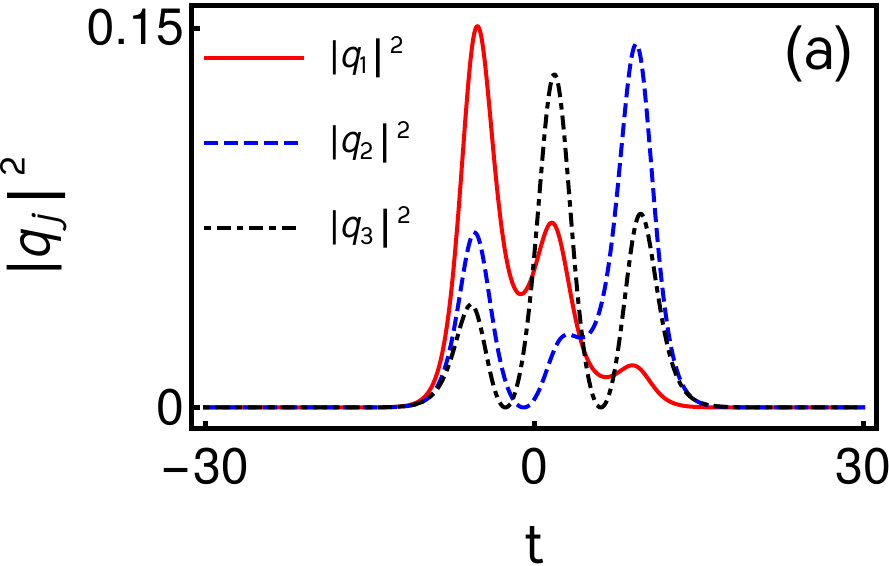}~	\includegraphics[width=5.0 cm]{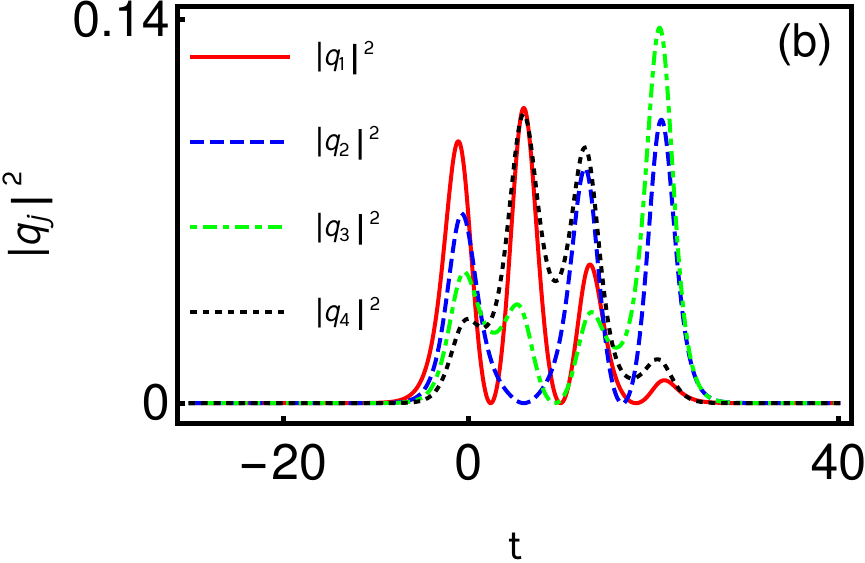}
			\caption{(a) denotes triple-hump profiles of nondegenerate fundamental soliton in the $3$-CNLS system and (b) represents a quadruple-humped nondegenerate soliton profiles in $4$-CNLS system.  \label{fig9}}
		\end{center}
	\end{figure}
	
	In general, to derive nondegenerate $N$-soliton solution of the $N$-CNLS system, we have to consider a more general form of the starting solutions $g_1^{(j)}=\sum_{l,j=1}^{N}\alpha_l^{(j)}e^{\eta_l^{(j)}}$, $\eta_l^{(j)}=k_l^{(j)}t+ik_l^{(j)2}z$ to the lowest order set of  $N$ linear PDEs $ig_{1,z}^{(j)}+g_{1,tt}^{(j)}=0$, $j=1,2,...,N$. This choice of initial seed solutions yield a very complicated nondegenerate $N$-soliton solution. We do not provide the details of such intricate form here for brevity and it will be published elsewhere.
	%%%%%%%%%%%%%%%%%%%%%%%%%%%%%%%%%%%%%%%%%%%%%%%%%%%%%%%%%%%%%%%%%%%%%%%%%%%%
\end{paracol}
%%%%%%%%%%%%%%%%%%%%%%%%%%%%%%%%%%%%%%%%%%
%%%%%%%%%%%%%%%%%%%%%%%%%%%%%%%%%%%%%%%%%%%%% 
\begin{paracol}{2}
	%\linenumbers
	\switchcolumn
	%%%%%%%%%%%%%%%%%%%%%%%%%%%%%%%%%%%%%%%%%%%%%%%%%%%%%
	\section{Nondegenerate and degenerate bright solitons in mixed 2-CNLS system}
	This section is essentially devoted to show the existence of nondegenerate fundamental bright solitons in the mixed 2-CNLS system or Eq. (\ref{cnls}) with $\sigma_1=+1$ and $\sigma_2=-1$. In this section, we also point out how the degenerate fundamental bright soliton can be captured from the obtained nondegenerate one-soliton solution and indicate its energy sharing collision. In order to write down the analytical form of nondegenerate fundamental soliton solution, one has to follow the same procedure that has been adopted to derive such a solution in the case of the Manakov system. Since the solution construction methodology has been extensively described in Refs. \cite{stalin-prl,stalin-pla,ramakrishnan-pre} and in the earlier section, here we immediately present the explicit form of nondegenerate fundamental soliton solution of the mixed $2$-CNLS system. It reads as
	\bes
	\begin{eqnarray}
	q_{1}=\frac{g_1^{(1)}+g_3^{(1)}}{1+f_2+f_4}=\frac{1}{D}(\alpha_{1}^{(1)} e^{\eta_{1}}+e^{\eta_{1}+\xi_{1}+\xi_{1}^*+\Delta_{1}^{(1)}}), \label{40a}\\ 
	q_{2}=\frac{g_1^{(2)}+g_3^{(2)}}{1+f_2+f_4}=\frac{1}{D}(\alpha_{1}^{(2)} e^{\xi_{1}}+e^{\eta_{1}+\eta_{1}^*+\xi_{1}+\Delta_{1}^{(2)}}).
	\label{40b}
	\end{eqnarray}\ees
	Here $D=1+e^{\eta_{1}+\eta_{1}^{*}+\delta_{1}}+e^{\xi_{1}+\xi_{1}^{*}+\delta_{2}}+e^{\eta_{1}+\eta_{1}^{*}+\xi_{1}+\xi_{1}^{*}+\delta_{11}}$, $e^{\Delta_{1}^{(1)}}=-\frac{(k_{1}-l_{1})\alpha_{1}^{(1)}|\alpha_{1}^{(2)}|^2}{(k_{1}+l_{1}^*)(l_{1}+l_{1}^*)^{2}}$, $e^{\Delta_{1}^{(2)}}=\frac{(k_{1}-l_{1})|\alpha_{1}^{(1)}|^2\alpha_{1}^{(2)}}{(k_{1}+k_{1}^*)^{2}(k_{1}^*+l_{1})}$,  $e^{\delta_{1}}=\frac{|\alpha_{1}^{(1)}|^2}{(k_{1}+k_{1}^{*})^{2}}$, $e^{\delta_{2}}=-\frac{ |\alpha_{1}^{(2)}|^2}{(l_{1}+l_{1}^*)^{2}}$ and 
	$e^{\delta_{11}}=-\frac{|k_{1}-l_{1}|^{2} |\alpha_{1}^{(1)}|^{2}|\alpha_{1}^{(2)}|^{2}}{(k_{1}+k_{1}^*)^{2}|k_{1}+l_{1}^*|^2(l_{1}+l_{1}^*)^{2}}$. Like in the Manakov system, the two complex parameters $\alpha_{1}^{(j)}$'s, $j=1,2$, and the two wave numbers $k_1$, and $l_1$ describes the behaviour of the above general form of one-soliton solution (\ref{40a})-(\ref{40b}). By rewriting the solution (\ref{40a})-(\ref{40b}) in hyperbolic form, as it has been done in Eqs. (\ref{17a}) and (\ref{17b}), we find the amplitude, velocity and central position of the soliton in the first mode is $2k_{1R}$, $2k_{1I}$ and $\frac{\phi_1}{2l_{1R}}=\frac{1}{2l_{1R}}\log\frac{(l_1-k_1|\alpha_1^{(2)}|^2)}{(k_1+l_1^*)(l_1+l_1^*)^2}$, respectively. In the second mode, the amplitude, velocity and central position of the soliton are defined by $2l_{1R}$, $2l_{1I}$ and $\frac{\phi_2}{2k_{1R}}=\frac{1}{2k_{1R}}\log\frac{(k_1-l_1|\alpha_1^{(1)}|^2)}{(k_1^*+l_1)(k_1+k_1^*)^2}$, respectively. In the mixed $2$-CNLS system too, the nondegenerate fundamental soliton propagates in the two modes either  with identical velocity ($v_1=v_2=2k_{1I}$) or with non-identical velocity ($v_1=2k_{1I}\neq v_2=2l_{1I}$) depending on the restriction on the imaginary parts of the wave numbers $k_1$ and $l_1$. The solution (\ref{40a})-(\ref{40b}) always shows singular behaviour due to the presence of negative sign in the constant terms $e^{\delta_2}$ and $e^{\delta_{11}}$ except for $k_1=l_1$. This negative sign essentially arises because of the presence of defocusing nonlinearity of the mixed CNLS system. The singularity nature of the solution (\ref{40a})-(\ref{40b}) is depicted in Figure \ref{fig10} with the parameter values $k_1 = 1.25 + 0.45i$, $l_1 = -0.5 + 0.45i$, $\alpha_{1}^{(1)} = 0.3$ and $\alpha_{1}^{(2)} = i$. We note that the singular nature of the soliton has been recently discussed in the context of singular optics \cite{sakaguchi-pre}. The nondegenerate higher order bright solitons can also be obtained in a similar way and one can  analyse their collision dynamics.
	\begin{figure}[H]
		\begin{center}
			\includegraphics[width=9.0 cm]{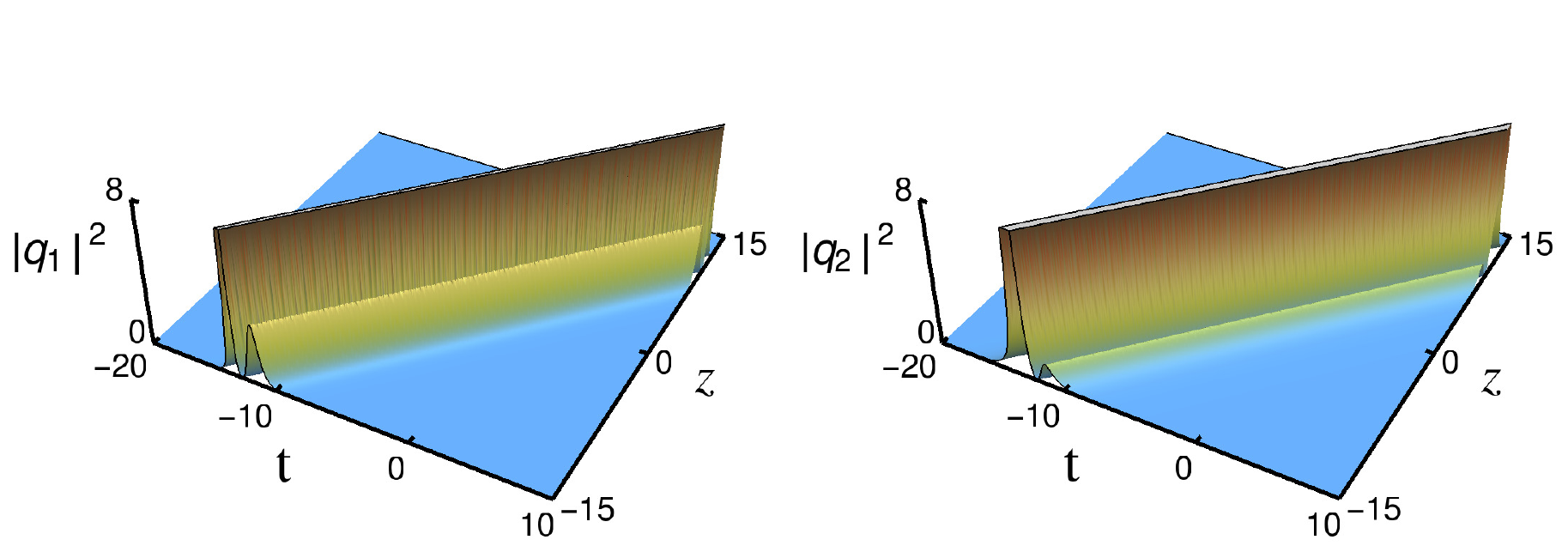}
			\caption{The singular double-hump profiles of the nondegenerate one-soliton solution (\ref{40a})-(\ref{40b}) of the mixed $2$-CNLS system.  \label{fig10}}
		\end{center}
	\end{figure}
	By imposing the limit $k_1=l_1$ in the solution (\ref{40a})-(\ref{40b}), one can capture following degenerate fundamental vector bright soliton solution of the mixed $2$-CNLS system, $q_{j}=k_{1R} \hat{A_{j}}e^{i\eta_{1I}}\sech(\eta_{1R}+\frac{R}{2})$,  where $\eta_{1R}=k_{1R}(t-2k_{1I}z)$, $\eta_{1I}=k_{1I}t+(k_{1R}^{2}-k_{1I}^{2})z$, $\hat{A_{j}}=\frac{\alpha_{1}^{(j)}}{\sqrt{(|\alpha_{1}^{(1)}|^2-|\alpha_{1}^{(2)}|^2)}}$, $e^{R}=\frac{(|\alpha_{1}^{(1)}|^2-|\alpha_{1}^{(2)}|^2)}{(k_{1}+k_{1}^*)^2}$, $j=1,2$. The latter degenerate bright soliton solution always admits the non-singular single-hump intensity profile when $|\alpha_{1}^{(1)}|>|\alpha_{1}^{(2)}|$. The degenerate multi-soliton solutions and their interesting collision property have been already discussed in \cite{kanna-pre2006}. The two-soliton solution of the mixed $2$-CNLS system can be easily obtained by replcing $B_{ij}$ as $B_{ij}=\kappa_{ji}=\ds{\frac{\left(\alpha_j^{(1)}\alpha_i^{(1)*}-\alpha_j^{(2)}\alpha_i^{(2)*}\right)}{(k_j+k_i^*)}}$, $i,j=1,2$ in the degenerate two-soliton solution (\ref{trans})-(\ref{2sol-cnls}) of the Manakov system.  However, here we indicate the special collision dynamics exhibited by the degenerate bright solitons only through a graphical demonstration as we illustrated below in Figure \ref{fig11} for the parametric choice $k_1=1-i$, $k_2=1.7+I$, $\alpha_{1}^{(1)}=1+i$, $\alpha_{2}^{(1)}=1-i$, $\alpha_{1}^{(2)}=0.5+0.3i$ and $\alpha_{2}^{(2)}=0.7$. From Figure \ref{fig11}, we identify that during the collision process of the degenerate two bright solitons $S_1$ and $S_2$ in the present mixed $2$-CNLS system, the intensity of the soliton $S_1$ is enhanced in all the modes. In contradiction to this, the intensity of the other soliton $S_2$ is suppressed in both the modes. Therefore, such a  special property of enhancement of the intensity of a given soliton always occurs in the mixed $2$-CNLS system. One may find the details of energy conservation in Ref. \cite{kanna-pre2006}. Additionally, we also observe the amplitude dependent phase shifts in each of the modes. This energy sharing collision is quite different from the shape changing collision of the Manakov system. The collision scenario is depicted in Figure \ref{fig11} can be viewed as a signal amplification process, in which the soliton $S_1$ refers as a signal wave and the soliton $S_2$ represents as a pump wave. During this amplification process, there is no external amplification medium is employed and is without the introduction of any noise \cite{kanna-pre2006}. We point out that the standard NLS soliton-like collision can be recovered by imposing the restriction $\frac{\alpha_{1}^{(1)}}{\alpha_{2}^{(1)}}=\frac{\alpha_{1}^{(2)}}{\alpha_{2}^{(2)}}$.
	\begin{figure}[H]
		\begin{center}
			\includegraphics[width=9.0 cm]{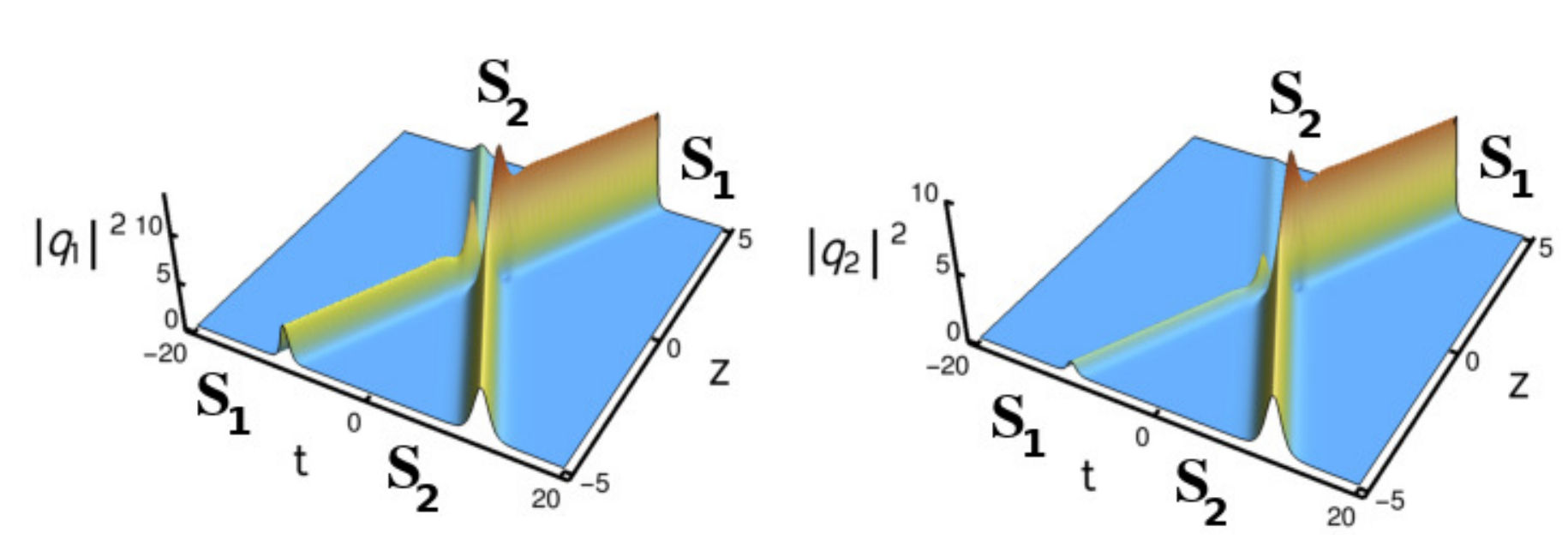}
			\caption{Energy sharing collision of degenerate two bright solitons of the mixed $2$-CNLS system \cite{kanna-pre2006}.   \label{fig11}}
		\end{center}
	\end{figure} 	
%%%%%%%%%%%%%%%%%%%%%%%%%%%%%%%%%%%%%%%%%%%%%%%%%%%%%%
\section{Existence of nondegenerate and degenerate bright solitons in two-component coherently coupled nonlinear Schr\"{o}dinger system}
	Now, we intend to derive a more general form of nondegenerate fundamental bright soliton solution of the two-component CCNLS system (\ref{2ccnlseqn}).  In this section, we also mention the already known degenerate one bright soliton solution and illustrate its fascinating energy switching collision property through a graphical demonstration. To obtain the explicit forms of the nondegenerate soliton solution, we adopt a non-standard bilinearization procedure in which an appropriate number of auxiliary functions have been introduced to match the number of bilinear equations with the number of bilinearizing variables. This procedure was developed by Gilson et al \cite{gilson-pre} for the Sasa-Satsuma higher order nonlinear Schr\"{o}dinger equations and by Kanna et al \cite{kanna-jpa-2010,kanna-jpa-2011} for the coherently coupled nonlinear Schr\"{o}dinger equations.  
	By adopting this technique, we get the following 
	correct bilinear equations of system (\ref{2ccnlseqn}) through the  bilinearizing transformation $
	q_j=\frac{g^{(j)}}{f}$, $j=1,2,$
	to equation (\ref{2ccnlseqn}) with the introduction of an auxiliary function $s$. The set of bilinear equations are
	\label{be}\bea
	D_1(g^{(j)}\cdot f) = \gamma s g^{(j)*}, ~j=1,2,~D_2(f \cdot f) = 2 \gamma \left(\sum_{j=1}^2 |g^{(j)}|^2\right), ~
	s\cdot f=\sum_{j=1}^2 (g^{(j)})^2, \label{3beq}
	\eea
	where $D_1=iD_z+D_t^2$ and $D_2=D_t^2$. Here $g^{(j)}$'s and $f$ are complex and real functions, respectively, $*$ denotes the complex conjugate. After the bilinearization, essentially we follow the procedure that has been described in \cite{kanna-jpa-2010} for the  degenerate case but now with the general forms of seed solutions $
	g_1^{(1)}=\al_1 e^{\eta_1}$, $g_1^{(2)}=\ba_1 e^{\xi_1}$, $\eta_1=k_1t+ik_1^2z$, $\xi_1=l_1t+il_1^2z$. While doing so, the series expansions get truncated as $g^{(j)}=\epsilon g_1^{(j)}+\epsilon^3 g_3^{(j)}+\epsilon^5 g_5^{(j)}+\epsilon^7 g_7^{(j)}$, $f=1+\epsilon^2 f_2+\epsilon^4 f_4+\epsilon^6 f_6+\epsilon^8 f_8$ and $s=\epsilon^2 s_2+\epsilon^4 s_4+\epsilon^6 s_6$. By substituting the obtained forms of the unknown functions in the appropriate places, we get the following a more general form of nondegenerate coherently coupled fundamental bright soliton solution of $2$-CCNLS system (\ref{2ccnlseqn}),
	\begin{eqnarray}
	\hspace{-1.1cm}q_{1}(z,t)&=&\frac{1}{f}\bigg(\alpha_{1}e^{\eta_{1}}+
	e^{2\eta_{1}+\eta_{1}^{*}+\Delta_{11}}+
	e^{\eta_1^*+2\xi_{1}+\Delta_{12}}+
	e^{\eta_{1}+\xi_{1}+\xi_{1}^{*}+\Delta_{13}}+
	e^{\eta_{1}+2(\eta_{1}^{*}+\xi_{1})+\Delta_{14}}\nonumber\\
	&&~~~+e^{\eta_{1}+2(\xi_{1}+\xi_{1}^{*})+\Delta_{15}}+e^{2\eta_{1}+\eta_{1}^{*}+\xi_{1}+\xi_{1}^{*}+\Delta_{16}}+e^{2(\eta_{1}+\xi_{1}+\xi_{1}^{*})+\eta_{1}^{*}+\Delta_{17}}\bigg),\nonumber\\
	\hspace{-0.5cm}q_2(z,t)&=&\frac{1}{f}\bigg(\beta_{1}e^{\xi_{1}}+
	e^{2\xi_{1}+\xi_{1}^{*}+\Delta_{21}}+e^{\xi_{1}^{*}+2\eta_{1}+\Delta_{22}}+e^{\xi_{1}+\eta_{1}+\eta_{1}^{*}+\Delta_{23}}+e^{\xi_{1}+2(\xi_{1}^{*}+\eta_{1})+\Delta_{24}}\nonumber\\
	&&~~~+e^{\xi_{1}+2(\eta_{1}^{*}+\eta_{1})+\Delta_{25}}+e^{2\xi_{1}+\xi_{1}^{*}+\eta_{1}+\eta_{1}^{*}+\Delta_{26}}+e^{2(\eta_{1}+\eta_{1}^{*}+\xi_{1})+\xi_{1}^{*}+\Delta_{27}}\bigg),\nonumber\\
	\hspace{-0.5cm}f&=&1+e^{\eta_{1}+\eta_{1}^{*}+\delta_{1}}+
	e^{\xi_{1}+\xi_{1}^{*}+\delta_{2}}+
	e^{2(\eta_{1}+\eta_{1}^{*})+\delta_{3}}+
	e^{2(\eta_{1}+\xi_{1}^{*})+\delta_{4}}+
	e^{2(\xi_{1}+\eta_{1}^{*})+\delta_{5}}\nonumber\\
	&&+e^{2(\xi_{1}+\xi_{1}^{*})+\delta_{6}}+
	e^{(\eta_{1}+\eta_{1}^{*}+\xi_{1}+\xi_{1}^{*})+\delta_{7}}+
	e^{2(\eta_{1}+\eta_{1}^{*})+\xi_{1}+\xi_{1}^{*}+\nu_{1}}\nonumber\\
	&&+e^{2(\xi_{1}+\xi_{1}^{*})+\eta_{1}+\eta_{1}^{*}+\nu_{2}}+
	e^{2(\eta_{1}+\eta_{1}^{*}+\xi_{1}+\xi_{1}^{*})+\nu_{3}}.
	\label{42}
	\end{eqnarray}
	The various constants which appear in the above solution are defined by
	\begin{eqnarray}
	&& \hspace{-0.5cm}	e^{\Delta_{11}}=\frac{\gamma \alpha_{1} |\alpha_{1}|^{2}}{2\kappa_{11}}, e^{\Delta_{12}}=\frac{\gamma \alpha_{1}^{*}\beta_{1}^2}{2 \theta_{1}^{*2}}, e^{\Delta_{13}}=\frac{\gamma \alpha_{1} |\beta_{1}|^{2} \rho_{1}}{ \theta_{1}l_{11}}, e^{\Delta_{14}}=\frac{\gamma^{2} \rho_{1}^{2}\alpha_{1}^{*}\beta_{1}^{2}|\alpha_{1}|^{2}}{4 \kappa_{11} \theta_{1}^{*4}},\nonumber\\ &&\hspace{-0.5cm}e^{\Delta_{15}}=\frac{\gamma^{2} \rho_{1}^{2}\alpha_{1} |\beta_{1}|^{4} }{4 l_{11}^{2} \theta_{1}^{2}}, e^{\Delta_{16}}=\frac{\gamma^{2}\rho_{1}^{2}\rho_{1}^{*}\alpha_{1}|\alpha_{1}|^{2}|\beta_{1}|^{2}}{2\kappa_{11}l_{11}\theta_{1}^{2}\theta_{1}^{*}}, e^{\Delta_{17}}=\frac{\gamma^{3}\rho_{1}^{4}{\rho_{1}^{*}}^{2}\alpha_{1}|\alpha_{1}|^{2}|\beta_{1}|^{4}}{8\kappa_{11}l_{11}^{2}\theta_{1}^{4}{\theta_{1}^{*}}^{2}},\nonumber	\\
	&&\hspace{-0.5cm}e^{\Delta_{21}}=\frac{\gamma \beta_{1} |\beta_{1}|^{2}}{2 l_{11}}, e^{\Delta_{22}}=\frac{\gamma \alpha_{1}^{2} \beta_{1}^{*}}{2\theta_{1}^{2}}, e^{\Delta_{23}}=-\frac{\gamma |\alpha_{1}|^{2} \beta_{1} \rho_{1}}{\theta_{1}^{*} \kappa_{11}}, e^{\Delta_{24}}=\frac{\gamma^{2}\rho_{1}^{2}\alpha_{1}^{2}|\beta_{1}|^{2}\alpha_{1}^{*}}{4l_{11}\theta_{1}^{4}},\nonumber\\
	&&\hspace{-0.5cm}e^{\Delta_{25}}=\frac{\gamma^{2}\rho_{1}^{2}|\alpha_{1}|^{4}\beta_{1}}{4\kappa_{11}^{2}\theta_{1}^{*2}}, e^{\Delta_{26}}=-\frac{\gamma^{2}\rho_{1}^{2}\rho_{1}^{*}\beta_{1}|\alpha_{1}|^{2}|\beta_{1}|^{2}}{2\kappa_{11}l_{11}\theta_{1}\theta_{1}^{*2}}, e^{\Delta_{27}}=\frac{\gamma^{3}\rho_{1}^{4}\rho_{1}^{*2}\beta_{1}|\alpha_{1}|^{4}|\beta_{1}|^{2}}{8\kappa_{11}^{2}l_{11}\theta_{1}^{2}\theta_{1}^{*4}},\nonumber\\
	&&\hspace{-0.5cm} e^{\delta_{1}}=\frac{\gamma |\alpha_{1}|^{2}}{\kappa_{11}}, e^{\delta_{2}}=\frac{\gamma |\beta_{1}|^{2}}{l_{11}}, e^{\delta_{3}}=\frac{\gamma^{2} |\alpha_{1}|^{4}}{4 \kappa_{11}^{2}}, e^{\delta_{4}}=\frac{\gamma^{2}\alpha_{1}^{2} \beta_{1}^{*2}}{4 \theta_{1}^{4}}, e^{\delta_{5}}=\frac{\gamma^{2} \alpha_{1}^{*2} \beta_{1}^{2}}{4\theta_{1}^{*4}},\nonumber\\
	&&\hspace{-0.5cm}e^{\delta_{6}}=\frac{\gamma^{2} |\beta_{1}|^{4}}{4 l_{11}^{2}}, e^{\delta_{7}}=\frac{\gamma^{2} |\rho_{1}|^{2} |\alpha_{1}|^{2} |\beta_{1}|^{2}}{\kappa_{11} l_{11} |\theta_{1}|^{2}}, e^{\nu_{1}}=\frac{\gamma^{3}|\rho_{1}|^{4}|\alpha_{1}|^{4}|\beta_{1}|^{2}}{4\kappa_{11}^{2}l_{11}|\theta_{1}|^{4}},\nonumber \\
	&&\hspace{-0.5cm}e^{\nu_{2}}=\frac{\gamma^{3}|\rho_{1}|^{4}|\alpha_{1}|^{2}|\beta_{1}|^{4}}{4\kappa_{11}l_{11}^{2}|\theta_{1}|^{2}},~ e^{\nu_{3}}=\frac{\gamma^{4}|\rho_{1}|^{8}|\alpha_{1}|^{4}|\beta_{1}|^{4}}{16\kappa_{11}^{2}l_{11}^{2}|\theta_{1}|^{8}},   l_{11}=(l_1+l_1^*)^2,
	\nonumber\\
	&&\hspace{-0.5cm}\theta_{1}=(k_1+l_1^*),~\rho_{1}=(k_1-l_1),~\kappa_{11}=(k_1+k_1^*)^2.\nonumber
	\end{eqnarray}
	The auxiliary function $s(z,t)$is found to be,
	$s=\alpha_1^2e^{2\eta_1}+\beta_1^2e^{2\xi_1}+e^{2\eta_1+\xi_1+\xi_1^*+\phi_1}\\+e^{2\xi_1+\eta_1+\eta_1^*+\phi_2}+e^{2(\eta_1+\eta_1^*+\xi_1)+\phi_3}+e^{2(\eta_1+\xi_1^*+\xi_1)+\phi_4}$, $e^{\phi_1}=\frac{\gamma\rho_{1}^2\alpha_1^2|\beta_1|^2}{\theta_{1}^2l_{11}}$, $e^{\phi_2}=\frac{\gamma\rho_{1}^2\beta_1^2|\alpha_1|^2}{\theta_{1}^{*2}\kappa_{11}}$, $e^{\phi_3}=\frac{\gamma^2\rho_{1}^4\beta_1^2|\alpha_1|^4}{4\theta_{1}^{*4}\kappa_{11}^2}$, $e^{\phi_4}=\frac{\gamma^2\rho_{1}^4\alpha_1^2|\beta_1|^4}{4\theta_{1}^{4}l_{11}^2}$. The shape of the coherently coupled nondegenerate fundamental soliton solution (\ref{42}) is governed by the four complex parameters $k_1$, $l_1$,  $\alpha_1$ and $\beta_1$. Due to the presence of coherent coupling among the two fields $q_1$ and $q_2$ (or four-wave mixing effect) and the additional wave number, the solution  (\ref{42}) admits rich geometrical structures, such as a breather, a quadruple-hump, a triple-hump, a double-hump, a flattop and a single-hump profiles under a suitable choice of parameter values. We display a novel non-trivial breathing nondegenerate fundamental soliton profile in Figure \ref{fig12}. To draw this figure, we have fixed the parametric values as $\gamma=4$, $k_1=2.5+0.5i$, $l_1=1.65+0.5i$, $\alpha_1=0.5+0.5i$ and $\beta_1=1-i$. The breathing nature of the multi-hump profile of the nondegenerate soliton in the present $2$-CCNLS system cannot be observed in the degenerate case \cite{kanna-jpa-2010,kanna-jpa-2011} as described below. We note that one can also derive the nondegenerate multi-soliton solutions to the $2$-CCNLS system. However, the resultant expressions will be cumbersome due to the presence of the four-wave mixing effect. 
	
	In order to obtain the degenerate one-soliton solution, one has to impose the wave number restriction $k_1=l_1$ in Eq. (\ref{42}). This results in the following explicit degenerate bright one-soliton solution,
	\bea
	q_1=\frac{\alpha_1 e^{\eta_1}+e^{2\eta_1+\eta_1^*+\Delta_{1}}}{1+e^{\eta_1+\eta_1^*+R_1}+e^{2\eta_1+2\eta_1^*+\delta_{11}}},~
	q_2=\frac{\beta_1 e^{\eta_1}+e^{2\eta_1+\eta_1^*+\Delta_{2}}}{1+e^{\eta_1+\eta_1^*+R_1}+e^{2\eta_1+2\eta_1^*+\delta_{11}}}, \label{43}
	\eea
	where the auxiliary function is reduced to the form 
	$
	s=(\alpha_1^2+\beta_1^2) e^{2\eta_1}$.
	Here,  
	$\eta_1=k_1(t+i k_1 z)$, $
	e^{\Delta_{1}}=\frac{\gamma \alpha_1^* (\alpha_1^2+\beta_1^2)}{2 (k_1+k_1^*)^2}$, $e^{\Delta_{2}}=\frac{\gamma \beta_1^* (\alpha_1^2+\beta_1^2)}{2 (k_1+k_1^*)^2}$,
	$e^{R_1}=\frac{\gamma (|\alpha_1|^2+|\beta_1|^2)}{(k_1+k_1^*)^2}$,
	$e^{\delta_{11}}=\frac{\gamma^2(\alpha_1^2+\beta_1^2) (\alpha_1^{*2}+\beta_1^{*2})}{4 (k_1+k_1^*)^4}$. The above degenerate solution (\ref{43}) is characterized by only two complex parameters $\alpha_1$ and $\beta_1$ and a single complex wave number $k_1$. We point out that the degenerate solution (\ref{43}) is classified as coherently coupled bright soliton and incoherently coupled bright soliton depending on the presence/absence of the auxiliary function $s$ \cite{kanna-jpa-2010}. If the restriction, $\alpha_1^2+\beta_1^2=0$ is imposed, where the auxiliary function $s$ becomes zero, in the solution (\ref{43}), then the resultant solution is called as ICS \cite{kanna-jpa-2010}. Due to this restriction, the coherent coupling among the fields $q_1$ and $q_2$ vanishes. Under the latter restriction, the analytical form of ICS is reduced from the solution (\ref{43}) as
	\bea
	q_1= A_1 \sech(\eta_{1R}+\frac{R_1}{2})e^{i\eta_{1I}},~q_2=\pm q_1.
	\label{ics}
	\eea
	Here, 
	$A_1=\frac{\alpha_1}{2}e^{-\frac{R_1}{2}}$, $R_1=\mbox{log}\left(\frac{2\gamma|\alpha_1|^2}{(k_1+k_1^*)^2}\right)$, $\eta_{1R}=k_{1R}(t-2k_{1I}z)$ and $\eta_{1I}=k_{1I}{t}+(k_{1I}^2-k_{1R}^2){z}$. From the above solution, it is evident that the ICS always admits a `sech'-type intensity profile only. However, very interestingly, a novel double-hump profile arises in the degenerate case when the auxiliary function is non-zero. That is, for $\alpha_1^2+\beta_1^2\neq 0$ the coherent coupling among the optical fields is established. Thus, the solution (\ref{43}) admits double-hump profile as  demonstrated below in Figure \ref{fig13}. But, in the degenerate case, even the presence of single wave number $k_1$ and the four wave mixing effect can induce only the double-hump profile apart from a flattop profile. We do not present the degenerate two-soliton solution of the $2$-CCNLS system for brevity. However, the explicit form of the degenerate two-soliton solution has been given in \cite{kanna-jpa-2010,kanna-jpa-2011}.
	\begin{figure}[H]
	\begin{center}
		\includegraphics[width=10.0cm]{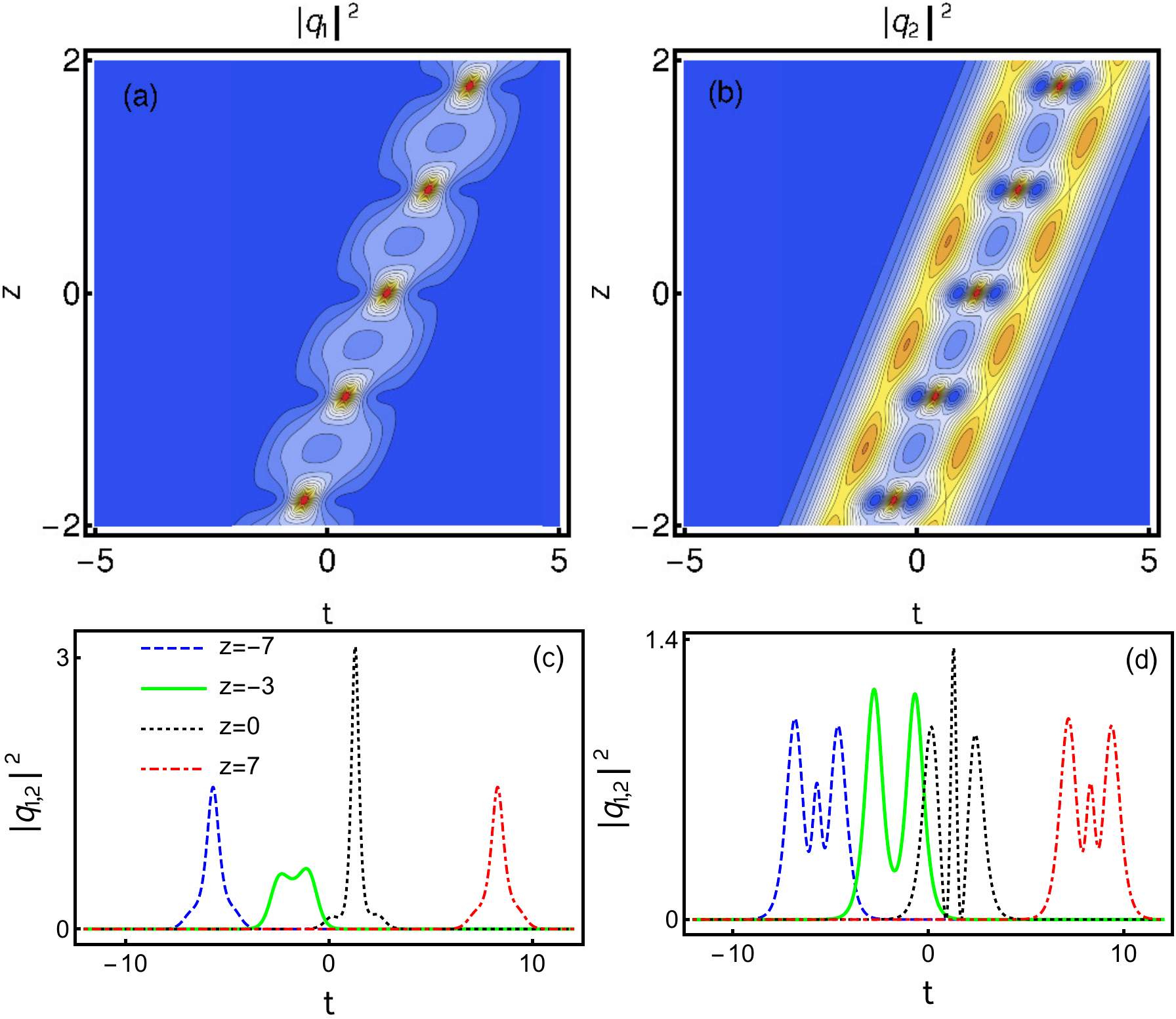}
		\caption{The figures (a) and (b) denote the contour plots of the breathing non-degenerate fundamental bright soliton of the $2$-CCNLS system and the corresponding line plots are drawn for various $z$ values in figures (c) and (d).  \label{fig12}}
	\end{center}
\end{figure}
\begin{figure}[H]	
	\widefigure
	\includegraphics[width=6cm]{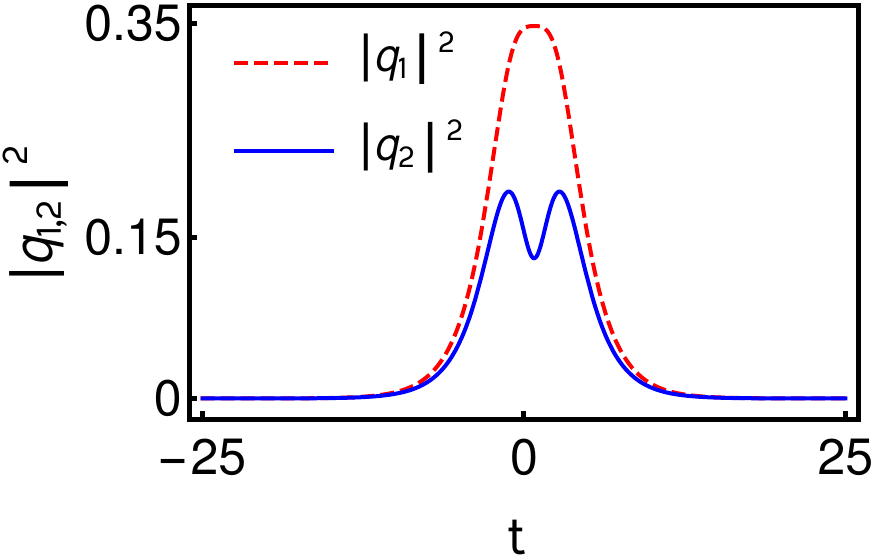}
	\caption{A typical degenerate bright soliton profiles  in the $2$-CCNLS system is drawn for the values $\gamma=2$, $k_1=0.5+0.5i$,$\alpha_1=0.72+0.5i$ and $\beta_1=0.5-0.42i$.    \label{fig13}}
\end{figure}

	In addition to the above, we wish to specify the fascinating shape changing collision of degenerate solitons in the $2$-CCNLS system. Especially, we discuss the collision between the coherently coupled soliton (\ref{43}) and incoherently coupled soliton (\ref{ics}). As an example, we illustrate such a novel collision scenario in Figure \ref{fig14}. In order to display both CCS and ICS in this figure we choose the parametric values as $\gamma=2$, $k_1=1.9+i$, $k_2=2.1-i$, $\alpha_1=0.5i$, $\alpha_2=0.5+0.5i$, $\beta_1=1.5$ and $\beta_2=0.5-0.5i$. In Figure \ref{fig14}, we refer the soliton $S_1$ as CCS and the soliton $S_2$ as ICS. This figure clearly explains that the CCS $S_1$ encounters intensity/energy switching in all the modes. In contradiction to this, the ICS  $S_2$ undergoes elastic collision with a finite phase shift as specified in \cite{kanna-jpa-2010}. Consequently, the CCS $S_1$ switches its double-hump intensity profile to the single-hump profile in the first component and  it is reversed in the second component without affecting the structure of ICS $S_2$. In this type of energy switching collision scenario, the energy in the individual component is not conserved. However, the total energy, $\int_{-\infty}^{+\infty}(|q_1|^2+|q_2|^2)dt$, is conserved. The detailed discussion on this collision scenario and its asymptotic analysis has been carried out   in \cite{kanna-jpa-2011}. We also note that elastic collision always occurs during the collision among the two coherently coupled solitons and it is true in the case of collision between two incoherently coupled solitons too. We remark that the generalization of the above outcome for the multi-component CCNLS system has been established in \cite{kanna-jpa-2011} with exciting results.
	\begin{figure}[H]
		\begin{center}
			\includegraphics[width=9.5 cm]{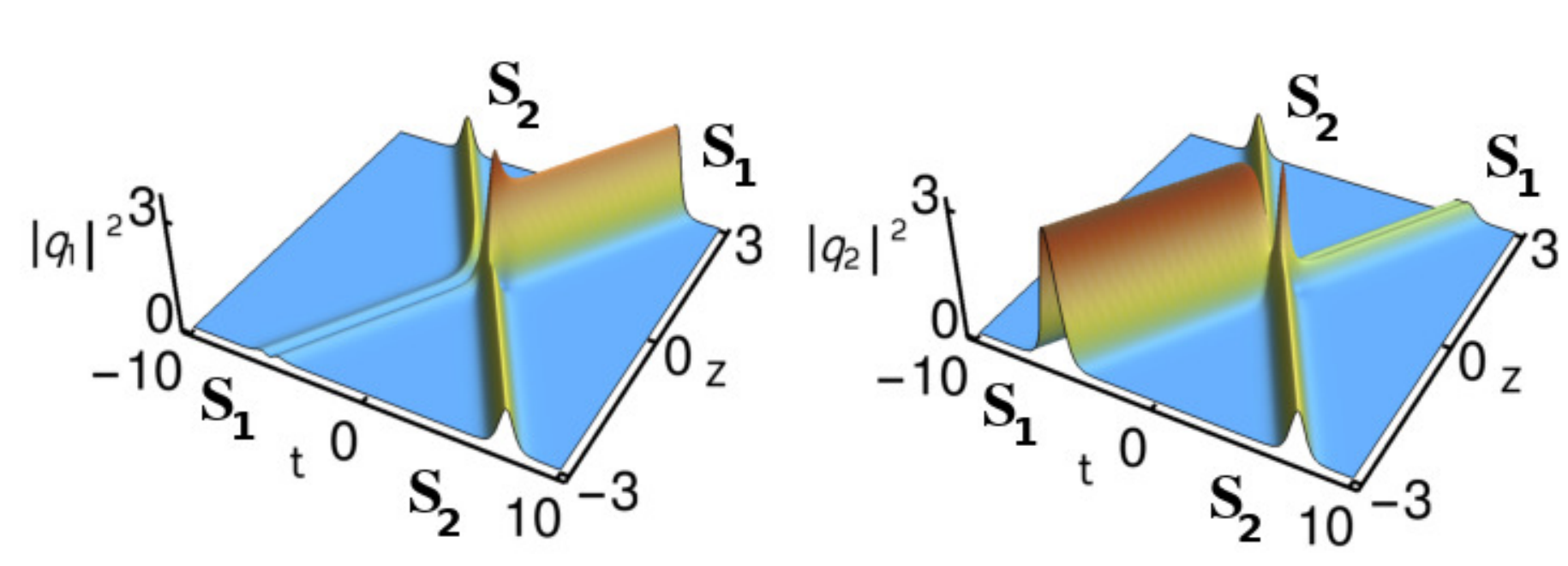}
			\caption{Energy switching collision between CCS and ICS in 2-CCNLS system \cite{kanna-jpa-2010,kanna-jpa-2011}.
				\label{fig14}}
		\end{center} 
	\end{figure} 
	%%%%%%%%%%%%%%%%%%%%%%%%%%%%%%%%%%%%%%%%%%%%%%%%%%%%%%
	\section{Fundamental vector bright solitons in GCNLS system}
	To construct both the nondegenerate and degenerate fundamental vector bright soliton solutions of the GCNLS system (\ref{11a})-(\ref{11b}), we consider the bilinear forms, $(iD_z+D^2_t)g^{(j)} \cdot f=0$, $j=1,2$, $D^2_t f \cdot f=2 (ag^{(1)}g^{(1)*}+cg^{(2)}g^{(2)*}+bg^{(1)}g^{(2)*}+b^*g^{(1)*}g^{(2)})$, which are result by sustituting the dependent variable transformation $q_j=\frac{g^{(j)}(z,t)}{f(z,t)}$, $j=1,2$, to Eqs. (\ref{11a}) and (\ref{11b}). Here $g^{(j)}$'s are complex functions and $f$ is a real function.
	By following the same procedure that has been outlined in section 4.1, we get the general form of nondegenerate fundamental bright soliton solution of the GCNLS system (\ref{11a})-(\ref{11b}) as \cite{ramakrishnan-jpa2}
	\bes\begin{eqnarray}
	&&\hspace{-1.5cm}q_{1}=\frac{g_1^{(1)}+g_3^{(1)}}{1+f_2+f_4}=\frac{1}{D}(\alpha_{1}^{(1)} e^{\eta_{1}}+e^{\eta_{1}+\xi_{1}+\xi_{1}^*+\nu_{11}}+e^{\eta_{1}+\eta_1^*+\xi_{1}+\nu_{12}}), \label{45a}\\ 
	&&\hspace{-1.5cm}q_{2}=\frac{g_1^{(2)}+g_3^{(2)}}{1+f_2+f_4}=\frac{1}{D}(\alpha_{1}^{(1)} e^{\eta_{1}}+e^{\eta_{1}+\xi_{1}+\xi_{1}^*+\nu_{21}}+e^{\eta_{1}+\eta_1^*+\xi_{1}+\nu_{22}}),\label{45b}\\
	&&\hspace{-1.5cm}D=1+e^{\eta_{1}+\eta_{1}^{*}+\delta_{1}}+e^{\eta_{1}+\xi_{1}^{*}+\delta_{2}}+e^{\eta_{1}^*+\xi_{1}+\delta_{2}^*}+e^{\xi_{1}+\xi_{1}^{*}+\delta_{3}}+e^{\eta_{1}+\eta_{1}^{*}+\xi_{1}+\xi_{1}^{*}+\delta_{4}}.\nonumber
	\end{eqnarray}\ees
	Here, $\eta_1=k_1(t+i k_1 z)$,  $\xi_1=l_1(t+i l_1 z)$,
	$e^{\nu_{11}}=\frac{c(k_{1}-l_{1})\alpha_{1}^{(1)}|\alpha_{1}^{(2)}|^2}{(k_{1}+l_{1}^*)(l_{1}+l_{1}^*)^{2}}$,
	$e^{\nu_{12}}=\frac{b^*(k_{1}-l_{1})\alpha_{1}^{(2)}|\alpha_{1}^{(1)}|^2}{(k_{1}+k_{1}^*)(l_{1}+k_{1}^*)^{2}}$,
	$e^{\nu_{21}}=-\frac{b(k_{1}-l_{1})\alpha_{1}^{(1)}|\alpha_{1}^{(2)}|^2}{(l_{1}+l_{1}^*)(k_{1}+l_{1}^*)^{2}}$,
	$e^{\nu_{22}}=-\frac{a(k_{1}-l_{1})\alpha_{1}^{(2)}|\alpha_{1}^{(1)}|^2}{(k_{1}+k_{1}^*)^2(l_{1}+k_{1}^*)}$, $e^{\delta_{1}}=\frac{a|\alpha_{1}^{(1)}|^2}{(k_{1}+k_{1}^{*})^{2}}$,    $e^{\delta_{2}}=\frac{b\alpha_{1}^{(1)}\alpha_{1}^{(2)*}}{(k_{1}+l_{1}^{*})^{2}}$, $e^{\delta_{3}}=\frac{c |\alpha_{1}^{(2)}|^2}{(l_{1}+l_{1}^*)^{2}}$ and 
	$e^{\delta_{4}}=\frac{|k_{1}-l_{1}|^{2} |\alpha_{1}^{(1)}|^{2}|\alpha_{1}^{(2)}|^{2}(ac|k_{1}+l_{1}^*|^2-|b|^2(k_{1}+k_{1}^*)(l_{1}+l_{1}^*))}{(k_{1}+k_{1}^*)^{2}(k_{1}^*+l_{1})^2(k_{1}+l_{1}^*)^2(l_{1}+l_{1}^*)^{2}}$.
	Under the restrictions, ($a=c=1$, $b=0$) and   ($a=1$, $c=-1$, $b=0$) the solution (\ref{45a})-(\ref{45b}) of GCNLS system exactly coincides with the nondegenerate one-soliton solution of the Manakov system and mixed 2-CNLS system, respectively. In the present GCNLS system, the properties of the nondegenerate fundamental bright soliton solution (\ref{45a})-(\ref{45b})  is determined by the four complex parameters $\alpha_1^{(j)}$, $j=1,2$, $k_1$ and $l_1$ apart from the system parameters $a$ (SPM), $c$ (XMP) and $b$ (four wave mixing effect). The nondegenerate one-soliton solution admits singularity whenever either one of the sign of SPM ($a$) and XPM ($c$)  is negative or both are negative . Additinally, the condition $(ac|k_{1}+l_{1}^*|^2-|b|^2(k_{1}+k_{1}^*)(l_{1}+l_{1}^*))>0$
	should also be maintained to obtain regular soliton solution of the GCNLS system. The solution exhibits a double-hump or a single-hump intensity profile for suitable choices of parameter values. Very surprisingly, like in the case of the $2$-CCNLS system, the presence of a four-wave mixing term and an additional wave number induces breather formation in the structure of nondegenerate fundamental soliton. A typical breathing behaviour along the $z$ direction is displayed in Figure \ref{fig15}. This kind of breathing soliton is not observed in the Manakov and mixed CNLS cases.
	\begin{figure}[H]
		\begin{center}
			\includegraphics[width=9.5 cm]{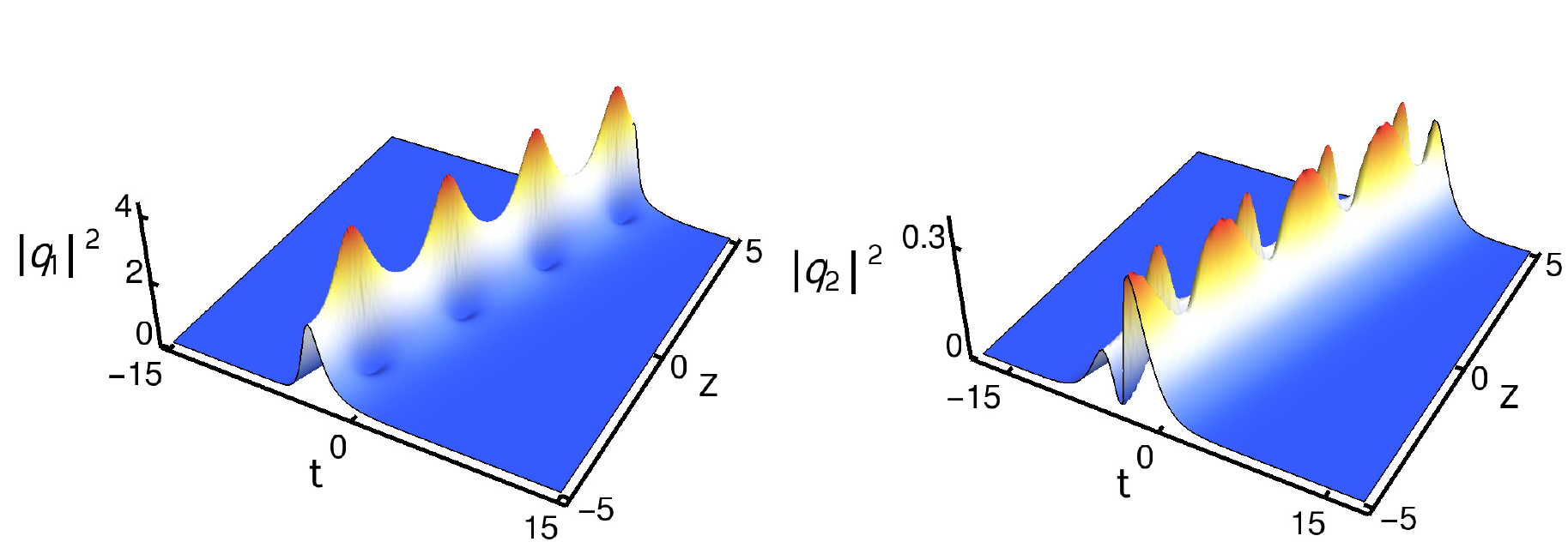}
			\caption{Breathing nondegenerate fundamental soliton in the GCNLS system. Here the parameters are $k_1=1.65+0.5i$, $l_1=0.45+0.5i$, $\alpha_{1}^{(1)}=0.35+0.35+i$, $\alpha_{1}^{(2)}=0.5+0.5i$, $a=c=1$ and $b=0.5-0.5i$.
				\label{fig15}}
		\end{center} 
	\end{figure}
	
	The degenerate bright soliton solution is recovered by incorporating the limit $k_1=l_1$ in the solution (\ref{45a})-(\ref{45b}). It leads to the following expressions of the degenerate bright soliton solution \cite{vpriya-cnsns}, $q_j= A_j k_{1R}\sech(\eta_{1R}+\frac{R_1}{2})e^{i\eta_{1I}}$, $A_j=\frac{\alpha_1^{(j)}}{(a|\alpha_1^{(1)}|^2+c|\alpha_1^{(2)}|^2+b\alpha_1^{(1)}\alpha_1^{(2)*}+b^*\alpha_1^{(1)*}\alpha_1^{(2)})^{1/2}}$, $e^{R_1}=\frac{(a|\alpha_1^{(1)}|^2+c|\alpha_1^{(2)}|^2+b\alpha_1^{(1)}\alpha_1^{(2)*}+b^*\alpha_1^{(1)*}\alpha_1^{(2)})}{(k_1+k_1^*)^2}$, $\eta_{1R}=k_{1R}(t-2k_{1I}z)$, $\eta_{1I}=k_{1I}t+(k_{1R}^2-k_{1I}^2)z$.
	The latter expressions ensure that the degenerate fundamental soliton always admits single-hump profile characterized by three complex constants $k_1$ and $\alpha_1^{(j)}$'s. The degenerate two-soliton solution can be easily obtained by replacing the form of $B_{ji}=\kappa_{ij}=\frac{(a\alpha_i^{(1)}\alpha_j^{(1)*}+c\alpha_i^{(2)}\alpha_j^{(2)*}+b\alpha_i^{(1)}\alpha_j^{(2)*}+b^*\alpha_j^{(1)*}\alpha_i^{(2)})}{(k_i+k_j^*)}$, $i,j=1,2$, in Eqs. (\ref{trans})-(\ref{2sol-cnls}).  With arbitrary values of $b$, the degenerate two solitons undergo two types of shape changing collisions corresponding to two different choices: (i) Manakov type shape changing collision for $a,c>0$, (ii) mixed 2-CNLS type shape changing collision for $a>0$, $c<0$. We do not provide the corresponding collision plots for brevity. We wish to point out that the degenerate bright solitons also undergo a special collision scenario, where the two degenerate solitons in each of the components do not pass through each other whereas they bounce off each other when they start to collide. This type of bright soliton collision scenario is referred to as soliton reflection  in the literature \cite{wang-stud,agalarov}.
	%%%%%%%%%%%%%%%%%%%%%%%%%%%%%%%%%%%%%%%%%%
	\section{Nondegenerate and degenerate bright solitons in two component LSRI system}
	Finally, we intend to construct the nondegenerate fundamental soliton solution for the two-component long-wave short-wave resonance interaction system, namely the 2-component Yajima-Oikawa system \cite{oikawa-ptp,stalin-pla}.  To derive the nondegenerate one-soliton solution we again  bilinearize Eq. (\ref{lsri}) through  the following dependent variable transformations, $ S^{(l)}(x,t)=\frac{g^{(l)}(x,t)}{f(x,t)}$, $l=1,2$, $L=2\frac{\partial^2}{\partial x^2}\ln f(x,t)$. By doing so, we get the following bilinear equations: 
	\bea
	D_1 g^{(l)}\cdot f=0, l=1,2, ~D_2 f\cdot f=\sum_{n=1}^{2} |g^{(n)}|^2,\label{12}
	\eea
	where $D_1\equiv iD_t+D_x^2$ and $D_2\equiv D_xD_t$. With the modified forms of seed solutions $
	g_1^{(1)}=\al_1 e^{\eta_1}$, $g_1^{(2)}=\ba_1 e^{\xi_1}$, $\eta_1=k_1x+ik_1^2t$, $\xi_1=l_1x+il_1^2t$, we find that the series expansions that are given in \cite{stalin-pla} get terminated as $g^{(l)}=\epsilon  g^{(l)}_1+\epsilon^3  g^{(l)}_3$, $f=1+\epsilon^2 f_2+\epsilon^4 f_4$. The explicit forms of the unknown functions lead to the following nondegenerate fundamental soliton solution, 
	\bes\begin{eqnarray}
	&&S^{(1)}=\frac{g_1^{(1)}+g_3^{(1)}}{1+f_2+f_4}=\frac{\alpha_{1}e^{\eta_{1}}+e^{\eta_{1}+\xi_{1}+\xi_{1}^{*}+\mu_{11}}}
	{1+e^{\eta_{1}+\eta_{1}^{*}+R_{1}}+e^{\xi_{1}+\xi_{1}^{*}+R_{2}}+e^{\eta_{1}+\eta_{1}^{*}+\xi_{1}+\xi_{1}^{*}+R_{3}}},\label{47a}\\
	&&S^{(2)}=\frac{g_1^{(2)}+g_3^{(2)}}{1+f_2+f_4}=\frac{\beta_{1}e^{\xi_{1}}+e^{\xi_{1}+\eta_{1}+\eta_{1}^{*}+\mu_{12}}}
	{1+e^{\eta_{1}+\eta_{1}^{*}+R_{1}}+e^{\xi_{1}+\xi_{1}^{*}+R_{2}}+e^{\eta_{1}+\eta_{1}^{*}+\xi_{1}+\xi_{1}^{*}+R_{3}}},\label{47b}\\
	&&L=\frac{2}{f^2}\bigg((k_1+k_1^*)^2e^{\eta_{1}+\eta_{1}^{*}+R_{1}}+(l_1+l_1^*)^2e^{\xi_{1}+\xi_{1}^{*}+R_{2}}+e^{\eta_{1}+\eta_{1}^{*}+\xi_{1}+\xi_{1}^{*}+R_{4}},\nonumber\\
	&&~~~~~~~~~+e^{2(\eta_{1}+\eta_{1}^{*})+\xi_{1}+\xi_{1}^{*}+R_1+R_{3}}+e^{\eta_{1}+\eta_{1}^{*}+2(\xi_{1}+\xi_{1}^{*})+R_2+R_{3}}\bigg),\label{47c}\\
	&&f={(1+e^{\eta_{1}+\eta_{1}^{*}+R_{1}}+e^{\xi_{1}+\xi_{1}^{*}+R_{2}}+e^{\eta_{1}+\eta_{1}^{*}+\xi_{1}+\xi_{1}^{*}+R_{3}})},	\nonumber
	%L=2\frac{\partial^2}{\partial x^2}\ln (1+e^{\eta_{1}+\eta_{1}^{*}+R_{1}}+e^{\xi_{1}+\xi_{1}^{*}+R_{2}}+e^{\eta_{1}+\eta_{1}^{*}+\xi_{1}+\xi_{1}^{*}+R_{3}}),\label{14}
	\end{eqnarray} \ees
	where  $ e^{\mu_{11}}=\frac{i\al_1|\ba_1|^2(l_1-k_1)}{2(k_1+l_1^*)(l_1-l_1^*)(l_1+l_1^*)^2}$, ~ $e^{\mu_{12}}=\frac{i\ba_1|\al_1|^2(k_1-l_1)}{2(k_1^*+l_1)(k_1-k_1^*)(k_1+k_1^*)^2}$,~
	$e^{R_1}=\frac{|\al_1|^2}{2i(k_1+k_1^*)^2(k_1-k_1^*)}$, $e^{R_2}=\frac{|\ba_1|^2}{2i(l_1+l_1^*)^2(l_1-l_1^*)}$, $e^{R_3}=-\frac{|\al_1|^2|\ba_1|^2|k_1-l_1|^2}{4|k_1+l_1^*|^2(k_1-k_1^*)(l_1-l_1^*)(k_1+k_1^*)^2(l_1+l_1^*)^2}$, $e^{R_4}=-2(k_1+k_1^*)(l_1+l_1^*)(e^{R_1+R_2}-e^{R_3})+((k_1+k_1^*)^2+(l_1+l_1^*)^2)(e^{R_1+R_2}+e^{R_3})$. The above nondegenerate one-soliton solution in the two-component LSRI system is also governed by the four arbitrary complex parameters $k_1$, $l_1$, $\alpha_1$ and $\beta_1$. The solution (\ref{47a})-(\ref{47c}) admits both regular and singular solutions. To get the non-singular solution, the quantities $e^{R_1}$, $e^{R_2}$ and $e^{R_3}$ should be positive definite. Consequently, the imaginary parts of the wave numbers $k_1$ and $l_1$ get restricted as $k_{1I}, l_{1I}<0$. Due to this reason, the nondegenerate soliton in the present LSRI system always propagate in the same direction. It has been shown in \cite{stalin-pla} that the velocity of the soliton is described by the imaginary parts of wave numbers $k_1$ and $l_1$. Then, the amplitudes of the nondegenerate soliton in the short-wave components $S^{(1)}$ and $S^{(2)}$ are found to be $4k_{1R}A_1\sqrt{k_{1I}}$ and $4l_{1R}A_2\sqrt{l_{1I}}$, respectively, where $A_1=\frac{i\alpha_1^{1/2}}{\alpha_1^{1/2*}}$ and $A_2=\frac{i\beta_1^{1/2}}{\beta_1^{1/2*}}$. From the expressions for the amplitudes we find that the nondegenerate one-soliton in the present LSRI system (\ref{lsri}) exhibits amplitude dependent velocity property like the KdV-soliton. The solution (\ref{47a})- (\ref{47b}) exhibits a double-hump, a flattop and a single-hump profiles depending on the appropriate choices of parameters. A typical asymmetric double-hump profile is illustrated in Figure \ref{fig16} with the parameter values $k_1=0.35-0.5i$, $l_1=0.315-0.5i$, $\alpha_{1}=0.5+i$, $\beta_1=0.45+0.5i$.
	\begin{figure}[H]
		\begin{center}
			\includegraphics[width=5.5 cm]{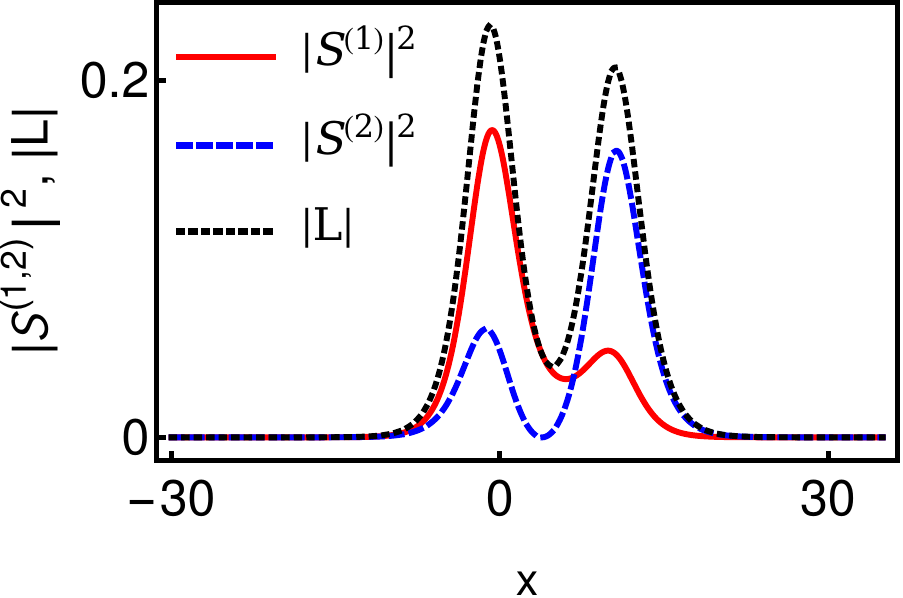}
			\caption{Asymmetric double-hump profile of the nondegenerate fundamental soliton in 2-component LSRI system. 
				\label{fig16}}
		\end{center} 
	\end{figure}
	
	We wish to point out that the explicit compact forms of higher-order nondegenerate soliton solutions have also been very recently obtained by us \cite{stalin-jpa}. Like in the Manakov system, we also find that the nondegenerate solitons in the present two-component LSRI system (\ref{lsri})  also in general exhibit three kinds of elastic collisions, namely shape preserving collision with zero phase shift and shape altering and shape changing collisions with a finite phase shifts. Remarkably, during the shape preserving collision, the two nondegenerate solitons pass through one another without any change in phase shift. In contrast to this collision scenario, the alteration in phase shift leads to a change in the profile structure of the solitons after collision. However, as we have demonstrated in the case of the Manakov system, the shape of the solitons will be restored after considering appropriate time shifts. In addition, the unity condition of the transition intensities also validates that both shape altering and shape changing collisions also belong to the case of elastic collision \cite{stalin-jpa}.   
	As in the case of the Manakov equation, here also we can identify partially nondegenerate two soliton, when the wave numbers satisfy the condition $k_1 = l_1$ and $k_2 \neq l_2$, as an example, and the collision of the nondegenerate soliton with the degenerate soliton exhibits novel energy exchange collision as demonstrated in \cite{kanna-pre2013}.
	
	We capture the degenerate soliton solution of Eq. (\ref{lsri}) by substituting the limit $k_1=l_1$ in Eq. (\ref{47a})-(\ref{47c}). This results in the following degenerate fundamental soliton forms \cite{kanna-pre2013}: $S^{(l)}=2A_lk_{1R}\sqrt{k_{1I}}e^{i(\eta_{1I}+\frac{\pi}{2})}\sech(\eta_{1R}+\frac{R}{2})$, $L=2k_{1R}^2\sech^2(\eta_{1R}+\frac{R}{2})$, $l=1,2$. Here $A_1=\frac{\alpha_1}{(|\alpha_1|^2+|\beta_1|^2)^{1/2}}$, $A_2=\frac{\beta_1}{(|\alpha_1|^2+|\beta_1|^2)^{1/2}}$, $\eta_{1R}=k_{1R}(t+2k_{1I}z)$, $\eta_{1I}=k_{1I}t+(k_{1R}^2-k_{1I}^2)z$, $e^R=\frac{-(|\alpha_1|^2+|\beta_1|^2)}{16k_{1R}^2k_{1I}}$.
	The degenerate soliton always admits a single-hump profile in both the SW components as well as in the LW component. The amplitude of the soliton in the SW and LW components are $2A_lk_{1R}\sqrt{k_{1I}}$, $2k_{1R}^2$, respectively. Their 
	velocity and the central position are identified as $2k_{1I}$ and $\frac{R}{2k_{1R}}$, respectively. From this, it is known that the degenerate bright soliton also exhibits the amplitude dependent velocity property, since the velocity explicitly appears in the amplitude part of the soliton. The explicit expression for the degenerate two bright soliton solution of the $2$-LSRI system (\ref{lsri}) can be identified from \cite{kanna-pre2013}.
	\begin{figure}[H]
		\begin{center}
			\includegraphics[width=13.0 cm]{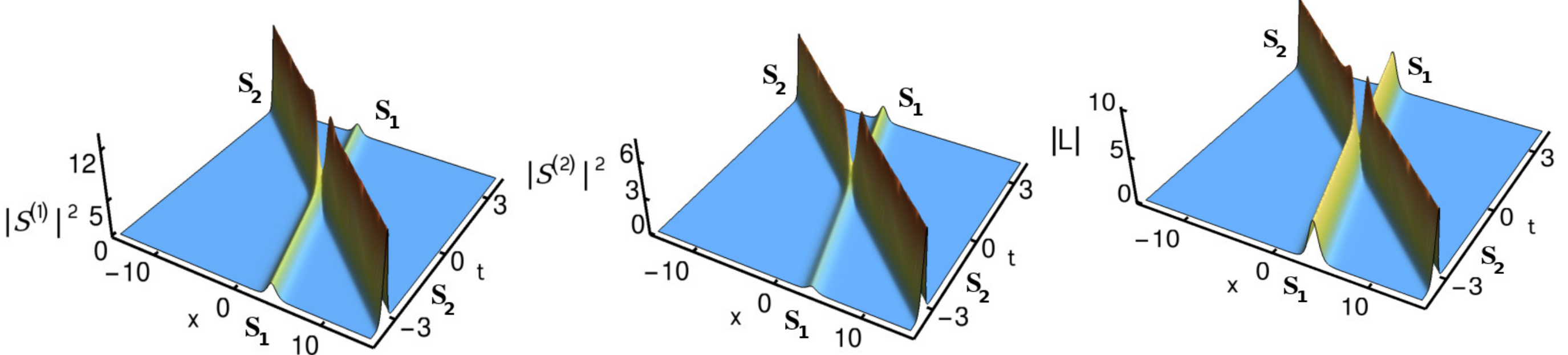}
			\caption{Energy sharing collision among the degenerate solitons in 2-component LSRI system. The parameter values are $k_1=1.5-0.5i$, $k_2=2-2i$, $\alpha_1^{(1)}=2.5$, $\alpha_1^{(2)}=1.2$, $\alpha_2^{(1)}=0.95$ and $\alpha_2^{(2)}=0.6$.
				\label{fig17}}
		\end{center} 
	\end{figure}
	
	As it has been demonstrated in Ref. \cite{kanna-pre2013}, the degenerate bright solitons undergo energy sharing collision through energy redistribution among the SW components. We demonstrate such energy sharing collision in Figure \ref{fig17}. It is evident from this figure, the intensity of the soliton $S_1$ is suppressed in the $S^{(1)}$ component after collision with the soliton $S_2$. And it gets enhanced in the second SW component $S^{(2)}$.  In order to hold the conservation of energy, the intensity of the soliton $S_2$ is enhanced in $S^{(1)}$ SW component and it gets suppressed in  $S^{(2)}$ SW component. However, in the degenerate case, the solitons in the LW component always undergo elastic collision. The standard elastic collision can occur in  both the SW components for the choice $\frac{\alpha_{1}}{\alpha_2}=\frac{\beta_1}{\beta_2}$ \cite{kanna-pre2013}.
	%%%%%%%%%%%%%%%%%%%%%%%%%%%%%%%%%%%%%%%%%%
	\section{Summary and outlook}
	
	In summary, we have shown that the coupled nonlinear Schr\"{o}dinger family of equations, namely the Manakov system or 2-CNLS system, $N$-CNLS system, mixed 2-CNLS system, 2-CCNLS system, GCNLS system and the $2$-component LSRI system, can admit a more general form of fundamental bright soliton solutions with non-identical propagation constants. In these systems, the obtained nondegenerate one-soliton solution admits novel geometrical structures which are not possible in the degenerate counterparts. Very surprisingly, the nondegenerate fundamental soliton in the $N$-CNLS system exhibits a novel intricate $N$-hump intensity profile. Then we have elucidated that the nondegenerate bright solitons possess novel collision properties. In particular, they exhibit  shape preserving, shape altering and shape changing collisions. However, by performing a careful asymptotic analysis, we found that all these three types of collision scenarios can be viewed as an elastic collision. For appropriate choices of parameters, they also exhibit energy sharing collision properties. Further, we have demonstrated that the degenerate vector bright solitons of all the CNLS systems can be captured by imposing appropriate constraints on the wave numbers. In addition to the above, we have also explained the various intriguing energy sharing collisions that occur between the degenerate vector bright solitons through graphical demonstration and analytical calculations. From the application point of view, the multi-hump nature of the nondegenerate solitons will be useful to enhance the flow of data  in multi-level optical communication
	applications. On the other hand, the energy sharing collision properties of the degenerate vector solitons are utilized to construct all the optical logic gates and it is also useful in optical switching device applications.          
	
We also wish to note here that the light pulse spread naturally occurs while it propagates in an optical fiber due to the intrinsic properties of the fibers.  This spreading or limitation usually occurs due to various fiber losses and fiber deformations. Practically, one cannot completely achieve stable propagation of information in laboratories. To overcome this difficulty a number of schemes have been proposed in the literature. Recently, the usage of dispersion managed solitons in optical communication has also been described to address this problem. In addition, the concept of soliton molecules and multi-soliton complexes have also been suggested to improve the data flow in optical fibers. In view of these facts, the multi-hump nature of the nondegenerate vector solitons is expected to be useful in enhancing the data flow in multi-level communication applications and in overcoming practical limitations.
	
	Although the existence of nondegenerate vector bright solitons have been pointed out in several CNLS family of equations, much remains to be uncovered, especially with higher-order nonlinear effects, such as third order dispersion, self-steepening and stimulated Raman scattering and so on. It is evident from our study much work is needed to study the collision properties associated with the newly derived vector solitons. From the current level of research activity, we believe that the area of nondegenerate vector solitons will continue to develop in future.
	
	%%%%%%%%%%%%%%%%%%%%%%%%%%%%%%%%%%%%%%%%%%
	%\section{Patents}
	
	%This section is not mandatory, but may be added if there are patents resulting from the work reported in this manuscript.
	
	%%%%%%%%%%%%%%%%%%%%%%%%%%%%%%%%%%%%%%%%%%
	\vspace{6pt} 
	
	%%%%%%%%%%%%%%%%%%%%%%%%%%%%%%%%%%%%%%%%%%
	%% optional
	%\supplementary{The following are available online at \linksupplementary{s1}, Figure S1: title, Table S1: title, Video S1: title.}
	
	
	%%%%%%%%%%%%%%%%%%%%%%%%%%%%%%%%%%%%%%%%%%
	\authorcontributions{Conceptualization, S.S., R.R. and M.L.; methodology, S.S. and R.R.; validation, S.S, R.R. and M. L.; 
		writing-original draft preparation, S.S. and R.R.; writing-review and editing, S.S.and M.L.; supervision, M.L.; project administration, M.L. ; funding acquisition, M.L. All the
		authors have read and agreed to the published version of the manuscript.}
	
	\funding{This research received no external funding.}

	\acknowledgments{The works of SS, RR and ML are supported by the DST-SERB Distinguished Fellowship program to ML under the Grant No. SB/DF/04/2017. RR is also grateful to Council of Scientific and Industrial Research, Government of India, for their support in the form of a Senior Research Fellowship (09/475(0203)/2020-EMR-I).}
	
	\conflictsofinterest{The authors declare no conflict of interest. } 
	
	%% Optional
	%\sampleavailability{Samples of the compounds ... are available from the authors.}
	
	%%%%%%%%%%%%%%%%%%%%%%%%%%%%%%%%%%%%%%%%%%
	%% Only for journal Encyclopedia
	%\entrylink{The Link to this entry published on the encyclopedia platform.}
	
	%%%%%%%%%%%%%%%%%%%%%%%%%%%%%%%%%%%%%%%%%%
	%% Optional
	%\abbreviations{The following abbreviations are used in this manuscript:\\
		
	%%%%%%%%%%%%%%%%%%%%%%%%%%%%%%%%%%%%%%%%%%
	%% Optional
	\appendixtitles{no} % Leave argument "no" if all appendix headings stay EMPTY (then no dot is printed after "Appendix A"). If the appendix sections contain a heading then change the argument to "yes".
	\appendixstart
	\appendix
	\section{Constants that appear in the asymptotic expressions in Section 4.4.1}
	%\subsection{}
	The various constants which arise in the asymptotic analysis of collision between degenerate and nondegenerate solitons in Section 4.4.1 are given below.
	\begin{eqnarray}
	&&\hspace{-1cm}e^{\Lam_1}=\frac{i\alpha_{1}^{(1)}(k_1-k_2)^{\frac{1}{2}}(k_1-l_2)^{\frac{1}{2}}(k_1^*+k_2)^{\frac{1}{2}}(k_1+k_1^*)(k_2+l_2^*)^{\frac{1}{2}}|k_1+l_2^*|^2}{\alpha_{2}^{(1)}(k_1^*-l_2^*)^{\frac{1}{2}}(k_2^*-l_2^*)^{\frac{1}{2}}}e^{R_5^*+\frac{R_3-R_6}{2}},\nonumber\\
	&&\hspace{-1cm}e^{\Lam_2}=\frac{(k_1-k_2)^{\frac{1}{2}}(k_2^*+l_2)^{\frac{1}{2}}(k_1+k_2^*)\hat{\Lam}_1\hat{\Lam}_2}{(k_1^*-k_2^*)^{\frac{1}{2}}(k_2^*-l_2^*)^{\frac{1}{2}}(k_1^*+k_2)},~ e^{\Lam_3}=\frac{|\alpha_{1}^{(1)}||\alpha_{1}^{(2)}|(k_1+k_1^*)(k_2+k_2^*)(l_2+l_2^*)}{|k_2-l_2|},\nonumber\\
	&&\hspace{-1cm}e^{\Lam_4}=(|\alpha_{1}^{(1)}|^2+|\alpha_{1}^{(2)}|^2)^{1/2}(|\alpha_{1}^{(1)}|^2|k_1-k_2|^2|k_1+l_2^*|^2+|\alpha_{1}^{(2)}|^2|k_1-l_2|^2|k_1+k_2^*|^2)^{1/2},\nonumber\\
	&&\hspace{-1cm}e^{\Lam_5}=\frac{|k_2+l_2^*|}{|k_2-l_2|}(|\alpha_{1}^{(1)}|^2|k_1+l_2^*|^2+|\alpha_{1}^{(2)}|^2|k_1-l_2|^2)^{1/2}(|\alpha_{1}^{(1)}|^2|k_1-k_2|^2+|\alpha_{1}^{(2)}|^2|k_1+k_2^*|^2)^{1/2},\nonumber\\
	&&\hspace{-1cm}e^{\Lam_6}=\frac{(k_1-l_2)^{\frac{1}{2}}(k_2+l_2^*)^{\frac{1}{2}}(k_1+l_2^*)\hat{\Lam}_3\hat{\Lam}_4}{(k_1^*-l_2^*)^{\frac{1}{2}}(k_2^*-l_2^*)^{\frac{1}{2}}(k_1^*+l_2)},~\hat{\Lam}_1=(|\alpha_{1}^{(1)}|^2(k_1-k_2)-|\alpha_{1}^{(2)}|^2(k_1^*+k_2))^{1/2},\nonumber\\
	&&\hspace{-1cm}e^{\Lam_7}=\frac{\alpha_{1}^{(2)}(k_1-k_2)^{\frac{1}{2}}(k_1-l_2)^{\frac{1}{2}}(k_1^*+l_2)^{\frac{1}{2}}(k_1+k_1^*)(k_2^*+l_2)^{\frac{1}{2}}|k_1+k_2^*|^2}{\alpha_{2}^{(2)}(k_1^*-k_2^*)^{\frac{1}{2}}(k_2^*-l_2^*)^{\frac{1}{2}}}e^{R_2^*+\frac{R_6-R_3}{2}},\nonumber	\\
	&&\hspace{-1cm}\hat{\Lam}_2=(|\alpha_{1}^{(1)}|^2(k_1-k_2)|k_1+l_2^*|^2-|\alpha_{1}^{(2)}|^2|k_1-l_2|^2(k_1^*+k_2))^{1/2},\nonumber\\
	&&\hspace{-1cm}\hat{\Lam}_4=(|\alpha_{1}^{(1)}|^2|k_1-k_2|^2(k_1^*+l_2)-|\alpha_{1}^{(2)}|^2(k_1-l_2)|k_1+k_2^*|^2)^{1/2},\nonumber\\
	&&\hspace{-1cm}\hat{\Lam}_3=(|\alpha_{1}^{(2)}|^2(k_1-l_2)-|\alpha_{1}^{(1)}|^2(k_1^*+l_2))^{1/2}, \nonumber\\
	&&\hspace{-1cm}e^{\frac{\Phi_{21}-\Del_{21}}{2}}=\frac{|\alpha_{2}^{(1)}|(k_1-k_2)(k_2^*-k_1^*)^{\frac{1}{2}}(k_2-l_2)^{\frac{1}{2}}}{(k_1+k_2^*)(k_2+k_2^*)(k_2+k_1^*)^{\frac{1}{2}}(k_2^*+l_2)^{\frac{1}{2}}},~e^{\frac{\lam_2-\lam_1}{2}}=\frac{|\alpha_{2}^{(2)}||k_1-l_2|(k_2-l_2)^{\frac{1}{2}}\hat{\Lam}_2}{(k_2+l_2^*)^{\frac{1}{2}}|k_1+l_2^*|^2(l_2+l_2^*)\hat{\Lam}_1}, \nonumber\\
	&&\hspace{-1cm}e^{\frac{\lam_5-R}{2}}=\frac{|k_1-k_2||k_1-l_2||k_2-l_2|\hat{\Lam}_5}{|k_1+k_2^*|^2|k_1+l_2^*|^2|k_2+l_2^*|(|\alpha_{1}^{(1)}|^2+|\alpha_{1}^{(2)}|^2)^{1/2}}e^{\frac{R_3+R_6}{2}},\nonumber\\
	&&\hspace{-1cm}e^{\frac{\vth_{12}-\varphi_{21}}{2}}=\frac{(k_2-k_1)^{\frac{1}{2}}(k_1^*-l_2^*)^{\frac{1}{2}}(k_2^*+l_2)^{\frac{1}{2}}}{(k_2+l_2^*)^{\frac{1}{2}}(k_2^*-k_1^*)^{\frac{1}{2}}(k_1-l_2)^{\frac{1}{2}}}e^{\frac{R_2^*+R_5-(R_2+R_5^*)}{2}},~e^{\frac{\lam_3-\lam_4}{2}}=\frac{|k_1-k_2|\hat{\Lam}_6|k_1+l_2^*|^2e^{\frac{R_3-R_6}{2}}}{|k_1+k_2^*|^2|k_1-l_2|\hat{\Lam}_7},\nonumber\\
	&&\hspace{-1cm}e^{\frac{\Gamma_{21}-\ga_{21}}{2}}=\frac{(k_2-l_2)^{\frac{1}{2}}(k_1-l_2)(k_1^*-l_2^*)^{\frac{1}{2}}}{(k_2+l_2^*)^{\frac{1}{2}}(k_1+l_2^*)(k_1^*+l_2)^{\frac{1}{2}}}e^{\frac{R_6}{2}},~e^{\frac{\lam_7-\lam_6}{2}}=\frac{(k_1-k_2)(k_2-l_2)^{\frac{1}{2}}\hat{\Lam}_4}{|k_1+k_2^*|^2(k_2^*+l_2)^{\frac{1}{2}}\hat{\Lam}_3}e^{\frac{R_3}{2}},\nonumber\\
	&&\hspace{-1cm}\hat{\Lam}_5= (|\alpha_{1}^{(1)}|^2|k_1-k_2|^2|k_1+l_2^*|^2+|\alpha_{1}^{(2)}|^2|k_1-l_2|^2|k_1+k_2^*|^2)^{1/2},\nonumber\end{eqnarray}\begin{eqnarray}
	&&\hspace{-1cm}e^{\frac{R'-\vsa_{22}}{2}}=\frac{|k_1-k_2||k_1-l_2|\hat{\Lam}_5}{|k_1+k_2^*|^2|k_1+l_2^*|^2(k_1+k_1^*)},~e^{\frac{\vsa_{22}}{2}}=\frac{|k_2-l_2|}{|k_2+l_2^*|}e^{\frac{R_3+R_6}{2}},~e^{\frac{R_3-R_6}{2}}=\frac{|\alpha_{2}^{(1)}|(l_2+l_2^*)}{|\alpha_{2}^{(2)}|(k_2+k_2^*)},\nonumber\\
	&&\hspace{-1cm}\hat{\Lam}_6=(|\alpha_{1}^{(1)}|^2|k_1-k_2|^2+|\alpha_{1}^{(2)}|^2|k_1+k_2^*|^2)^{1/2},~\hat{\Lam}_7=(|\alpha_{1}^{(1)}|^2|k_1+l_2^*|^2+|\alpha_{1}^{(2)}|^2|k_1-l_2|^2)^{1/2},\nonumber\end{eqnarray}\begin{eqnarray}
	&&\hspace{-1cm}e^{\frac{\Lam_{22}-\rho_1}{2}}=\frac{(k_2-l_2)^{\frac{1}{2}}}{(k_2+l_2^*)^{\frac{1}{2}}}e^{\frac{R_6}{2}},~e^{\frac{\mu_{22}-\rho_2}{2}}=\frac{(l_2-k_2)^{\frac{1}{2}}}{(k_2^*+l_2)^{\frac{1}{2}}}e^{\frac{R_3}{2}},~e^{R_1}=\frac{|\alpha_{1}^{(1)}|^2}{(k_1+k_1^*)^2},~e^{R_2}=\frac{\alpha_{1}^{(1)}\alpha_{2}^{(1)*}}{(k_1+k_2^*)^2},\nonumber\\
	&&\hspace{-1cm}e^{R_3}=\frac{|\alpha_{2}^{(1)}|^2}{(k_2+k_2^*)^2},~e^{R_4}=\frac{|\alpha_{1}^{(2)}|^2}{(k_1+k_1^*)^2},~e^{R_5}=\frac{\alpha_{1}^{(2)}\alpha_{2}^{(2)*}}{(k_1+l_2^*)^2},~e^{R_6}=\frac{|\alpha_{2}^{(2)}|^2}{(l_2+l_2^*)^2}.\nonumber
	\end{eqnarray}

	%%%%%%%%%%%%%%%%%%%%%%%%%%%%%%%%%%%%%%%%%%
\end{paracol}
\reftitle{References}

\end{document}